# Project Severe Weather Archive of the Philippines (SWAP). Part 2: Baseline Climatology of Close Proximity Soundings in Hailstorm Environments across Luzon, Philippines

Generich H. Capuli*,[1,2]

(1) Department of Science and Technology – Philippine Atmospheric, Geophysical and Astronomical Services Administration, Research & Development and Training Division – Numerical Modelling Section, Quezon City, Philippines
(2) Project Severe Weather Archive of the Philippines (SWAP), Quezon City, Philippines

Article history: received November 01, 2025; accepted January 17, 2026

## Abstract

The environments of severe thunderstorms that produced hail were examined using 171 proximity soundings (2005 2024) archived in the 3rd Data Release of Project SWAP. These soundings were categorized based on their geographical occurrence into three hail prone environments across Luzon, Philippines. For each case, key parameters describing instability, vertical wind shear, and moisture were calculated to assess the environmental conditions for hail production. The probability of hail occurrence, expressed as a function of $W_{MAX}$ ($\sqrt{2} \times CAPE$) and 0 6 km bulk shear (DLS), revealed patterns distinct from those reported in other regions. Hail events in Luzon were most likely under high CAPE conditions, where boundary-layer moisture was sufficient, mid- and low-level lapse rates were steep, and lifting condensation levels were high. Surprisingly, weak DLS was common across Luzon hail environments, diverging from existing severe weather climatologies, yet large DCAPE indicated environments conducive to damaging wind events. When DLS was replaced with the shear magnitude between the cloud base and equilibrium level, the probability of hail occurrence increased, better aligning with global severe weather climatologies. This finding is supported by hodograph analyses, which show largely unidirectional wind profiles: strong speed shear aloft but weak directional shear in the low levels. Parameters such as $W_{MAX}SHEAR$, $W_{MAX}SHEAR_{LCL\ EL}$, and $BWD_{LCL\ EL}$ emerge as potential discriminators between non severe and severe thunderstorms capable of producing hail, and as useful metrics for assessing convective storm severity in Luzon and possibly countrywide. Finally, two recurring severe setups conducive to hail were identified: (1) an easterly regime associated with trade winds, and (2) a westerly regime linked to the Asian summer monsoon.

Keywords: Severe Weather Events (SWE) – Hail Events; Convective; Thermodynamics; Kinematics; Baseline Climatology

# 1. Introduction

Whether thunderstorms produce any severe weather event (SWE) such as large hail, severe wind gusts, extreme rainfall, and tornadoes is an extremely important challenge for weather forecasters. These forecasts require knowledge of the environment of the storms, which can be obtained from radiosonde measurements or numerical weather prediction (NWP) models. Out of all these SWEs, hail stands out as the most economically costly, posing significant threats to both safety and property (Johnson and Sugden, 2014; Blair et al., 2017). Compared to tornadoes, literature on hail and its dependency on the near-storm environment is less abundant, especially in the Philippines.

The thermodynamic profile has been regarded as a fundamental component in hail forecasting. Hail growth critically depends on the presence of a favorable temperature regime (Nelson, 1983; Browning and Foote, 1976; Miller et al., 1988; Knight and Knight, 2001), as convective updrafts that fail to penetrate sub-freezing layers aloft are incapable of producing hail. The temperature layer between 0 °C and −20 °C, often referred to as the hail growth zone (HGZ), is considered optimal for hail formation (Nelson, 1983; Browning and Foote, 1976; Miller et al., 1988; Knight and Knight, 2001; Pilorz et al., 2022). This zone typically resides in the mid-troposphere, though its altitude varies across different storm types and environments. Both numerical cloud simulations (Brimelow et al., 2002; Kumjian and Lombardo, 2020) and observational studies (Knight and Knight, 2005) consistently demonstrate that hail growth occurs almost exclusively within sub-freezing layers. As hailstones descend into warmer air, melting ensues; an effect that disproportionately impacts smaller hailstones due to their slower terminal velocities and higher surface-area-to-mass ratios (Rasmussen and Heymsfield, 1987).

As stipulated by Johns and Doswell (1992), a straightforward way to characterize the environment of these hail-bearing and deep, moist convective storms is to assess the following three ingredients; (i) sufficient and quality low-level moisture, (ii) conditionally unstable temperature lapse rates in the midtroposphere, and (iii) sufficient lift to transport a potentially buoyant air parcel to its level of free convection. The presence of low-level moisture and midlevel conditional instability ensures that the lifted parcel has sufficient buoyancy to sustain a convective updraft, whereas the lift is required to initiate the storm. To further sustain storms and make them long-lived, strong vertical wind shear; particularly deep-layer shear (DLS), can be regarded as an additional, fourth, ingredient, especially important to well-organized storms such as supercell. This was demonstrated both in numerical studies (e.g., Weisman and Klemp, 1982), and in studies of storm environments (Rasmussen and Blanchard, 1998; Thompson et al., 2003, 2012). Smith et al. (2012) have shown that such well-organized storms are responsible for the vast majority of significant severe weather in the United States.

Three of the ingredients (i.e., all except lift), can be analyzed using radiosonde data. The combined presence of the ingredients low-level moisture and mid-level conditional instability through mid-level lapse rates results in convective available potential energy (CAPE), a proxy for updraft buoyancy. Previous proximity sounding climatology studies have indeed confirmed that severe weather probability increases with increasing vertical wind shear and with increasing CAPE, both across the United States (Rasmussen and Blanchard, 1998; Craven and Brooks, 2004; Brooks, 2009) and across Europe (Groenemeijer and van Delden, 2007; Brooks, 2009; Johnson and Sugden, 2014; Púčik et al., 2015; Taszarek et al., 2017, 2020) and have also shown some skill in hail forecasting as shown in these aforementioned studies. However, due to the limitations of parcel theory, CAPE can be an unrealistic approximation. And, the environmental conditions conducive to a particular hazard differ from one another.

While significant progress has been made in understanding thunderstorm and hail formation (Allen et al., 2020; Kumjian et al., 2021), substantial gaps remain in our knowledge of the dynamical and microphysical processes that govern storm evolution, intensification, and large hail production. These knowledge gaps continue to limit forecasting skills. Much of the current conceptual framework is derived from studies conducted in the United States, particularly over the Great Plains (e.g., Kumjian and Ryzhkov, 2008; Picca and Ryzhkov, 2012; Kaltenboeck and Ryzhkov, 2013; Snyder et al., 2015; Van Den Broeke, 2020), where the atmospheric conditions differ markedly from those in Southeast Asia (SEA), including the Philippines. Emerging climatological studies in the mentioned region highlight this distinction. For example, across Thailand and Indonesia, Mahavik et al. (2025) and Sari and Lasher-Trapp (2025) respectively show that storm environments in the tropics are typically characterized by ample convective instability (CAPE > 2000 J $kg^{-1}$), weaker DLS (5-10 m $s^{-1}$), and presumably weaker low-level storm-relative inflow ($V_{SR}$). Similarly, Capuli (2025), in a case study of the 13 August 2021 hail event in the Philippines, documented a favorable configuration of sufficient CAPE within and below the HGZ alongside weak low-level inflow.

Such environments are consistent with the framework proposed and results by Nixon et al. (2023), who emphasized the need for a balance between instability and the kinematic profile (both DLS and $V_{SR}$) to support hail growth in mid-level updrafts. They found that large CAPE below the HGZ enhances hail growth only when low-level storm-relative winds remain weak, and vice versa. Complementary observational studies also point to large-hail environments being associated with relatively weak low-level shear (LLS) within the lowest 1 km (Johnson and Sugden, 2014; Kumjian et al., 2019; Taszarek et al., 2020; Gutierrez and Kumjian, 2021; Nixon and Allen, 2022). On the other hand, idealized simulations suggest that excessive CAPE can generate unfavorable storm trajectories, reducing hailstone residence time within the HGZ (Lin and Kumjian, 2022). Therefore, these findings underscore the importance of environmental balance in hail production and suggest that such a configuration may be directly applicable to the tropics, including the Philippines, where it may serve as a baseline for characterizing the environments of hail-producing storms useful for discriminating against thunderstorms that produce hail or not.

Severe weather research goes back near the end of 1920s by Selga (1929) in the Philippines. Hail observations in particular were extremely sparse at this time, but some frequency of hail was noted near Northern Luzon and Greater Metro Manila were subject to such hail activity. In fact, hail reports within these areas are the highest across the archipelago and is the hotspot for such activity according to recent severe weather climatology constructed by Severe Weather Archive of the Philippines and its Part 1 study (SWAP; Capuli, 2024). It was also recently found out that the climatology of severe weather activity, including hail storm events, in the Philippines are found to have an annual cycle similar to the United States, with the bulk of hail activity occurring during the spring and summer; hot, dry season from March to May extending to the Asian summer monsoon (i.e., southwest monsoon) period (Capuli, 2024). These tend to occur in the afternoon and early evening; a common diurnal pattern across the tropics. It was also seen that orographic effects can compensate for weakly-sheared environments that generate hailstorms in the European continent (Allen et al., 2014). And it can be the same in the Philippines given that the Equator’s horizontal temperature gradient is very small, and upper-level air circulation is comparatively weak.

Given these considerations, the development of a baseline climatology is essential to support both operational thunderstorm forecasting and research, particularly for environments conducive to hail-producing severe storms. Sounding climatologies have long been employed to characterize storm environments and assess conditions favorable for specific thunderstorm types, including hail-bearing storms (e.g., Rasmussen and Blanchard, 1998; Púčik et al., 2015; Taszarek et al., 2020; Zhou et al., 2021). In operational practice, measures of convective instability such as CAPE are often described qualitatively as “marginal,” “large,” “extreme,” or “ample” etc. However, in the Philippines, no comprehensive baseline climatology exists to objectively support these quantifications either for traditional parameters or for more advanced, recently developed ones. This gap is particularly evident when considering the balanced instability-shear condition thought to be crucial for hail growth. Rather, they generally are based on the subjective experience and ‘‘mental calibration’’ of the weather forecasters. Motivated by these limitations and conducted as Part II of Project SWAP, this study establishes, for the first time, a baseline climatology of sounding-derived parameters relevant to hail-producing severe thunderstorms over the Luzon landmass. Specifically, the study aims to:

1) Characterize the types of hail environments across the Luzon landmass, accounting for variations in topographic setting,
2) Assess the climatological occurrence and distribution of physically important parameters relevant to operational meteorology, particularly hail forecasting,
3) Identify the climatologically large or extreme values of thermodynamic and kinematic parameters (including the introduction of new parameters) that can be useful in forecasting, and
4) Examine the thermodynamics and hodograph shapes associated with hail-supportive environments.

Within this framework, the study pursues three broader objectives: (1) to document the environments of severe thunderstorms across Luzon using proximity soundings, thereby identifying potential predictors for hail-producing storms; (2) to place these environments in the context of hail “hotspots” identified in Part I of Project SWAP; and (3) to compare the Philippine results with proximity-sounding studies conducted in the United States and Europe. A subsequent Part III of Project SWAP is envisioned to extend this climatological framework to tornadic environments across the Philippines.

The paper is organized as follows; Section 2 outlines the materials and methods, including the reanalysis and thunderhour dataset, the atmospheric profile parameters used in the study. In section 3, the results are introduced, with subsections dealing with individual parameters or their combinations. Section 4 is dedicated to the discussion of some of the results, while in section 5 I briefly summarize them.

# 2. Methodology

## 2.1 Hail Event Data

This study uses the 3rd Data Release (DR3) by the Project SWAP (Capuli, 2024) which amasses hail events from multiple sources, whether official and unofficial documentary sources, to establish a baseline climatology for hail-producing storms. Official sources include reports from the DOST-PAGASA, the National Disaster Risk Reduction and Management Council (NDRRMC), the Department of Social Welfare and Development-Disaster Response Operations Management, Information and Communication (DSWD-DROMIC), and various local government units. On the other hand, unofficial reports that include national and local news media, eyewitness accounts, and publicly available social media posts, contribute a substantial portion of the dataset. These sources are especially valuable in capturing hail events that are often undocumented by official agencies. In the community project, each report undergoes a three-stage verification process: (i) classification by source type, (ii) validation through consistency checks (time and location), and (iii) metadata tagging, which includes source link, platform, geo-coordinates, and any available visual documentation. Reports lacking photographic or videographic evidence are retained only if they attained at least one additional independent and credible source.

The imperfect nature of reporting will always affect report-based observational datasets such as those used in this study. Humans do not always accurately report hail times, locations, or sizes (Schaefer and Galway, 1982; Amburn and Wolf, 1997; Baumgardt, 2011; Allen et al., 2015; Allen and Tippett, 2015). In addition, the probability and density of hail reports is biased toward population density, where more people are available to make them (Dobur, 2005; Blair and Leighton, 2012; Groenemeijer et al., 2017; Allen et al., 2020). This can impact environmental parameters by leading to oversampling where observations are more frequent, potentially biasing against events which occur in more remote areas. Reports are also more likely during the day when more people are awake, which may exclude representative samples of nocturnal cases (Ashley et al., 2008; Blair et al., 2017). Hail reporting may also be deferred during life-threatening events (such as a tornado), possibly causing it to go unreported during tornadoes (Warren et al., 2021), but this has not yet been investigated in published literature.

Despite certain limitations, this citizen science approach offers a valuable, structured dataset that helps bridge a key observational gap in the Philippines. Similar frameworks have proven effective in severe weather research elsewhere e.g., Community Collaborative Rain, Hail and Snow network (CoCoRaHS; Cifelli et al., 2005, Reges et al., 2016) and Meteorological Phenomena Identification Near the Ground project (MPING; Elmore et al., 2014, 2022), demonstrating that, with proper verification and metadata tagging, community-contributed observations can meaningfully support climatological and severe weather studies, especially in data-sparse regions (Tan et al., 2022).

## 2.2 Environmenal Reanalysis Dataset

Environmental reanalysis data for this study were obtained from the fifth-generation European Centre for Medium-Range Weather Forecast (ECMWF) reanalysis (ERA5) accessed through the Climate Data Store (CDS; Hersbach et al., 2020). Reanalysis datasets, such as ERA5, are widely employed to diagnose atmospheric conditions conducive to severe weather such as tornadoes, damaging winds, and large hail. They are particularly useful for calculating instability indices and other parameters essential to SCS forecasting (Taszarek et al., 2020). Thus, has demonstrated good performance in depicting vertical profiles of convective environments, especially across the United States and Europe (Coffer et al., 2020; Taszarek et al., 2021a, Pilguj et al., 2022).

Any reanalysis product will have biases, due to its underlying model, the observations that have been assimilated, etc. However, the biases in ERA5 are in general reduced compared to those in ERA-Interim (Hersbach et al., 2020), Still, some known biases persist, particularly in the boundary layer. These include discrepancies in low-level parcel characteristics and vertical shear parameters, especially near surface boundaries and/or can be influenced by geographic location and surface elevation (King et al., 2019). Furthermore, Virman et al. (2021) did find an increase in the temperature and moisture bias of ERA5 versus ERA-Interim compared to observed rawinsonde soundings over the tropics, with ERA5 being too cold (~0.2-0.6 K) and too moist (~2-6%) in the low to mid-levels (~500-800-hPa). Despite these limitations, ERA5 remains among the most reliable and accessible datasets for investigating severe convective environments, globally (Coffer et al., 2020; Taszarek et al., 2021a, Taszarek et al., 2021b). Thus, while systematic ERA5 biases should be borne in mind, previous work supports ERA5 being a suitable reanalysis product for investigating severe weather environments, particularly when performing spatial averaging, which will decrease the impact of unbiased noise.

Vertical profiles were obtained from hybrid sigma-pressure level ERA5 data. To ensure comparable vertical resolutions upon compositing, profiles used in composites were interpolated to a common 250 m. This spacing was chosen to ensure a balance between computational cost and resolution of the environmental features relevant for hail. Many obstacles to an advanced understanding of severe thunderstorm environments exist when considering 'close' proximity soundings. Primarily, there is the nontrivial question of which time and space scales are most appropriate to represent the storm environment (Brooks et al., 1994). For example, Thompson et al. (2003) who used model soundings from Rapid Update Cycle-2 (RUC-2) analyses define a close proximity sounding as it was taken 40 km away from the storm and within 3-hr of a standard sounding time (i.e., 00 and 12 UTC, including special soundings on 06 and 18 UTC). These authors took into account the influence of severe thunderstorms such as supercells in the low-level environmental profile up to 30 km as demonstrated both by Beebe (1958) and Weisman et al. (1998). Here, I employed the approach of Nixon et al. (2023) to define close proximity soundings, extracting vertical profiles from the ERA5 grid point closest to the hail event. Because individual storms often generate widespread reports (hence with more latitude-longitude pairs), I 'consolidated' these into a single event location by determining the centroid of the cluster of report coordinates, ensuring a single reference point for the environmental analysis. This study does not employ a design as suggested in Potvin et al. (2010) to minimize the impacts of convective contamination using spatial offsetting, but it does aim for pre-convective conditions by using the analysis hour rounded down to the top of the hour preceding each report (in accordance with Thompson et al., 2012). For instance, a report at 2045 UTC would be represented with a 2000 UTC ERA5 sounding. All derived thermodynamic and kinematic variables used in this study were calculated from their respective raw hybrid sigma-pressure level ERA5 profiles using SounderPy (Gillett, 2025).

The parcel profiles were computed assuming a non-entraining, irreversible adiabatic process, following the recent formulation by Peters et al. (2022). Compared to the pseudoadiabatic ascent, which assumes all condensate is assumed to fall out of an air parcel immediately (Emanuel, 1994), the parcel calculation now accounts for the layer of mixed-phase condensate in which liquid and ice are present just below the triple point temperature. Additionally, an entraining, irreversible adiabatic ascent was also employed to compute the Entraining CAPE (ECAPE; Peters et al., 2023). ECAPE accounts for dilution of updraft plumes due to environmental entrainment, particularly under conditions of limited storm-relative inflow and/or mid- to upper-tropospheric dryness. Thus, providing a more realistic estimate of updraft intensity than the conventional, undiluted CAPE, which can overestimate buoyancy in environments where entrainment is significant. All undiluted Skew-T profiles are calculated using virtual temperature correction (Doswell and Rasmussen, 1994).

The thermodynamic variables used in this study use most-unstable (MU) and mixed-layer (ML) parcel profiles. CAPE (J $kg^{-1}$), level of free convection (LFC), and lifted condensation level (LCL) all assume the MU parcel, since this can be used in both surface-based and elevated storm scenarios and detects the degree of low-level stability, and the lowest potential cloud base, respectively, in both surface-based and elevated storm scenarios (e.g., in the case of a storm elevated above a stable layer, most-unstable CAPE is necessary to better approximate the maximum potential energy available to an updraft, but most-unstable CIN may considerably underestimate the CIN that this updraft experiences pulling parcels from the surface). Other significant heights include the freezing level (FZL) demarcating the HGZ and depth between the LCL-HGZ which can represent the CAPE below the HGZ. All heights are measured in meters unless otherwise specified. The 0 and −20 °C levels are used to calculate CAPE within the HGZ. This study also examines the relative humidity of the ambient air between 1-3 km ($RH_{13}$) and 1-6 km ($RH_{16}$), a proxy for lower-tropospheric moisture above the cloud base where entrainment matters most (Peters et al., 2019), as a percentage from 0 to 100%.

Kinematically, storm-relative winds were subject to assumptions on storm motion since neither observed storm motions, supercell type (right-moving or left-moving) nor storm mode (supercell, multicell cluster, squall line, etc.) were attained. Thus, estimated storm motions were calculated using the Bunkers ID Method (B2K; Bunkers et al., 2000), also an important component of ECAPE. For northern hemisphere countries like the Philippines, the right-moving vector was assumed (storm moving towards the right of non-pressure weighted mean wind). Other kinematics such as vertical wind shear parameters (i.e., 0-1, 0-6 km AGL, LCL-EL shear) were computed as is throughout the hodograph. A list of meteorological parameters and their meanings were presented in Table 1.

**Table 1.** Parameters used in the study, including their abbreviations and units.

| Parameter Definition | Abbreviation | Units |
|---|---|---|
| *Thermodynamic Parameters* | | |
| Most unstable Convective Available Potential Energy of any parcel in the lowest 400-hPa | MUCAPE | $J\ kg^{-1}$ |
| Most unstable CAPE between 0 °C and −20 °C | $MUCAPE_{HGZ}$ | $J\ kg^{-1}$ |
| Most unstable CAPE below 0 °C | $MUCAPE_{<HGZ}$ | $J\ kg^{-1}$ |
| Undiluted Updraft Velocity, most unstable parcel | $W_{MAX}$ | $m\ s^{-1}$ |
| 100-hPa Mixed-layer CAPE | MLCAPE | $J\ kg^{-1}$ |
| 100-hPa Mixed-layer CAPE between 0 °C and −20 °C | $MLCAPE_{HGZ}$ | $J\ kg^{-1}$ |
| 100-hPa Mixed-layer CAPE below 0 °C | $MLCAPE_{<HGZ}$ | $J\ kg^{-1}$ |
| Entraining CAPE, most unstable parcel | ECAPE | $J\ kg^{-1}$ |
| Downdraft CAPE, lowest 400-hPa | DCAPE | $J\ kg^{-1}$ |
| Lifted Condensation Level, most unstable parcel | LCL | m |
| Level of Free Convection, most unstable parcel | LFC | m |
| 0-3 km Temperature Lapse Rate; Low-level Lapse Rate | $LR_{03}$ (LLR) | $°C\ km^{-1}$ |
| 3-6 km Temperature Lapse Rate; Mid-level Lapse Rate | $LR_{36}$ (MLR) | $°C\ km^{-1}$ |
| *Moisture Parameters* | | |
| Mean 1-3 km Relative Humidity | $RH_{13}$ | frac/% |
| Mean 1-6 km Relative Humidity | $RH_{16}$ | frac/% |
| Precipitable Water from surface to 400-hPa | PWAT | mm |
| Freezing Layer | FZL | m |
| Difference between the FZL and LCL | ΔFZL | m |
| *Wind Parameters* | | |
| 0-1 km Bulk Wind Shear; Low-level Shear | $BWD_{01}$ (LLS) | $m\ s^{-1}$ |
| 0-6 km Bulk Wind Shear; Deep-layer Shear | $BWD_{06}$ (DLS) | $m\ s^{-1}$ |
| Cloud-layer Bulk Wind Shear | $BWD_{LCL\text{-}EL}$ | $m\ s^{-1}$ |
| Mean 0-1 km Storm-relative Wind | $V_{SR}$ | $m\ s^{-1}$ |
| *Composite Parameters* | | |
| Hail Size Index | HSI | cm |
| Undiluted Updraft Velocity × $BWD_{06}$ | $W_{MAX}SHEAR$ | $m^2\ s^{-2}$ |
| Undiluted Updraft Velocity × $BWD_{LCL\text{-}EL}$ | $W_{MAX}SHEAR_{LCL\text{-}EL}$ | $m^2\ s^{-2}$ |

## 2.3 Thunder Hour Data

A thunder hour is defined as an hour during which thunder can be heard (e.g., Jayaratne and Ramachandran, 1998; Bourscheidt et al., 2012). Traditionally, thunder hours would be recorded by trained human observers. Unfortunately, human observation of thunder hours is spatially sporadic and inconsistent, and impossible in high density on a global scale. Further, the distance at which thunder can be heard by a human observer is dependent on environmental conditions, as well as the training of the human. In this study, the thunder hour observations were derived using the observations gathered from the Earth Networks Global Lightning Network (ENGLN). The ENGLN dataset consists of lightning observations from nearly 2,000 Earth Networks Total Lightning Network (ENTLN) lightning sensors along with observations from over 70 World Wide Lightning Location Network (WWLLN) sensors. ENTLN utilizes ground-based broadband (1 Hz to 12 MHz) electric field change sensors to detect and locate both intra-cloud (IC) and cloud-to-ground (CG) lightning up to 1,500 km from the sensors. Meanwhile, WWLLN consists of ground-based very low-frequency (VLF) electric field sensors to locate primarily CG lightning on a global scale (Rodger et al., 2006; Hutchins et al., 2012).

Provided by DiGangi et al. (2022), the thunder hour calculation was made using a 0.05° × 0.05° latitude-longitude grid. For each grid point and for each hour, if at least two lightning pulses were located within 15 km of that point, the grid value was set to true (or 1), otherwise the grid value was false (or 0). The two-pulse criteria helps prevent spurious detections from having a significant impact on the results, although the results were not strongly dependent on this criteria. The climatology of thunderstorm activity and its computation was made for all grid points and all UTC and local solar time (LST) hours from 1 January 2014 to 31 December 2024. The gridded data were then aggregated by calendar month. For each hour in a calendar month for all 11 years analyzed, the probability of thunder being observed at each grid cell was calculated. This is simply $N_{\mathrm{TH}}/N_{\mathrm{TOT}}$, where $N_{\mathrm{TH}}$ is the number of observations of thunder in a given hour of a given calendar month, and $N_{\mathrm{TOT}}$ is the total number of observations of the same hour of that calendar month.

## 2.4 Definition of Hail Environments in Luzon

In Project SWAP Part I, Capuli (2024) identified the hotspots of hail events across Luzon. Given this basis, I created key areas to study the environmental setups capable of hail-bearing storms. I simply call this as Hail Environments with prefixes 1, 2, and 3 (HE1, HE2, and HE3).

HE1 is composed of the entire high plains of Northern Luzon. Geographically, it was defined by mountainous regions in the Cordillera Mountains and the upper-section of the Sierra Madre Mountain Range (SMMR) with elevation of more than 1000 mASL. At the tail end of the sector, the coastal province of Pangasinan that opens along the Lingayen Gulf was included and considered generally as flat land. Meanwhile, HE2 encompasses the entire Central Luzon provinces down towards the Metropolis (or the National Capital Region). These areas lie between the chain of mountains known as the Zambales Mountain Range to the west with the Manila Bay opening to the West Philippine Sea and the SMMR to the east. It is worth noting that the tornado hotspot was also saddled in this area as found in the initial part of this study (Capuli, 2024). The sector also includes major river basins and mountainous boundaries that influence local weather patterns, storm dynamics, and flood susceptibility (Lagmay et al., 2015). Lastly, HE3 encompasses Southern Luzon, with notable hotspots across the provinces of Cavite and Batangas. The geographic features of this sector include a blend of semi-rural and peri-urban lowlands, extensive coastal plains, and elevated upland areas such as Tagaytay City. At its core lies Taal Volcano, set within Taal Lake, which serves as a prominent geologic feature influencing the region.

Some of the key provinces in HE2 and HE3 are closely interconnected and collectively referred to as the Greater Metro Manila Region (GMMR). This area is distinguished by its high population density, rapid urban expansion, and complex land-use dynamics. Serving as the nation's primary economic and political hub, the GMMR is characterized by sprawling metropolitan centers interwoven with peri-urban communities and remaining agricultural landscapes. The combination of dense urban development and concentrated population makes this sector particularly vulnerable to severe convective hazards, with hail events observed in notable abundance (Capuli, 2024). While the climatological environment contributes to the frequency of these occurrences (as we shall see), the extensive population density amplifies both the exposure and the documentation of hail impacts in this region.

## 2.5 Statistical Evaluation

For each identified hail environment, convective parameters derived from Skew-T profiles (as detailed in Section 2.2) were evaluated to characterize their statistical distributions. To ensure a robust analysis of variability, key percentiles, specifically the $10^{th}$, $25^{th}$, median, $75^{th}$, and $95^{th}$, were calculated. These distributions were visualized using 'Raincloud' plots, a hybrid approach that combines Kernel Density Estimation (KDE), boxplots, and jittered raw data to minimize the loss of distributional detail often obscured by standard boxplots alone. Parallel to this statistical evaluation, angle-averaged composite Skew-T profiles were generated to distinguish the representative vertical thermodynamic and kinematic structures associated with each located hail environment.

Furthermore, to quantify the shape of these distributions beyond simple measures of central tendency, the skewness ($g_1$) and kurtosis ($g_2$) were computed. These higher-order moments provided critical insights into the asymmetry and 'tailedness' of the data, highlighting significant distributional differences between the hail environments in Luzon. In particular, $g_1$ is computed as the Fisher-Pearson coefficient of skewness:

$$g_1 = \frac{m_3}{m_2^{3/2}}$$

where

$$m_1 = \frac{1}{N}\sum_{n=1}^{N}(x[n] - \bar{x})^i$$

is the biased sample $i$ central moment, and $\bar{x}$ is the sample mean.

The interpretation of $g_1$ follows standard conventions: a positive value ($g_1 > 0$) indicates a right/positively-skewed distribution, while a negative value ($g_1 < 0$) indicates a left/negatively-skewed distribution. Distributions with skewness values falling within the range of $-0.5 < g_1 < 0.5$ were classified as approximately symmetric. Distributions were further classified based on the peakedness of the distribution relative to a normal distribution: values greater than 0.5 ($g_2 > 0.5$) indicated a Leptokurtic distribution (sharp peak with heavy tails), while values less than $-0.5$ ($g_2 < -0.5$) indicated a Platykurtic distribution (flat peak with light tails). Values falling within the range of $-0.5 < g_2 < 0.5$ were categorized as Mesokurtic, representing a normal-like distribution. A summary of the statistical characteristics e.g., skewness and kurtosis statistics, for each climatological convective distribution is presented in Table 2.

# 3. Results

## 3.1 Spatio-Temporal Analysis of Hail Events

In order to establish a baseline climatology of hail environments, it is required first to understand the spatial and temporal domain of these identified sectors. As an update to climatology of SWEs done in Part 1 and the initiatives of Project SWAP, Fig. 1 shows the climatological spatial distribution of (a) Luzon hail events in each hail environment and (b) Thunder Hours across Luzon landmass. The population density effect on hail and other forms of severe weather activity detection is well known (Anderson, 2007) and seen in this case where many cases were reported between sub-urbanized and highly-urbanized areas (Fig. 1a), similar to the prior work e.g., Southern Tagalog encompassing both the provinces of Cavite and Batangas, Metro Manila, Bulacan province, Pampanga province, Benguet province (Capuli, 2024). In addition to the population factor, the improvement on the internet coverage and smartphone devices can explain the increase in the number of reported hail in recent years. This situation implies that the actual number of cases, especially when going back several years ago, has been underestimated.

In relation to these hail events, it coincides with high thunderstorm activity climatologically, as parameterized using Thunder Hours, along the Luzon landmass and in these hail environments (Fig. 1b). DiGangi et al. (2022) asserted that the Philippines, alongside other neighboring states in the southeast asian region, is one of the most active lightning regions around the globe. Sure enough, the western portion of Northern Luzon down towards the entire

**Table 2.** Statistical evaluation of convective parameters' climatological distribution in this study.

| Parameter | HE1 | | HE2 | | HE3 | |
|---|---|---|---|---|---|---|
| | $g_1$ | $g_2$ | $g_1$ | $g_2$ | $g_1$ | $g_2$ |
| MUCAPE | −0.51 | 0.24 | −0.91 | 0.71 | 0.56 | −0.92 |
| $MUCAPE_{03}$ | −0.07 | 0.15 | −0.67 | 0.43 | 0.49 | 0.22 |
| $MUCAPE_{HGZ}$ | −0.38 | −0.11 | −1.00 | 1.22 | 0.66 | 0.30 |
| MLCAPE | −0.63 | 0.21 | −0.44 | −0.12 | −0.18 | −0.27 |
| $MLCAPE_{03}$ | 0.23 | −0.46 | 0.38 | 1.24 | 0.67 | 1.39 |
| $MLCAPE_{HGZ}$ | −0.19 | −0.42 | −0.44 | 0.36 | −0.01 | 0.51 |
| ECAPE | −0.15 | −0.72 | −0.53 | −0.02 | 0.06 | −0.71 |
| DCAPE | 0.16 | −0.16 | 0.43 | −0.59 | 0.69 | 0.74 |
| $W_{MAX}$ | −0.88 | 1.34 | −1.34 | 2.20 | 0.44 | −1.00 |
| LCL | 0.83 | 0.42 | 0.80 | 0.48 | 1.25 | 1.61 |
| $LR_{03}$ (LLR) | −0.10 | 0.60 | 0.02 | 0.52 | −0.30 | 0.19 |
| $LR_{36}$ (MLR) | 0.09 | −0.77 | 0.03 | 0.82 | 0.27 | −0.68 |
| $BWD_{01}$ (LLS) | 3.12 | 16.04 | 1.83 | 5.57 | 0.70 | −0.09 |
| $BWD_{06}$ (DLS) | 0.72 | 0.08 | 2.42 | 10.95 | 0.53 | −0.42 |
| $BWD_{LCL-EL}$ | 0.55 | −0.39 | 0.50 | −0.69 | 0.42 | −0.58 |
| $V_{SR}$ | −0.31 | 0.73 | 0.67 | 6.34 | 0.28 | 0.39 |
| $RH_{13}$ | −0.91 | 1.52 | −0.42 | 0.39 | −0.50 | 0.33 |
| $RH_{16}$ | −0.06 | −0.14 | −0.39 | 0.39 | 0.09 | −0.49 |
| PWAT | −0.55 | −0.32 | −0.53 | 0.57 | −0.05 | −0.96 |
| FZL | −0.86 | −0.43 | −0.81 | 0.08 | −0.94 | 0.90 |
| ΔFZL | −0.32 | −0.49 | −1.18 | 2.12 | 0.03 | −0.16 |
| HSI | 0.57 | 1.01 | 2.68 | 15.60 | 0.52 | 0.22 |

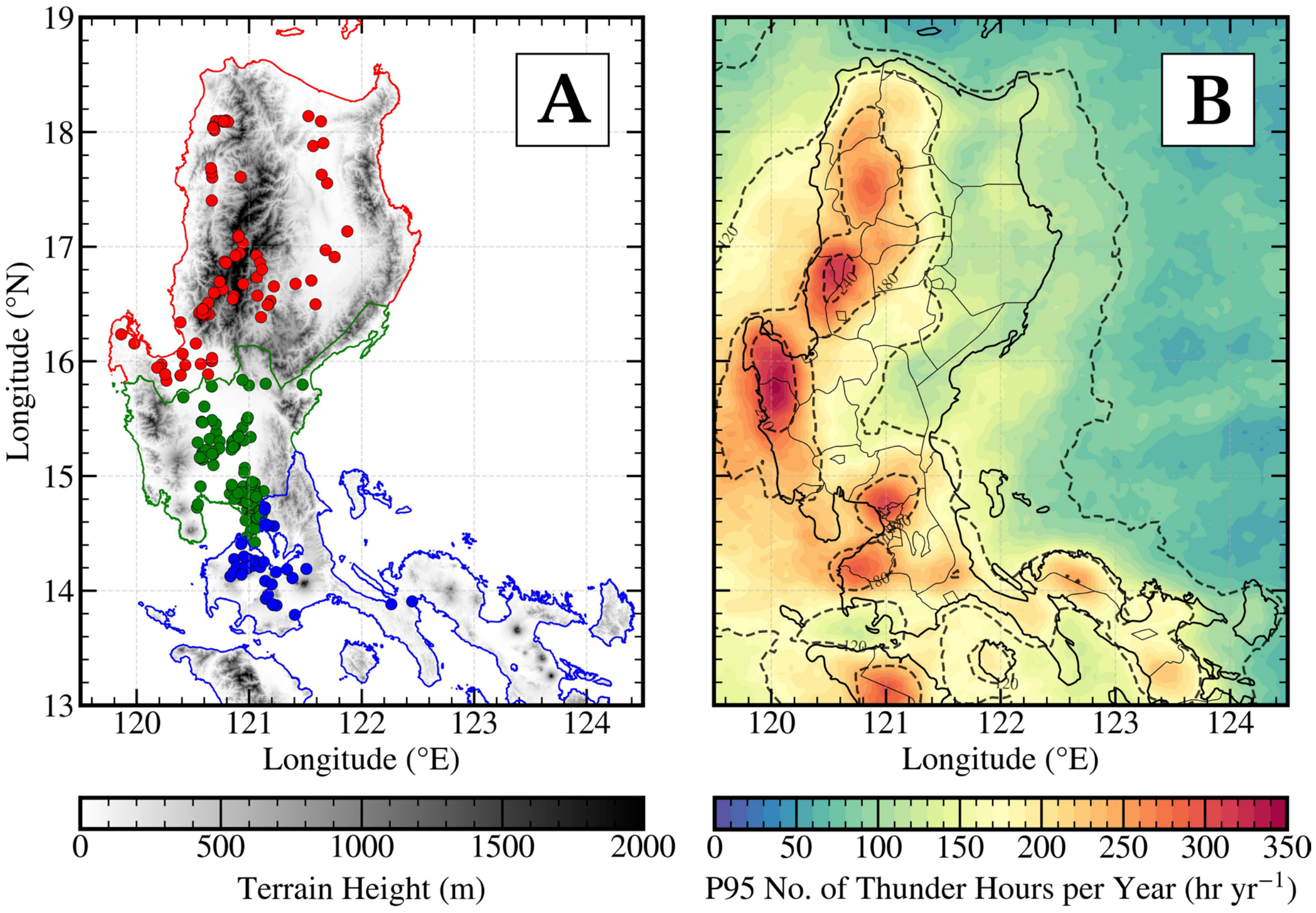

**Figure 1.** (a) Spatial Distribution of Hail Events across Luzon, Philippines. Red circles are events for HE1, Green circles are associated with HE2, and Blue circles are for HE3. (b) 12 yr Climatology of Thunder Hours (hr yr$^{-1}$) at P95 (contour) and average (dashed lines).

Central and Southern Luzon, including Metro Manila, all experience annual thunder hours > 100 hr yr$^{-1}$. There are several hotspots of thunderstorm activity (including severe ones) identified across the Luzon landmass whose average and 95$^{th}$ Percentile (P95) annual thunder hours reach close to more than 200 hr yr$^{-1}$; (1) the entire stretch of Northern High Plains of Cordilleras, (2) western coast of Zambales/Pangasinan area along the Mountain range of Zambales, and (3) GMMR encompassing the province of Bulacan, Cavite, and Batangas, and National Capital Region. Unsurprisingly, these areas overlay our hail environments identified and so as where the hail events occur in prior Fig. 1a. They follow the typical sea breeze convective pattern year round, with probabilities peaking over land during the day. Land-sea breeze oscillation combined with complex terrain features and localized convergence zones determines which direction these storms propagate (DiGangi et al., 2022; Capuli, 2025), although large-scale flow such as southwesterlies or in close proximity of a tropical cyclone can influence the storm motion.

In the temporal domain, increasing hailstorm activity is seen with peak time activity in the afternoon (06-08 UTC) as shown in Fig. 2. Severe weather season entailed to hail events tend to occur during months of March all the way to October, with primary peak by May and a minor peak in October supporting previous Part 1 of the study (Capuli, 2024). The aforementioned author also indicated that these hail events can be associated with the prevailing wind pattern during that time of year i.e., easterlies. The increase in hail activity across the Luzon landmass can also be attributed to the increase in thunderstorm activity, including those of severe ones, in each identified sector of the hail environment (Fig. 3). In general, the uptick in $N_{TH}$ from mid-noon up to the evening with a clear peak in activity between 06-08 UTC ($N_{TH}$ > 1000 counts) was recorded inline to observational and temporal analysis above. Significant increase in monthly $N_{TH}$ was depicted from $N_{TH}$ < 5000 counts by month of April to $N_{TH}$ > 8000 counts by month of May to September which subtly peaks by June then clears off by October. The monthly temporal distribution of $N_{TH}$ coincides as well with the peak of monthly hail events in each identified environment.

Detailed inspection, as revealed in Fig. 4, of the entire hail events shows the severe weather season related to hail activity that peaks in the month of May, similar to prior analysis in Fig. 2. Along with the hail activity, the climatology of thunder hours also reveals an increase in average thunderstorm activity by previously-mentioned month at 30.40 hr mon$^{-1}$ (P95: 46.1 hr mon$^{-1}$) which peaks by month of June at 32.25 hr mon$^{-1}$ (P95: 49.74 hr mon$^{-1}$). In HE1, the hail activity subtly increases by April then slowly terminates by July with a small peak in October.

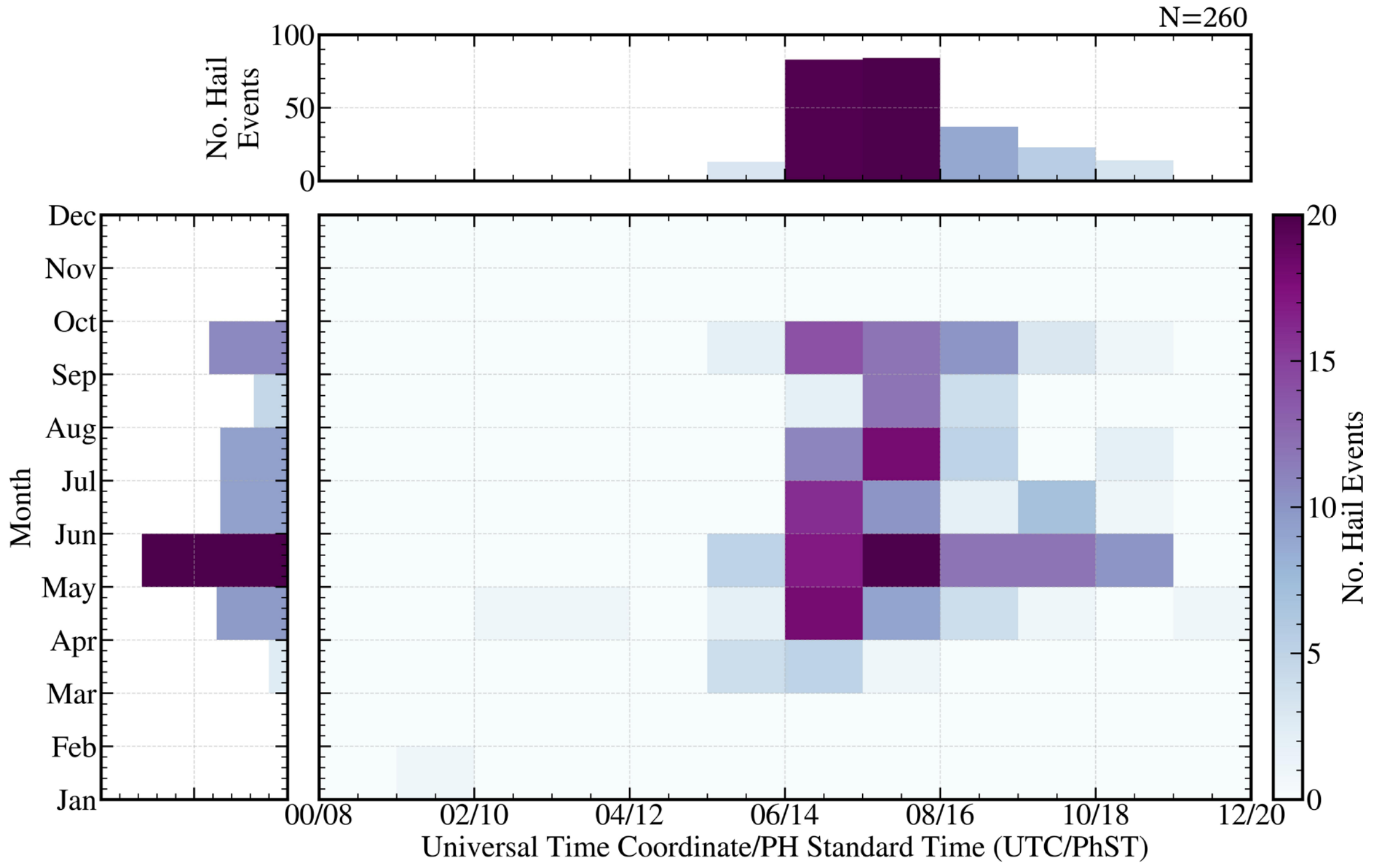

**Figure 2.** 2D Histogram (joint distribution) of Hail Events across Luzon, Philippines.

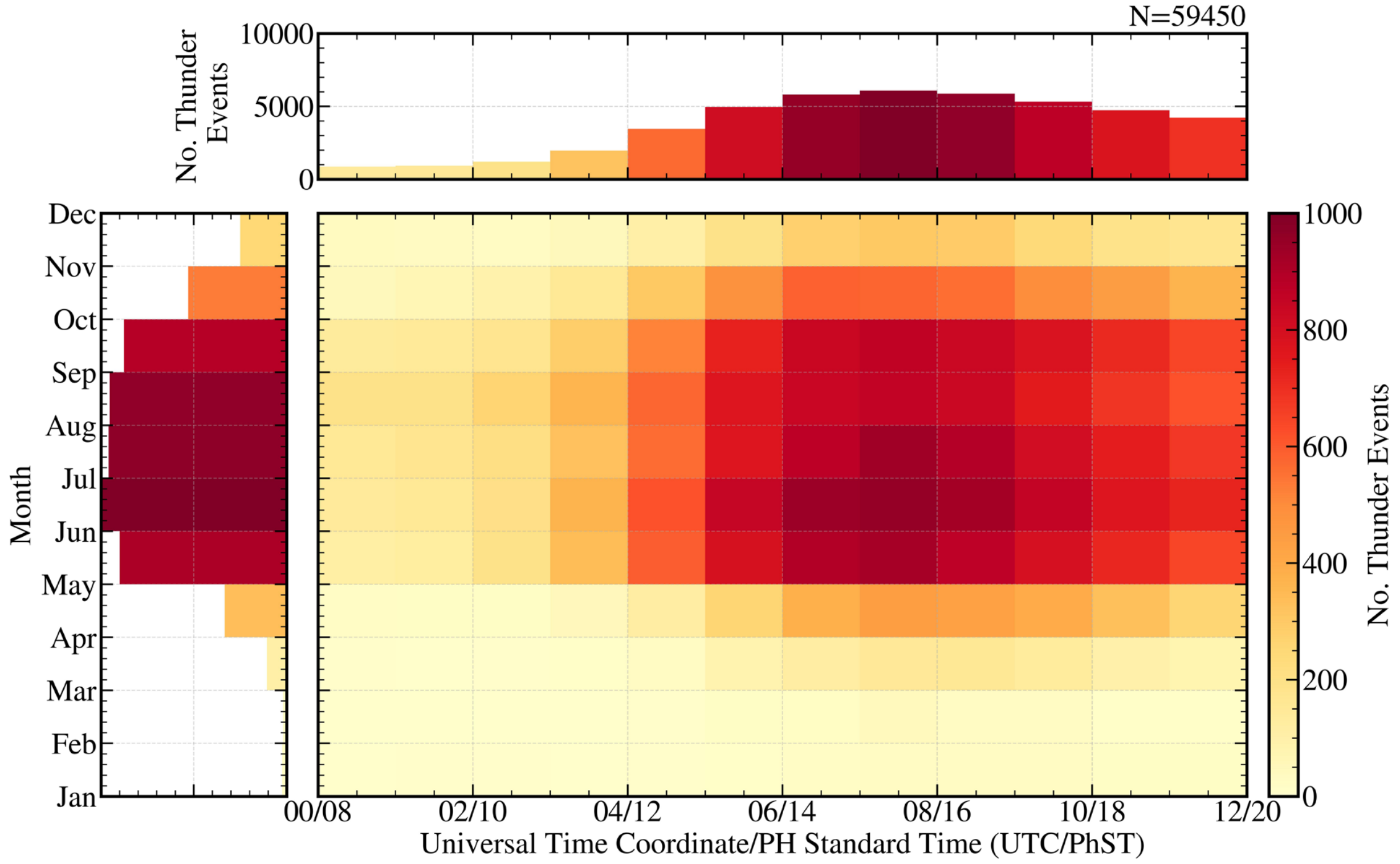

**Figure 3.** Same as Fig. 2, but showing thunder counts (a proxy for thunderstorm events) across Luzon, Philippines.

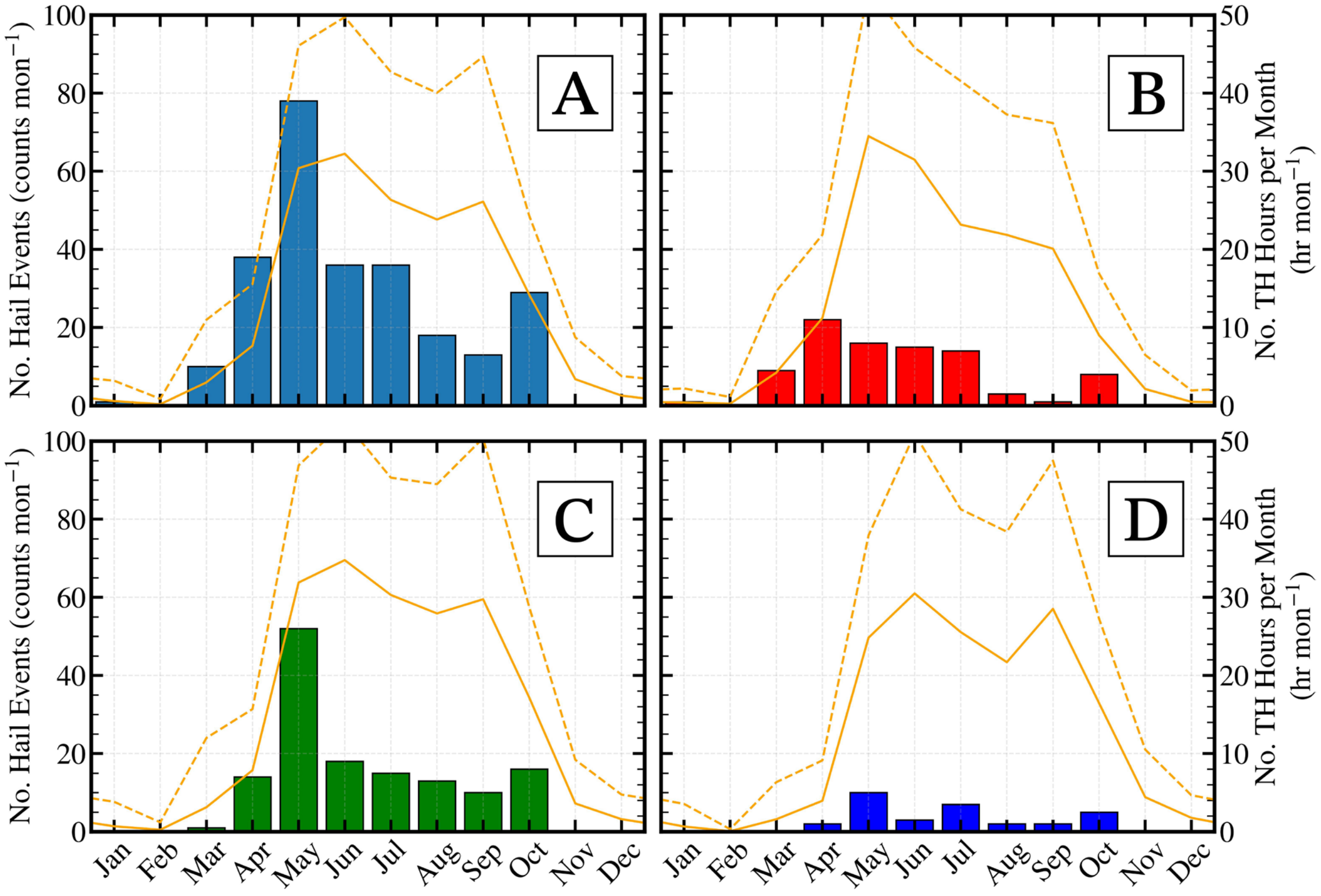

**Figure 4.** Monthly Distribution of Hail Events (counts mon$^{-1}$) and Thunder Hours (hr mon$^{-1}$) across Luzon, Philippines in different Environments associated in this study. Yellow lines correspond to the average (solid line) and P95 (dashed line) Thunder Hours per month. (a) Total Hail Events across Luzon, (b) HE1, (c) HE2, (d) HE3.

The average thunderstorm activity in April is 11.21 hr mon$^{-1}$ (P95: 21.87 hr mon$^{-1}$) while it peaks by May at 34.50 hr mon$^{-1}$ (P95: 53.44 hr mon$^{-1}$). On the other hand, HE2 had the highest reported hail events which peaks at 52 counts by May then also slowly terminates by October. Coincidingly, the average thunderstorm activity in May is at 31.89 hr mon$^{-1}$ (P95: 46.87 hr mon$^{-1}$), while the peak activity is by June (Average: 34.75 hr mon$^{-1}$; P95: 52.48 hr mon$^{-1}$) which also clears off by October. Lastly, HE3 had the least hail activity which shows a small peak by May, inline to HE2, although may suffer some sample limitation for this analysis. Still, the average thunderstorm activity for the month of May in this environment is at 24.82 hr mon$^{-1}$ (P95: 37.91 hr mon$^{-1}$), with similar peak to HE2 in June at 30.50 hr mon$^{-1}$ (P95: 50.92 hr mon$^{-1}$).

Overall, the alignment of hail activity with peak thunder hours further underscores the role of convective storm environments. Such as that thunderstorms that are accompanied by frequent lightning activity (a proxy for thunderstorm activity and for verification purposes) tend to be associated with high CAPE and steep lapse rates, while anything below than that (e.g., average mixing ratio $< 5$ g kg$^{-1}$; low-level lapse rates $<5$ °C km$^{-1}$) suggests that thunderstorms are less likely to convect (Taszarek et al., 2020). Importantly, the asymmetry in seasonal tails (earlier onset in HE1 versus prolonged persistence in HE2) points to the influence of consistent large-scale rising motion that enables convection forced primarily by differential diurnal heating or aided by topographic factors throughout the year (Albrecht et al., 2016; DiGangi et al., 2022).

## 3.2 Thermodynamic Parameters

### 3.2.1 CAPE (MU and ML)

Based on the results of prior research, one may expect an increase of CAPE for increasing severe weather intensity. Indeed, across all hail environments as shown in Fig. 5, they normally reach $> 2000$ J kg$^{-1}$, with HE3 MUCAPE values having a tighter distribution compared to other setups (Fig. 5a). Meanwhile, the other environments (HE1 and HE2) feature a very wide range of MUCAPE values suggesting that hail events are not restricted to the environments of high MUCAPE and given the distribution's negative skewness, particularly HE2. Statistically, HE1 has a median MUCAPE of 4145 J $kg^{-1}$, HE2 has 4421 J $kg^{-1}$, and HE3 has 4292 J $kg^{-1}$. Minor differences were observed for MUCAPE in the lowest 3 km AGL and within the HGL across all environmental setups (Fig. 5a2 and 5a3) with their respective median > 200 J $kg^{-1}$ and near 1000 J $kg^{-1}$. Particularly, the differences to the skewness of distribution were seen following the same as in MUCAPE where HE1 and HE2 are negatively skewed in favor of high $MUCAPE_{03}$ and $MUCAPE_{HGZ}$, while HE3 are positively skewed but overlaps the interquartile range (IQR) of other hail environments. Given these relatively large CAPE values in the lower section of the troposphere (and so as to the total CAPE), this suggests that the low-level moisture is in play that produces these CAPE measurements. Nixon et al. (2023) asserted that higher RH between the cloud base to the mid-levels is accompanied by stronger CAPE. This setup, along with weaker DLS, can combat entrainment (discussion to the entrainment and DLS is in the following subsection).

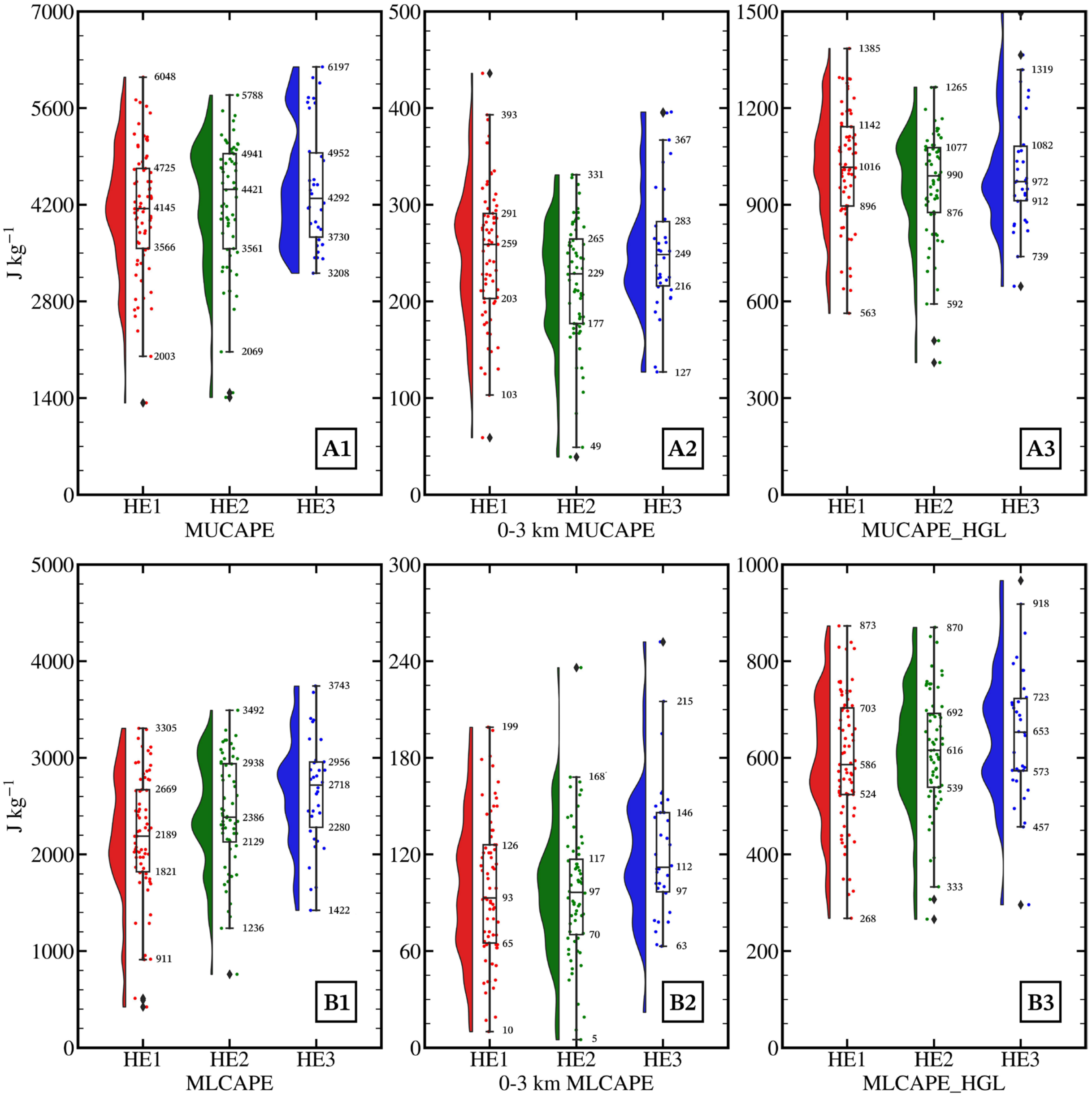

**Figure 5.** Raincloud diagrams for (a) MU-parcel based and (b) ML-parcel based profile. (1) CAPE, (2) 0-3 km CAPE, and (3) $CAPE_{HGL}$ associated with HE1 (Red), HE2 (Green), and HE3 (Blue). The box and whiskers cover the 25th-75th percentiles and the median values are marked by the horizontal line within each box.

For MLCAPE and its other derivatives (Fig. 5b), there is a noticeable increase of MLCAPE across different hail environments, with HE2 and HE3 attaining a nearly identical IQR, but with lower median for HE2. As expected, compared to MUCAPE, the MLCAPE values are modest (Fig. 5b1). Across all hail environments, the HE3 displays a higher median in $MLCAPE_{03}$ at 113 J $kg^{-1}$, as compared to HE1 and HE2 which were at < 100 J $kg^{-1}$. Regardless, this may imply that the presence of buoyancy/instability in the lowest 0-3 km layer reflects a well-mixed and moist boundary layer, providing favorable conditions for storm initiation and sustained convection (Fig. 5b2). In Fig. 5b3, the $MLCAPE_{HGL}$ also shows some degree of overlap across all hail environments in Luzon, with HE3 being positively skewed, whereas both HE1 and HE2 have a wide range of measurements and slightly inclined to the left of the distribution, hinting that these two environments favor high $MLCAPE_{HGL}$. This suggests that convective storms in HE1 and HE2 also rely on the availability of instability within and above the FZL. However, it should be noted, that regardless of how small or large CAPE is existing, it doesn't translate to larger or the potential for hail formation. But, residence time within the updraft matters.

### 3.2.2 Entraining CAPE (ECAPE)

As formulated by Peters et al. (2023), ECAPE accounts for the entrainment (as the name suggests) due to the dilution of buoyancy caused by either lack of storm inflow (usually in the first 1 km) and/or mid-tropospheric dryness. In the case of our hail environments, depicted in Fig. 6a, the climatology reveals that these hail-bearing storms are accompanied by median ECAPE values play at around 2000 J $kg^{-1}$; for HE1 its 2152 J $kg^{-1}$, for HE2 its 2283 J $kg^{-1}$, and for HE3 its 2186 J $kg^{-1}$. In addition, HE1 also features a wider range of IQR of entrainment-adjusted CAPE: roughly 1500-2500 J $kg^{-1}$, as compared to HE2 and HE3. This suggests that despite the maximum ambient instability (as represented by MUCAPE), the effective buoyancy available to storms is reduced once the effects of entrainment are included. In fact, the ECAPE climatology is shifted to the lower values by at least 50% relative to the MUCAPE highlighting the critical role of environmental moisture and low-level inflow in modulating storm intensity and hail growth potential. This adjustment aligns with the findings of both Lin and Kumjian (2022) and Nixon et al. (2023), who emphasize that excessive CAPE alone often overestimates storm vigor by neglecting entrainment-induced dilution and may be detrimental to hail growth.

Furthermore, Peters et al. (2023) posited that an entrainment ratio (or fractional entrainment) exceeding 0.5 ($\geq$50%) implies that a substantial fraction of the available convective instability is realized by the storm updraft and can be indicative of severe convection, particularly thunderstorms with supercellular mode. Across all hail-producing environments being examined, this threshold was exceeded implying that entrainment does not fully suppress updraft strength, but rather modulates storm organization and severity. This further supports the interpretation that while high MUCAPE may favor strong convection, it is the net CAPE available to an updraft that depends not only on the CAPE profile, but on the wind shear and relative humidity of the ambient environment. Thus, a balance between instability and entrainment (as captured by ECAPE) must exist to constrain the likelihood of hail-producing storms (Peters et al., 2019, 2023).

### 3.2.3 Downdraft CAPE (DCAPE)

In the Philippines, many hail events were also accompanied by strong wind gusts. DCAPE is maximized by a combination of steep lapse rates below 700-hPa and a very dry layer between 700- and 500-hPa. In fact, Craven and Brooks (2004) found that there is a stark difference in DCAPE between an ordinary cell environment and a severe hail/wind environment. A possible explanation is that higher DCAPE values permit stronger rear flank downdrafts (RFD), which could result in an outflow-dominated severe storm that undercuts the convective system.

As shown in Fig. 6b, all hail environments exhibit a similar distribution of DCAPE, with each setup having an IQR between 700 and 1000 J $kg^{-1}$. Statistically, HE1 has a median DCAPE of 817 J $kg^{-1}$, HE2 has 884 J $kg^{-1}$, and HE3 has 840 J $kg^{-1}$. These values are consistent with the findings of Craven and Brooks (2004), who reported a median DCAPE of ~900 J $kg^{-1}$ for significant hail and wind environments across the Great Plains. Large DCAPE values are typically associated with stronger downdrafts and outflow winds, and thus with severe convective wind events. Because DCAPE is directly proportional to low-level lapse rates i.e., the steeper the lapse rate, the greater the DCAPE, these results align with the observed 0-3 km AGL lapse rates (which will be discussed later). Therefore, DCAPE provides additional predictive value in diagnosing hail-prone environments, as higher DCAPE not only favors downdraft strength but also increases the probability of both severe hail and damaging winds. Not only that, but our hail environments serve as a proxy for understanding downdraft events.

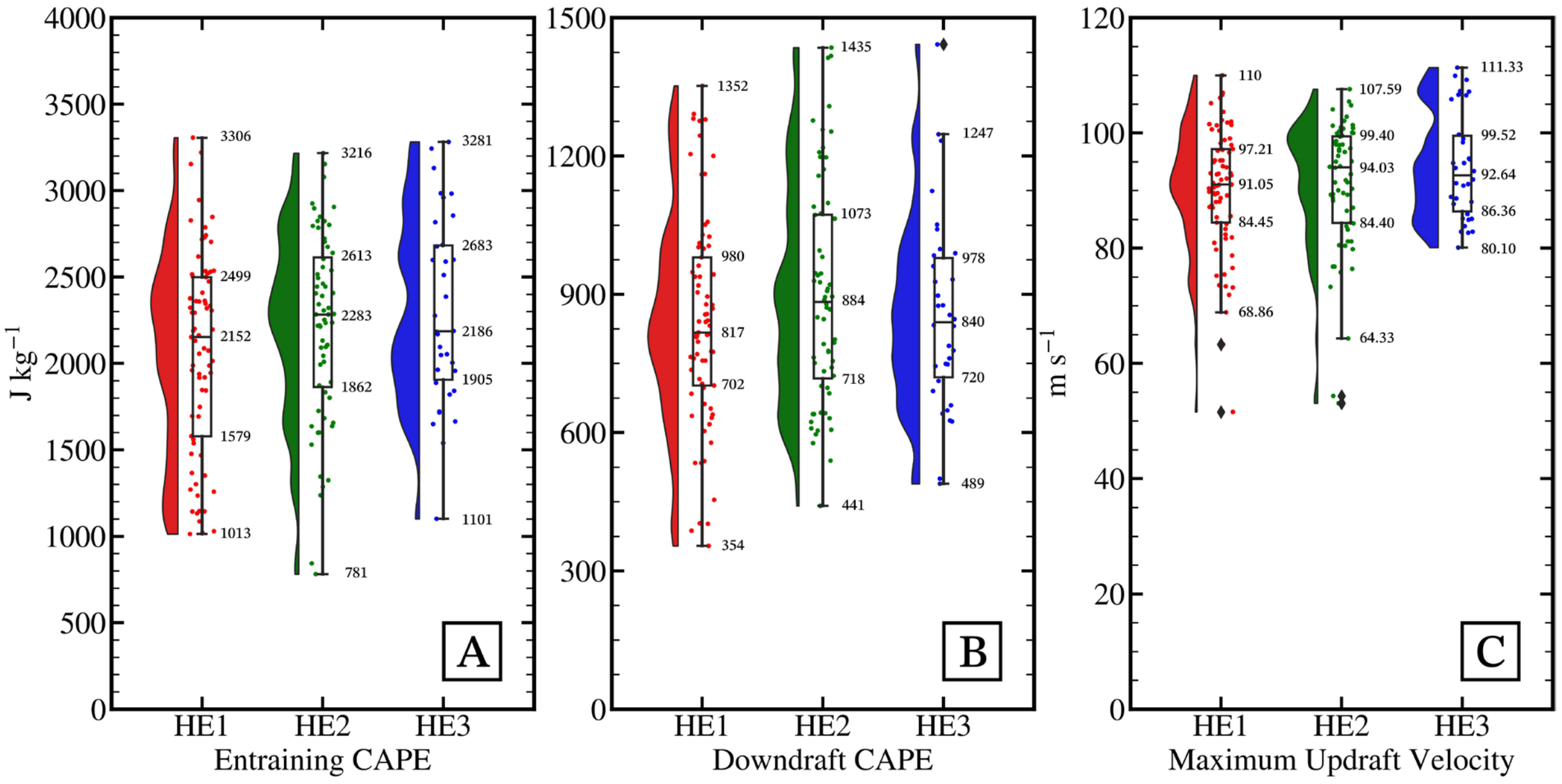

**Figure 6.** Same as in Fig. 5 except for (a) ECAPE, (b) DCAPE, and (C) $W_{MAX}$.

#### 3.2.4 Theoretical Maximum Updraft Velocity ($W_{MAX}$)

The undiluted velocity within the updraft can be calculated from ambient instability, such that it is proportional to the square root of twice the CAPE, representing the theoretical maximum vertical velocity achievable in a buoyant parcel ($W_{MAX} = \sqrt{2} \times$ MUCAPE). For instance, stronger updrafts produce larger hailstones than weaker updrafts (e.g., Browning, 1963), downdrafts and the associated damaging wind potential are dynamically coupled with updrafts (Marion and Trapp, 2019), precipitation production relates to vertical mass flux and consequently *w* (Doswell et al., 1996), and vertical accelerations in the lower part of updrafts also play a critical role in tornadogenesis (Coffer and Parker, 2015).

Calculated from a non-entraining most unstable parcel, the $W_{MAX}$ distributions (Fig. 6c) across hail environments display the same feature as in the climatology of MUCAPE of each hail setups. Both HE1 and HE2 feature negatively skewed in favor of high $W_{MAX}$ values with median undiluted updraft velocity of 91 m s$^{-1}$ and 94 m s$^{-1}$, respectively. Meanwhile, the HE3 exhibits a positively skewed distribution, although the IQR overlaps and shows no differences when compared to the rest of hail environments (HE1 and HE2) despite the higher 10$^{th}$ percentile at 80 m s$^{-1}$. This could imply that HE3, despite their overall different statistical profile compared to HE1 and HE2, is still capable of producing the strong updrafts necessary for hail production which can compensate for the lack of deep-shear magnitude.

#### 3.2.5 Lifted Condensation Level (LCL)

While LCL is typically used for forecasting tornadoes i.e., lower LCL is inclined to produce more tornadoes (Rasmussen and Blanchard, 1998; Thompson et al., 2003; Craven and Brooks, 2004; Groenemeijer and vanDelden, 2007; Kaltenböck et al., 2009; Taszarek and Kolendowicz, 2013), high LCLs are found to have linkage to produce hail (Púčik et al., 2015; Taszarek et al., 2017). This should not be surprising given the LCL's relationship with the depth of the mixed layer as high LCLs are indicative of a well-mixed low-level atmosphere, and thus indicative of steep low-level lapse rates, usually in the first 3 km AGL (which will be discussed in the next subsequent section).

The climatological distribution of LCLs (Fig. 7a) across the identified hail environments displays an asymmetry towards the positive axis. This reflects that the LCLs' IQR of these hail environments are in the range between and around 600-1000 m indicative of overlapping and similarities in the low-level thermodynamics. In particular, the median LCL is 705 m for HE1, 800 m for HE2, and 699 m for HE3. This reinforces the notion that hail events typically occur in higher cloud bases than the non-severe thunderstorms (Púčik et al., 2015).

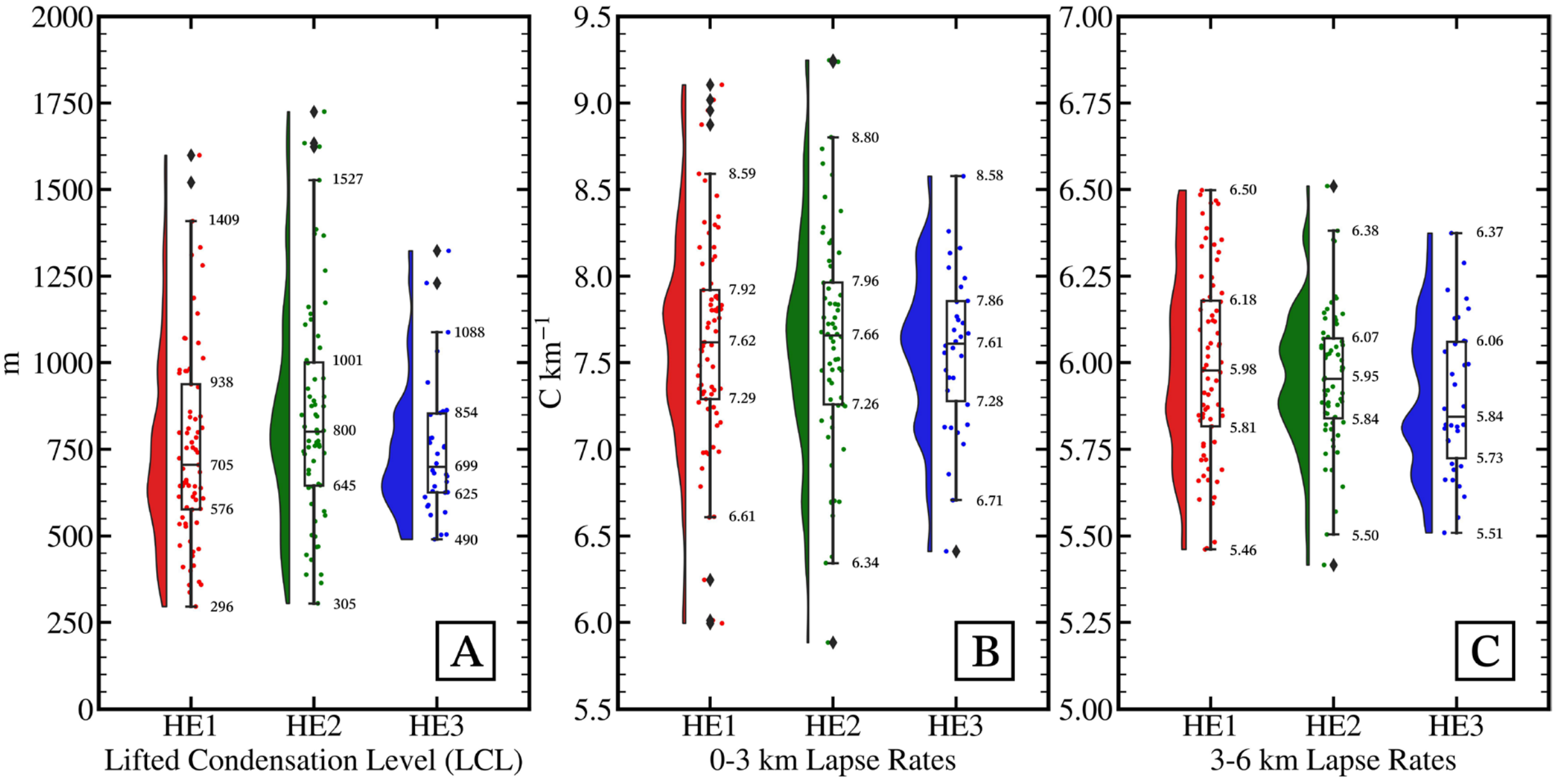

**Figure 7.** Same as in Fig. 5 except for (a) LCL, (b) 0-3 km Lapse Rates, and (c) 3-6 km Lapse Rates.

### 3.2.6 0-3 km Lapse Rates (LLR)

Accompanied by low-level moisture, both the low- and mid-level lapse rates (the latter discussed in the next section) are key contributors to CAPE. Unsurprisingly, they exhibit a pattern similar to MUCAPE across all hail environments. Figure 7b displays the climatological distribution of the LLR in per hail environment. Each identified hail environment displays overlapping distributions, with IQRs spanning roughly 7-8 °C km$^{-1}$. The median 0-3 km Lapse Rates are 7.62 °C km$^{-1}$ for HE1, 7.66 °C km$^{-1}$ for HE2, and 7.61 °C km$^{-1}$ for HE3.

These steep lapse rates likely reflect a well-mixed low-level troposphere and may indicate modest inverted-*V* thermodynamic profiles, where moisture gradually increases with height before diverging near the LCL or LFC. It can be concluded that hail events occur typically in higher 0-3 km lapse rates compared to non-severe thunderstorms, but as asserted by Púčik et al. (2015), this parameter has no relation to hail size.

### 3.2.7 3-6 km Lapse Rates (MLR)

As discussed in the LLR subsection, SWEs such as the environments of hail-producing storms feature typically higher mid-level lapse rates, which is unsurprising given their relation to CAPE and updraft strength. In Fig. 7c, each of the hail domains' climatological raincloud plots show relatively similar medians close to 6 °C km$^{-1}$ for the MLR, although considerable overlap in their IQR does exist. Specifically, the median MLR is 5.98 °C km$^{-1}$ for HE1, 5.95 °C km$^{-1}$ for HE2, and 5.84 °C km$^{-1}$ for HE3.

While these modest values are still classified as 'conditionally unstable atmosphere', the overlap in the IQR suggests that a specific, narrow range of MLR is not a complete pre-requisite for hail events in these types of environments, and other ingredients must be considered (such as the low-level and boundary-level moisture; Taszarek et al., 2017). While the mid-level instability is as important in producing the undiluted instability, the 0-3 km lapse rates may be able to compensate for the lack of steep MLR and of elevated mixed layer (EML) in these tropical environments given that the low-level lapse rates can approach dry adiabatic.

## 3.3 Kinematic Parameters

### 3.3.1 Low-level Shear (LLS)

For hail events, there is an overlap in the negatively skewed LLS distribution across all hail environments. Shown in Fig. 8a, in general, the values constitute weak 0-1 km vector shear magnitude in all hail environments, but the upper bounds of each distribution show a decreasing trend from HE1 to HE3. In fact, the median LLS of HE1, HE2, and HE3 are 2.21 m $s^{-1}$, 1.95 m $s^{-1}$, and 1.61 m $s^{-1}$, respectively. HE1 also had a slightly higher 10$^{th}$ percentile compared to HE2, and so as HE2 had a slightly higher 10$^{th}$ percentile compared to HE3. This can be attributed to latitudinal/geographical differences between each environment, especially with HE1 closer to the sub-tropics where frontal systems can impact the location. It is also worth to note that past studies such as in Craven and Brooks (2004) found that hail environments tend to be associated by weak LLS but higher LCLs, and the same is found in this baseline climatology of hail environments across Luzon. This suggests that strong LLS is not a primary ingredient for hail production across Luzon; instead, hail development is likely dominated by thermodynamic factors, particularly in the low-level buoyancy/instability profile, and by the depth of the updraft (including the shear vector difference between the cloud base and top level).

Further, the lack of LLS magnitude can be attributed to the low-level wind flow more aligned i.e., lack of turning with the DLS leading to unidirectional shear profile (such as that it creates a straight hodograph; Nixon et al., 2023). This is believed to be favorable for hail growth as it is also indicative of weak low-level storm-relative winds, which cannot impart excessive momentum on air parcels and associated hydrometeors as they approach the HGZ.

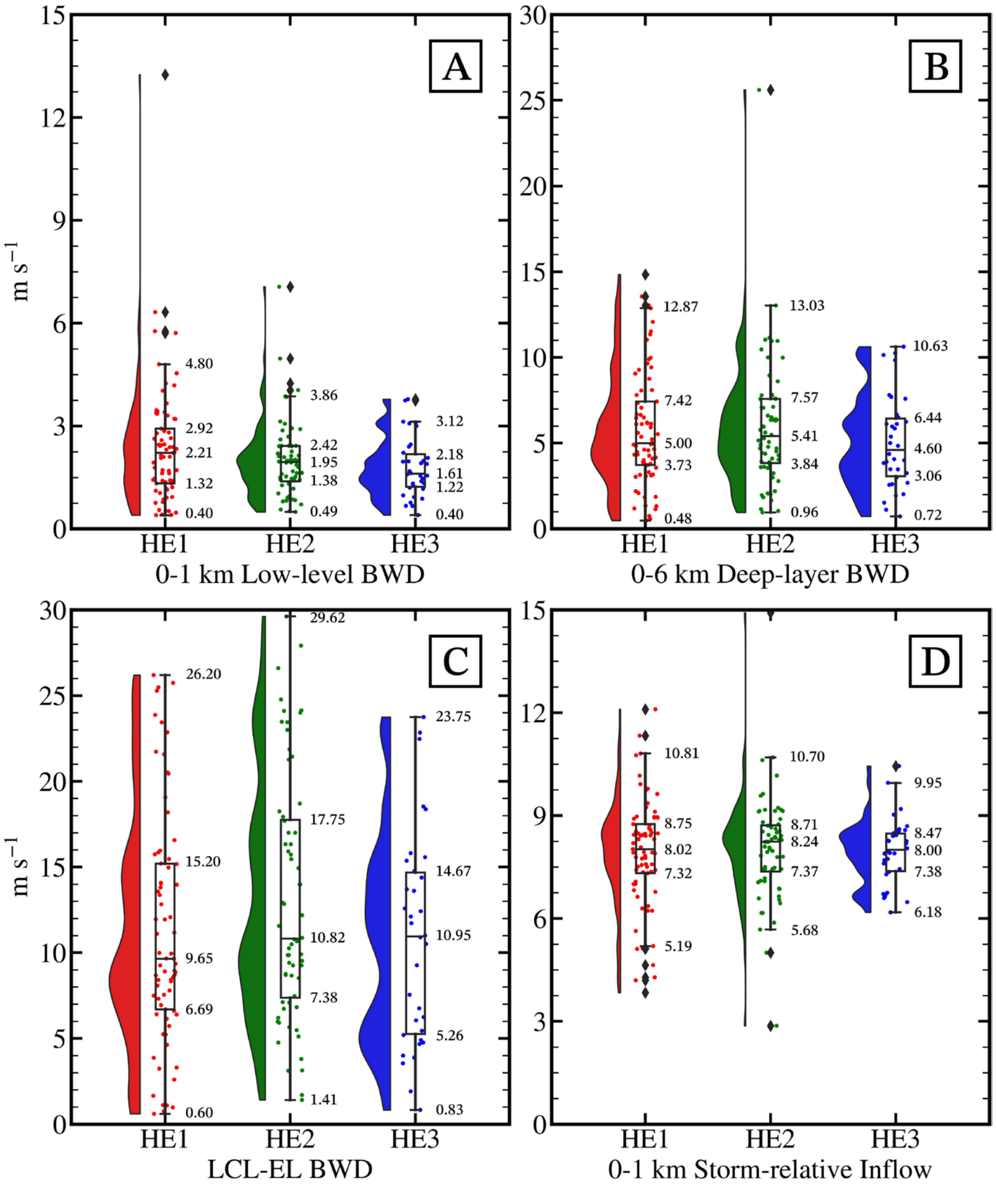

**Figure 8.** Same as in Fig. 5 except for (a) LLS, (b) DLS, (c) $BWD_{LCL-EL}$, and (d) $V_{SR}$.

Whereas, in cases of strong LLS, it can negatively impact the trajectories of hailstones through the HGZ (Dennis and Kumjian, 2017; Kumjian et al., 2021; Lin and Kumjian, 2022). However, while a weak 0-1 km shear magnitude profile is predominantly found across hail environments (e.g. Taszarek et al., 2020; Nixon and Allen, 2022), hail production is still possible even with strong LLS, based from the recent findings of Kumjian et al. (2021) and Homeyer et al. (2023) (and so as low-level inflow in the first 1 km).

### 3.3.2 Deep-layer Shear (DLS)

Several studies found that 0-6 km bulk wind shear (DLS) discriminates well between severe e.g., supercells and non-severe convection (Rasmussen and Blanchard, 1998; Thompson et al., 2003). Because severe storms are almost always accompanied by hazardous weather (Duda and Gallus, 2010; Smith et al., 2012), the changing distribution of DLS with different hail environments was studied.

In Fig. 8b, there is no pronounced difference between the distribution of DLS across all environments. The results were consistent with Sari and Lasher-Trapp (2025) who found weak DLS regimes across SEA, not extreme as compared to midlatitudes like in Europe and the United States where a pronounced increase in DLS is observed per severe category. That may be because large hail is almost exclusively related to supercells (Smith et al., 2012), whereas smaller hail may occur with other morphology of organized storms. Note here is that the median of DLS is close to 5 m s$^{-1}$ (HE1: 5.00 m s$^{-1}$; HE2: 5.41 m s$^{-1}$; HE3: 4.60 m s$^{-1}$), a value also close to those found by Sari and Lasher-Trapp (2025). This supports the notion that these hail events occur in weaker shear regime, compensated by the high buoyancy and CAPE (and so as high $W_{MAX}$). The compensation of the instability allows balance due to the lack of low-level and 0-6 km shear magnitudes that can support hail formation (Nixon et al., 2023). As we shall see below, I explored an additional facet of kinematic profile that can be used for hail prediction in the Philippines.

### 3.3.3 Cloud-layer Shear ($BWD_{LCL\text{-}EL}$)

Given the generally weak low-level shear (and correspondingly weak deep-layer shear) in the tropical hail environments I identified so far, and the characteristic wind profiles of hail-producing storms in other regions, it is reasonable to consider a new kinematic parameter that can mimic the influence of stronger shear environments. Capuli (2025) demonstrated that the bulk wind difference between the cloud base and the height of neutral buoyancy parameterized as Equilibrium Level (known as $BWD_{LCL\text{-}EL}$) shows utility in discriminating hail-supportive environments in the tropics. This is physically consistent, as stronger deep shear vectors promote updraft tilting, a process commonly linked to severe convective storms and can also lead to larger updraft area in the HGZ, permitting longer trajectories and thus larger sizes (Kumjian and Lombardo, 2020). Furthermore, Taszarek et al. (2020) showed that the convective cloud depth (difference between the height of LCL and EL) had some skill in discriminating hailstorms (including in size) and tornadic storms.

The climatological distribution of $BWD_{LCL\text{-}EL}$ (Fig. 8c) shows promising results with relatively strong vertical shear (10 m s$^{-1}$/20 knots) environments across all hail environments determined, somehow mirroring High CAPE, High Shear hail setups in the Great Plains. The median $BWD_{LCL\text{-}EL}$ for HE1 is 9.62 m s$^{-1}$, 10.82 m s$^{-1}$ for HE2, and 10.95 m s$^{-1}$ for HE3. More than half of the hail events associated in their respective environment occurred where $BWD_{LCL\text{-}EL} > 10$ m s$^{-1}$, a typical threshold for organized severe convection which parallels the role of DLS in midlatitude severe storm climatologies (Doswell and Evans, 2003). The increased shear magnitudes in this cloud depth, compared to the conventional DLS, significantly enhances hail production by elongating the updraft and increasing the volume over which hail growth processes occur. Thus, can serve as a relevant parameter for identifying environments supportive of organized convection and hail production in the tropics, where traditional bulk shear metrics often underperform. Additional inspection of this parameter has been undertaken in succeeding sections.

### 3.3.4 Storm-relative Inflow ($V_{SR}$)

In prior 0-1 km shear magnitude discussion (Section 3.3.3), storm-relative winds in the lowest 1 km AGL i.e., $V_{SR}$ dictates the width of the convective storm's updraft. If the storm inflow is stronger, it yields larger and wider updrafts due to the mass flux being ingested and vice versa (Peters et al., 2020). While this is supportive of hail production, it is a double edge sword where it can diminish the residence time of the hail embryos in the updraft's HGZ as strong $V_{SR}$ imparts excess momentum in and out of the storm thereby causing quick ejection and (Dennis and Kumjian, 2017; Kumjian and Lombardo, 2021; Lin and Kumjian, 2022).

As shown in Fig. 8d, the climatology of storm inflow across identified hail environments in Luzon show overlapping and similar features. The IQR of each sectors are small spanning from 7.3 m $s^{-1}$ to 8.7 m $s^{-1}$, accompanied by nearly identical median $V_{SR}$ (HE1: 8.02 m $s^{-1}$; HE2: 8.24 m $s^{-1}$; HE3: 8.00 m $s^{-1}$). From this result, it shows that all of our hail environments are entailed by weaker $V_{SR}$ (and so as LLS) but accompanied by ample, if not strong, instability within and below the HGZ, whether MU-parcel and ML-parcel was used. It was asserted by Nixon et al. (2023) that there's a correlation between $V_{SR}$ and the low-level thermodynamic profile such as that an environment defined by stronger $CAPE_{HGZ}$ (and/or $CAPE_{sub\text{-}HGZ}$) with weaker storm-relative winds in the lowest 1 km is likely to produce hail. The assertion was also supported by previous extensive simulations being conducted in hail-producing storms (Lin and Kumjian, 2022, and references therein).

## 3.4 Moisture Parameters

### 3.4.1 Relative Humidity between 1-3 km and 1-6 km ($RH_{13}$ and $RH_{16}$)

Recent studies have found out that increasing moisture content generally favors increasing hail probability, especially when low levels of the convective updraft are considered. Lin and Kumjian (2022) asserted that air parcels originating from the lowest altitudes can supply more moisture to the HGZ as they have greater vapor mixing ratios (i.e., more vapor content available to condense into liquid). Particularly, the 1-3 km layer, extending to the 6-8 km layer i.e., mid-levels, of the updraft is indicated to be critical in modulating the liquid water content (LWC) within the HGZ. At same time, these layers are a good proxy for lower-tropospheric RH above the cloud base as it is where changes to the amount of entrained dry free tropospheric air are seen (Peters et al., 2019; Nixon et al., 2023).

Both Figs. 9a and 9b display mean RH across different layers; from the cloud-base up to the lower-portion of the updraft, and extending to the mid-levels. One feature being evident is how the baseline climatology of each environment shrinks in terms of covering the range of $RH_{13}$ and $RH_{16}$, likely contributed by the difference in geographical location. Nonetheless, the IQR of $RH_{13}$ per each hail environment shows overlapping features, with median $RH_{13}$ of 78%, 75%, and 71% for HE1, HE2, and HE3, respectively. These distributions are skewed negatively, indicating that most of these hail environments occur with low-level moisture > 75-80%. Meanwhile, for $RH_{16}$, HE1 covers a greater range of values with distinct small bimodal bumps between 50-60% and 65%-75%, while HE2 is negatively skewed and HE3 is symmetrically distributed. The median $RH_{16}$ for each environment shows very little variance, hence tight to one another between 63-65%.

### 3.4.2 Precipitable Water (PWAT)

In conjunction with the moisture across several cloud depths (seen in Section 3.4.1), I established and supported previous studies that moisture is a key ingredient in hail formation, so its abundance seemingly would be conducive to hail production. Similarly, displayed in Fig. 9c, the amount of precipitable water available across the troposphere shows the same trend as in the previous section. HE1 climatological distribution covers a variety of measurements with some of the events having PWAT <40 mm leaning to the initial $10^{th}$ quartile, but can also increase to 55-60 mm in some cases. Given the wider range, the IQR spans from 43 to 55 mm with a median close to 50 mm. This suggests that HE1 can see more environmental variations whether dry or moist conditions aloft i.e., large or small dewpoint depressions, as compared to other hail environments in Luzon.

When compared to both HE2 and HE3, these sectors had tighter/smaller IQRs where the $25^{th}$ quartiles are above 50 mm indicative of abundant moisture in the troposphere. The median PWAT for these two hail environments are 54.04 and 54.25 mm, respectively. This result suggests that an abundance of tropical moisture is available throughout the troposphere for the convective storms. Perhaps, the dewpoint temperature of these hail events along these sectors show less discrepancies, especially aloft. Afterall, PWAT is sensitive to the quality of moisture resources (Kumjian et al., 2019), such as that large (small) dewpoint depressions can lead to lower (higher) PWAT. This, in turn, tends to promote inverted-*V* (moist low-level) profiles and facilitates evaporation (persistence) of raindrops (which is also depicted for HE1 in Fig. 9a and 9b). And these inverted-*V* profiles are the hall mark of excessive downdraft production, as measured previously in Fig. 7b. In addition, modelling efforts by Grant and van den Heever (2014) also found that the extent of mid-tropospheric dryness/moistness affects storm structure and hail-growth regions. Thus, highlighting that both RH across several cloud depths as diagnosed earlier and vertical moisture distribution fundamentally regulates both the extent of hail production and growth of the embryos, and storm thermodynamics consistent with recent studies as cited above.

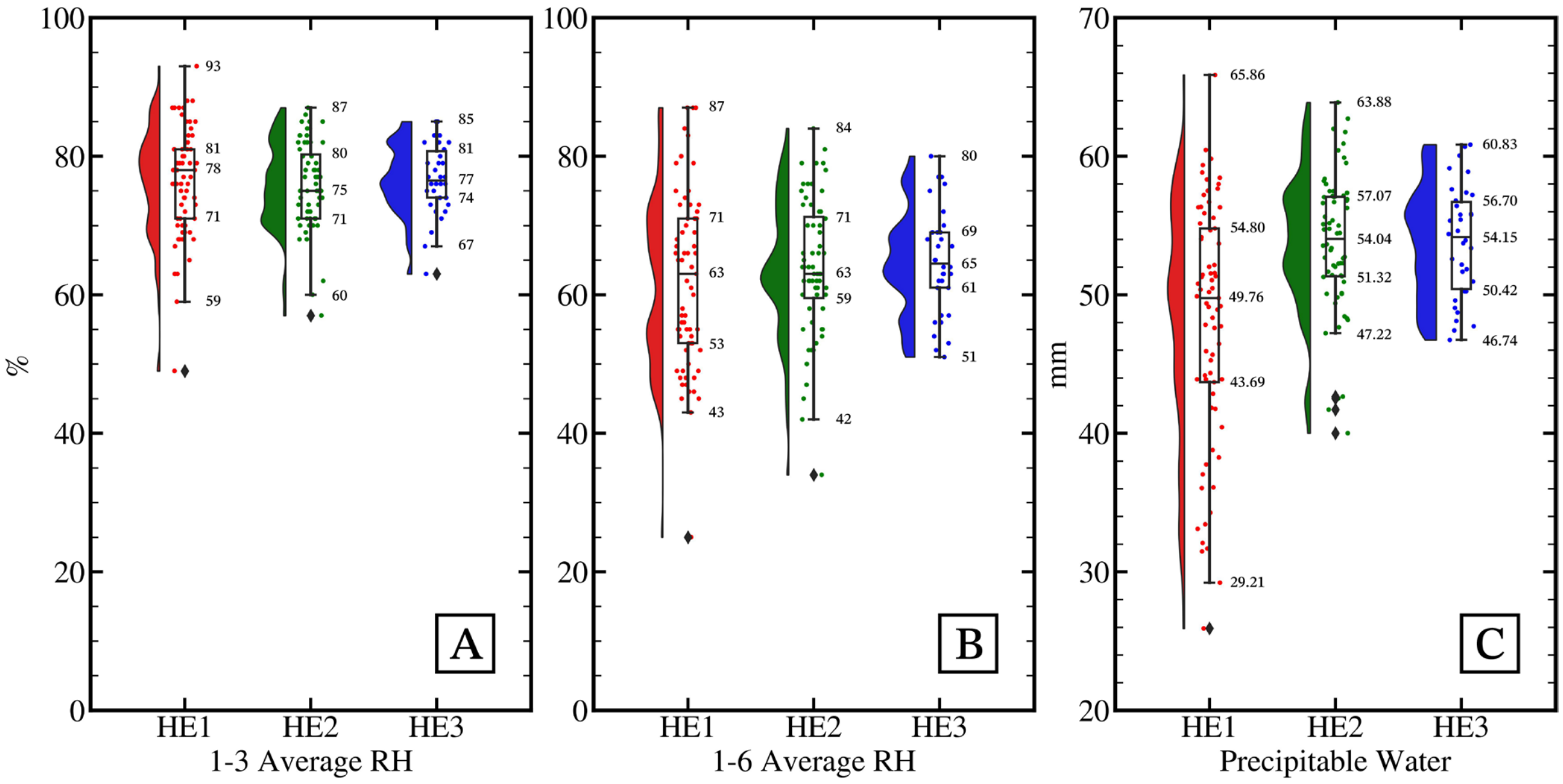

**Figure 9.** Same as in Fig. 5 except for (a) $RH_{13}$, (b) $RH_{16}$, and (c) Precipitable Water.

### 3.4.3 Freezing Layer (FZL)

As shown throughout, the large CAPE (either using MU, ML, and entraining parcel profile, and even storm depth) of these hail environments is correlated with the environmental temperature through latent heat release, such that environments with higher FZLs generally feature more moisture and stronger CAPE. The FZL, as a quick refresher, is the height of the 0 °C isotherm and the depth of the atmosphere that is above freezing which liquid undergoes phase change (Zhou et al., 2021).

Figure 10a shows the climatological distribution of FZLs across identified hail environments in Luzon. Similar to the baseline climatology related to moistures discussed above, the FZL distribution across hail environments showcase a shrinkage as we go down in geographical sectors. Just like before, HE1 features a variety of values and a bimodal distribution with the bulk of FZLs > 4500 m and extending <4000 m (minimum). The lower end of the distribution is likely contributed by the varying terrain heights, originating from the Benguet areas where hailstorms also occur at > 1000 mASL. After all, even with stronger low-level CAPE combined with the weak storm-relative winds in the lowest 1 km (Fig. 8d), lower FZLs can facilitate hail production more readily where buoyancy will essentially allow embryos to reach deep in the HGZ (Nelson, 1983; Browning and Foote, 1976; Miller et al., 1988; Knight and Knight, 2001; Nixon et al., 2023). The median FZL of this hail environment is at 4700 m.

In comparison, both HE2 and HE3 are relatively similar with overlapping across these environments, although HE3 can be limited by the sample size for this hail environment, hence without a median. But for HE2, the median FZL is at 5000 m. The higher FZL reflected hailstones will have more time to melt and vice versa, and these FZL values are typically found in southeast Asia (Chen and Chou, 2006; Zhou et al., 2021). This correlates well with the initial climatology being constructed in Part 1 (Capuli, 2024), where most of the hailstones being captured by netizens have sizes that are <1 cm (peanut size).

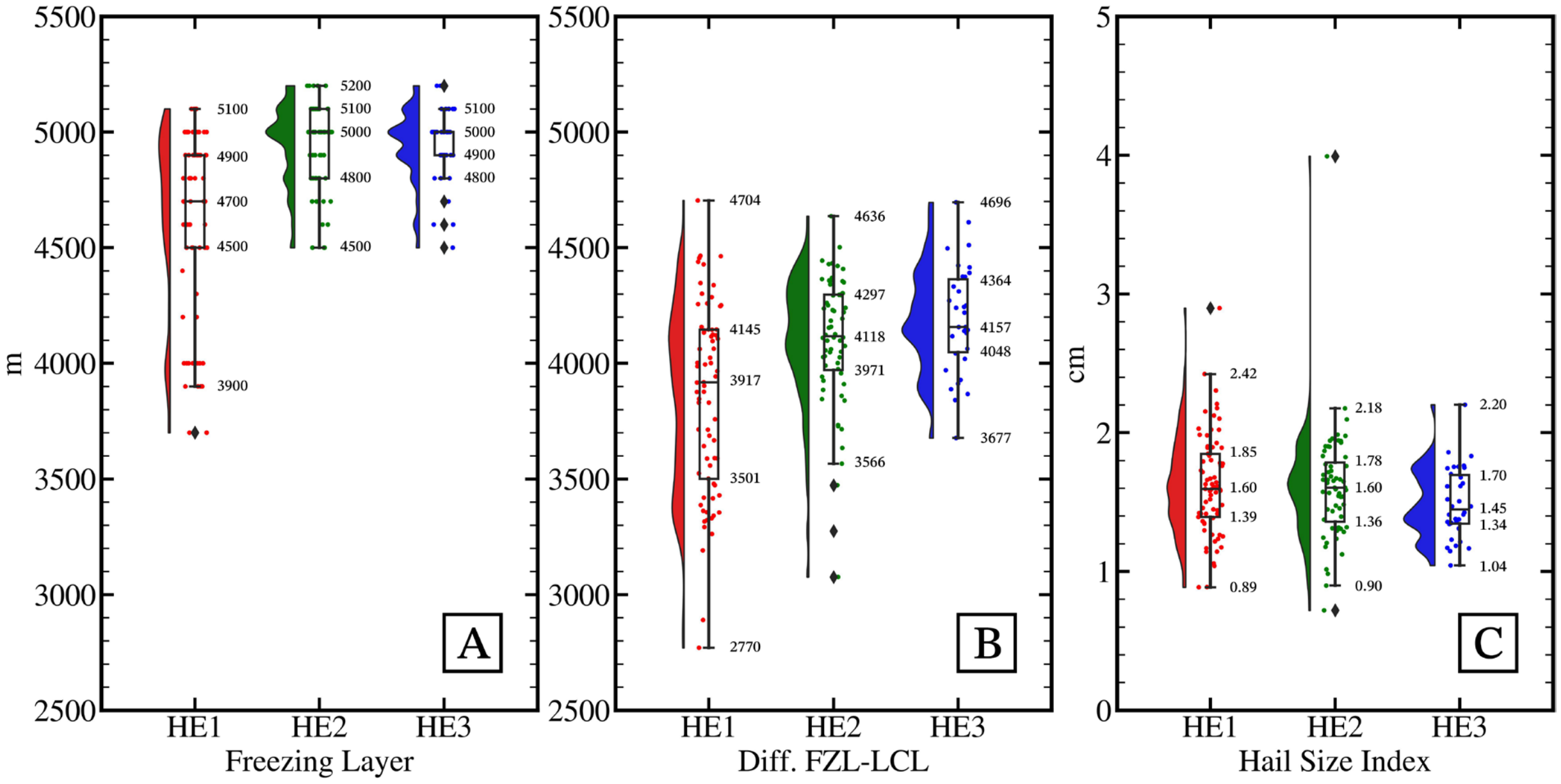

**Figure 10.** Same as in Fig. 5 except for (a) FZL, (b) ΔFZL and (c) Hail Size Index.

### 3.4.4 Depth between the LCL and FZL (ΔFZL)

The depth between the cloud base and the height of 0 °C isotherm (ΔFZL) can reflect the CAPE below HGZ. Recent studies such as in Nixon et al. (2023) found out stronger (weaker) CAPE below HGZ is unfavourable (necessary) for hail development, as long as the storm-relative wind in the lowest 1 km is weak (strong). Thus, low ΔFZL means that below the HGZ, there is a residing low CAPE and vice versa.

Figure 10b reveals the raincloud diagrams of ΔFZL associated with the hail environments in Luzon landmass. Similar to Figs. 10c and 11a, HE1 also features a bimodal distribution covering a wider range of ΔFZL. Particularly, there are setups whose ΔFZL are <3500 m while there are also environmental conditions that are accompanied by ΔFZL > 4000 m. The median ΔFZL for HE1 is at 3917 m. These two ranges likely represent the higher terrain and lower terrain sectors of the HE1. Afterall, a higher surface elevation (hence high LCL/LFC) typically means a lower FZL. Given the sensitivity of hail to FZL as explored above, this may explain the prevalence of large hail in the High Plains and the Appalachian Mountains (Allen and Tippett, 2015), as well as the Alps in Europe (Púčik et al., 2019). Meanwhile, HE2 and HE3 show identical distribution patterns which tend to overlap to one another. These two environments tend to have constricted IQRs with HE2 having a negatively skewed distribution while HE3 having a symmetric form. The tight IQRs' difference between the 25$^{th}$ and 75$^{th}$ percentile is around 300 m (spanning from 4000 to 4300 m). The median ΔFZL for HE2 and HE3 are 4118 m and 4157 m, respectively; thus nearly identical to one another. The difference between the ΔFZL of HE1 and both HE2 and HE3 may have been due to the varying quality of moisture and vertical distribution of PWAT, including that of the identified geographical sectors in Luzon (and so as the varying FZL). The extent of melting depends on the depth between the FZL and the cloud base (>0 °C), with larger hailstones required to survive prolonged exposure to above-freezing temperatures. Although, I would like to point out that the melting phase change process may not begin immediately upon a hailstone falling to the 0 °C level, due to evaporative cooling effects on the hailstone's surface (Rasmussen and Heymsfield, 1987) as contributed by the well-mixed low-level troposphere i.e., depending on the quality of RH both between 1-3 km and 1-6 km (Fig. 9a and 9b) that allows steep LLRs (Fig. 7b and 7c; Johnson and Sugden, 2014) and enhanced downdraft production (Fig. 6b). Hence, melting can be significantly delayed when falling into sub-saturated environments (RH < 80%).

Overall, when sufficient moisture supports ample instability; tempered by entrainment that reduces excess buoyancy, and the ambient kinematic environment (both in the low- and upper-levels) is favorable for severe convection, hail embryos can grow efficiently. A shallower depth between the sub-freezing and above-freezing levels (i.e., between the LCL and the 0 °C isotherm) minimizes CAPE below the HGZ, allowing hail embryos to be lofted more directly and deeply into the HGZ without experiencing limiting updraft trajectories. This promotes longer residence time within the HGZ and increases the likelihood of hailstones reaching the surface (Brimelow et al., 2017;

Prein and Heymsfield, 2020; Lin and Kumjian, 2022). Conversely, higher FZL and larger ΔFZL (>4000 m) give ascending air parcels more time to accelerate along the storm inflow, so they enter the HGZ with greater momentum. While this reduces their residence time in the HGZ, the enhanced CAPE below HGZ can partly offset this effect, especially when stronger CAPE below HGZ exists while at the presence of weak storm-relative winds in the first 1 km layer, as demonstrated in the idealized simulations of Lin and Kumjian (2022) and recent analysis undertaken by Nixon et al. (2023). Similar to the previously mentioned authors, I echo that these results suggest that the FZL's role as a proxy for the HGZ may be equally critical in diagnosing hail potential.

## 3.5 Composite Parameters

### 3.5.1 Hail Size Index (HSI)

Composite parameters utilize the interrelationship between a number of convective parameters to improve on the forecast skill of any individual parameter (Brooks et al., 2003). Here I consider the distributions of two commonly used operational forecasting parameters: HSI and $W_{MAX}$SHEAR. Common to each of these indices is inclusion of instability and shear parameters, often with calibrations to a set of known cases or conditional requirements. Each index features increasing values along with increasing probability of the convective hazards. First, as conceived for severe convective storms that produce (large) hail in the Europe, the Hail Size Index (HSI) is a composite thermodynamic and kinematic parameter that takes into account the instability, mid-level lapse rates, 0-6 km bulk shear, EL, cloud base height in LCL, and FZL to forecast expected maximum hail size in cm (Czernecki et al., 2019).

Figure 10c displays the climatological distribution of HSI across all hail environments in Luzon landmass. All profiles associated with these environments indicate that the IQR of HSI are relegated between 1.3 and 1.7 cm, while the $10^{th}$ percentiles of each hail environment are close to 1 cm and $90^{th}$ percentile above 2 cm. The distribution patterns are uniform for HE1 and HE2 with HE3 showing multimodal distribution likely due to sampling size limitation, though shares an overlapping (similar) features as mentioned above. The median HSI for both HE1 and HE2 are at 1.60 cm, while HE3 minorly lags behind at 1.45 cm. These results i.e., relatively small hail sizes, can be inclined to the relatively large ΔFZL for most of the hail environments identified which in turn produces large CAPE within and below HGZ. Coupled with the weak LLS and $V_{SR}$, yet strong storm-depth deep shear up to the EL, the dynamical support can favor sustained severe convection but with limited hail embryo recycling, thereby can restrict hail growth to smaller sizes despite favorable thermodynamics, inclined to the balance required as stipulated in Nixon et al. (2023). This interplay between thermodynamic buoyancy below the HGZ and kinematic organization aloft suggests that Luzon hail environments are more conducive to frequent but modest hail production rather than supporting giant hail occurrences.

### 3.5.2 Parameter Distribution

The combination of $W_{MAX}$ (or CAPE depending on the parcel formulation) and DLS has been used as a crude proxy for severe weather environments in climatological studies (e.g., Brooks, 2009, 2013; Diffenbaugh et al., 2013). Therefore, I would like to explore the joint distribution of severe events in the two-dimensional $W_{MAX}$-DLS parameter phase space, in addition to another shear vector parameter defined as $BWD_{LCL-EL}$. The phase space between the $W_{MAX}$ and shear magnitudes such as in DLS across all and each environment were depicted in Fig. 11.

Hail occurrence is greater with higher $W_{MAX}$, in this case at 90 m s$^{-1}$ (Fig. 11a). The probability is maximized with DLS between 2.5 to 10 m s$^{-1}$, which is counter-intuitive given that higher DLS between 20 to 30 m s$^{-1}$ is conducive for the occurrence of severe thunderstorms, particularly supercells (Doswell and Evans, 2003). Condition tail distributions of $W_{MAX}$SHEAR over the entire Luzon landmass exceed 450-500 m$^2$ s$^{-2}$, which is highly supportive for severe thunderstorms and consistent with prior work (Brooks et al., 2003; Gensini and Ashley, 2011; Li et al., 2020). This suggests that the probability of hail strongly increases toward the higher CAPE and DLS. Below those aforementioned results, the probability is less than 0.15.

A closer examination of individual hail environments highlights the role of $W_{MAX}$ as the primary modulator relative to DLS. In HE1 (Fig. 11b), DLS values are concentrated between 2.5 and 7.5 m s$^{-1}$, with most events associated with $W_{MAX}$ exceeding 85 m s$^{-1}$. In HE2 (Fig. 11c), the DLS range broadens to around 10 m s$^{-1}$, while events are typically characterized by $W_{MAX}$ near 90 m $s^{-1}$. HE3 (Fig. 11d), exhibits a similar DLS range to HE1 but with lower $W_{MAX}$ values clustered between 80 and 90 m $s^{-1}$. Correspondingly, the mean $W_{MAX}$SHEAR values were 522.39 $m^2$ $s^{-2}$ for HE1, 545.76 $m^2$ $s^{-2}$ for HE2, and 473.75 $m^2$ $s^{-2}$ for HE3. These results suggest that DLS alone may be a poor discriminator of hail potential, as its range remains on the weaker side (<10 m $s^{-1}$) across environments. Although, increases in $W_{MAX}$SHEAR appear more closely tied to hail occurrence, underscoring the importance of updraft intensity in modulating hail growth. Still, the relatively weak DLS values also indicate that storm organization is not solely controlled by shear vector at these heights. As a practical threshold, $W_{MAX}$SHEAR values near or above 500 $m^2$ $s^{-2}$ may serve as a baseline indicator of severe thunderstorms capable of producing hail in each hail environment. This points to the influence of additional wind profile factors shaping the hodograph, such as shear vector difference between the LCL and EL, which I previously identified as capable of mimicking high-shear setups. Thus, while $W_{MAX}$ provides a first-order control on hail growth, secondary parameters like cloud depth BWD may refine discrimination of hail-supporting environments. Accordingly, I extend our analysis to evaluate the role $BWD_{LCL\text{-}EL}$ in modulating hail environments.

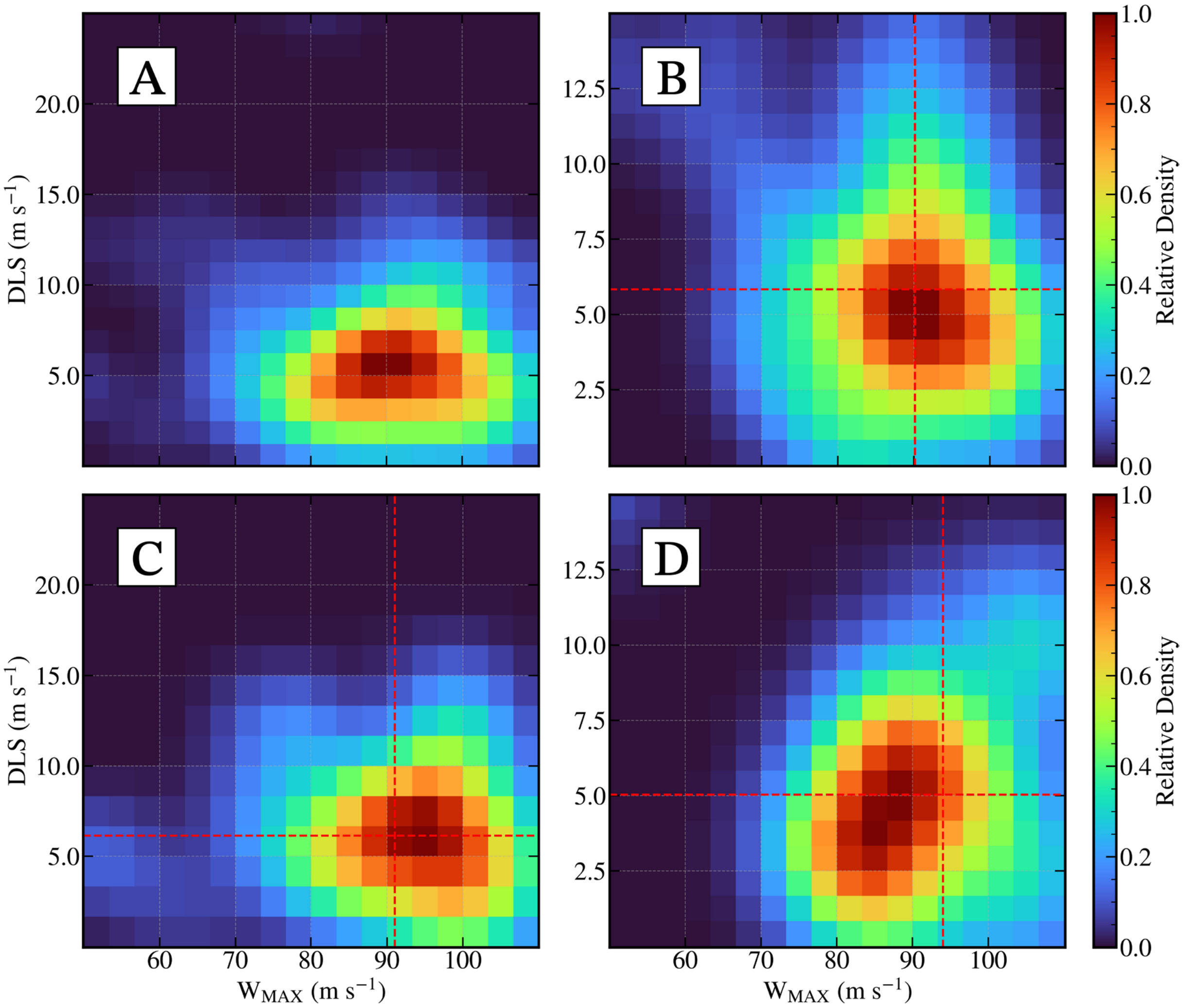

**Figure 11.** Kernel Density Estimate of hail-producing severe thunderstorms as a function of maximum undiluted updraft velocity and DLS for (a) across all hail environments, (b) HE1, (c) HE2, and (d) HE3. The probability density function was normalized from 0 to 1, while the red dashed lines in each identified hail environment are the average measurements.

Figure 12 shows the $W_{MAX}$ and shear vector magnitudes within the convective cloud depth across all hail environments. With this modification, the concentration of hail events' $W_{MAX}$ has shifted between around 90 m $s^{-1}$ and 100 m $s^{-1}$ accompanied with $BWD_{LCL\text{-}EL}$ focused around 7.5 m $s^{-1}$ at its lower bounds up to the 15 m $s^{-1}$, extending further to 20 m $s^{-1}$ at slightly lower probabilities. Per each hail environment, the average $W_{MAX}$SHEAR (or I should say,

$W_{MAX}SHEAR_{LCL\text{-}EL}$) of HE1 is 1019.91 $m^2$ $s^{-2}$, HE2 had 1188.97 $m^2$ $s^{-2}$, and for HE3, it is at 1011.52 $m^2$ $s^{-2}$. The presence of $W_{MAX}$ was important, but the increase in probability was now mainly due to the increase in $BWD_{LCL\text{-}EL}$. In fact, more than half of the hail events I analyzed turned out to have a $BWD_{LCL\text{-}EL} > 10$ m $s^{-1}$ (20 knots; purple circles). This relationship reveals that although probability typically increases in response to increasing instability, severe thunderstorms that are capable of producing hail are more likely to occur when deeper-layer shear is considered, not just conventional 0-6 km layer since the metric underperforms, is sufficient (e.g., at least 10-20 m $s^{-1}$), rather than when CAPE is large but shear is weak (<10 m $s^{-1}$), especially when the net CAPE was essentially halved due to entrainment. Meanwhile, high CAPE and weak shear regimes are presumably linked to damaging winds within microbursts (Atkins and Wakimoto, 1991).

Furthermore, I find that for hail events across these 3 environments that LCL contains additional information about the probability of hail besides $W_{MAX}SHEAR$ (Fig. 11) and $W_{MAX}SHEAR_{LCL\text{-}EL}$ (Fig. 12). To visualize, I constructed a phase space for hail by multiplying the square root of MUCAPE with DLS and $BWD_{LCL\text{-}EL}$. Displaying this parameter against LCL shows that the probability of hail increases as a function of either predictor (Fig. 13a and 13b). A practical limit of 500 $m^2$ $s^{-2}$ for the conventional $W_{MAX}SHEAR$ and 1000 $m^2$ $s^{-2}$ for the $W_{MAX}SHEAR_{LCL\text{-}EL}$ to mimic high shear storm setups across other severe storm regions globally seems a good benchmark for distinguishing hail-prone environments, as values near or above these thresholds consistently align with increased hail probability. These thresholds may therefore serve as practical proxies for environments supportive of organized convection capable of hail activity, even in cases where traditional DLS remains weak. Given any product of $W_{MAX}$ and shear vector magnitude between the cloud base and level of neutral buoyancy, a relatively high cloud bases (>700 m) in combination tend to favor and increase the probability of hail than non-severe ones as depicted in Fig. 8a and supported by previous studies (Púčik et al., 2015; Taszarek et al., 2017). Whereas, LCL below 500 m AGL and $W_{MAX}SHEAR$ below 500 $m^2$ $s^{-2}$ probability for hail drops.

Taken together, the two phase spaces suggest that the combined effect of shear and theoretical maximum updraft velocity (or CAPE), including LCL, provides a robust discriminator of hail-supporting environments. Similar findings were highlighted by Craven and Brooks (2004) and Púčik et al. (2015), who demonstrated the utility of coupling DLS with $W_{MAX}$ as a hail forecasting tool. In this case, however, I propose substituting DLS with $BWD_{LCL\text{-}EL}$, since it better captures/represents hodograph shape (as we shall see) and potentially storm organization in environments where traditional DLS is weak.

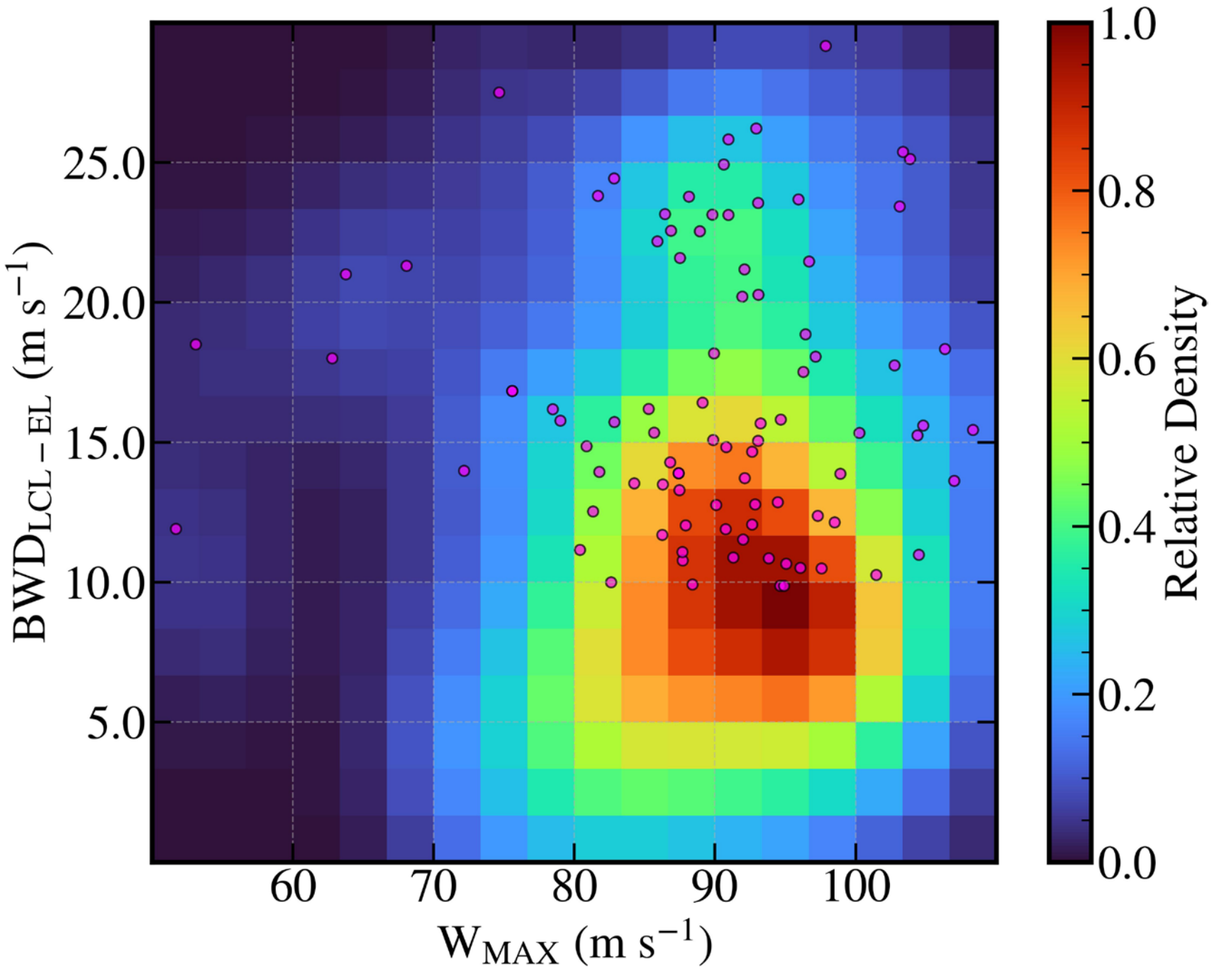

**Figure 12.** Kernel Density Estimate of hail-producing severe thunderstorms as a function of maximum undiluted updraft velocity and $BWD_{LCL\text{-}EL}$ across all hail environments. The probability density function was normalized from 0 to 1, while the pink scatters are the hail environments that have cloud-layer shear > 10 m $s^{-1}$ (20 kts).

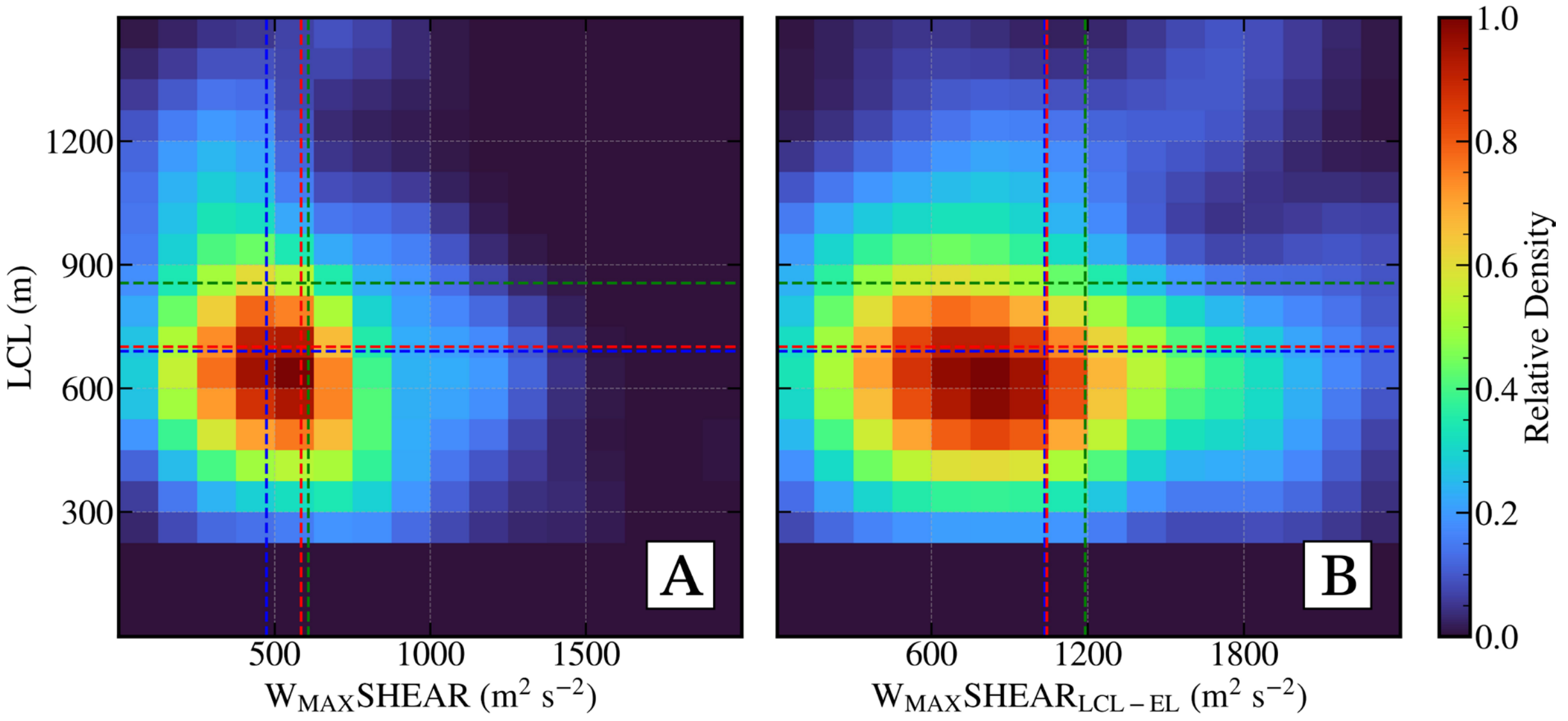

**Figure 13.** Kernel Density Estimate of hail-producing severe thunderstorms as a function of the product between the maximum undiluted updraft velocity and either of (a) DLS and (b) $BWD_{LCL\text{-}EL}$, and LCL parameter across all hail environments. The probability density function was normalized from 0 to 1, while the dashed lines correspond to each color coded hail environment (red-HE1; green-HE2; blue-HE3) and its average parameter.

## 3.6 Mean Thermo-Kinematic Profile of Hail-Producing Storms

### 3.6.1 Sounding Schematics

The angle-averaged instability profiles for each hail environment reveal slight differences in the temperature and moisture characteristics that favor hail growth (Fig. 14). Across all environments, however, the profiles consistently exhibit CAPE values exceeding 4000 J kg$^{-1}$, $RH_{13}$ values greater than 75%, and $RH_{16}$ above 60%, consistent with a sub-saturated regime. In HE1, the instability and moisture profiles show drier mid- and upper-level air, yet this environment is characterized by stronger CAPE and lower FZLs, as well as varying depths of LCL-FZL layer. The low FZLs primarily reflect the higher terrain elevation, while the strong CAPE arises from steeper lapse rates, both near the surface (7.71 °C km$^{-1}$) and in the mid-levels (6.02 °C km$^{-1}$). Notably, HE1 also exhibits larger variability in its temperature and moisture profiles compared to HE2 and HE3, especially above 700-hPa, where the spread becomes more pronounced. By contrast, HE2 and HE3 display more similar characteristics. Both are associated with higher FZLs and deeper LCL-FZL layers, but their lapse rates are less steep in the mid-levels (5.97 °C km$^{-1}$ and 5.91 °C km$^{-1}$, respectively). Despite this, the instability values remain high (HE2: 4180 J kg$^{-1}$; HE3: 4460 J kg$^{-1}$). Importantly, the differences in mid-level lapse rates across environments are modest, with all environments maintaining values near 6 °C km$^{-1}$, still within the steep range for the tropics, where lapse rates are known to support hail development (e.g., Johnson and Sugden, 2014; Tang et al., 2019).

A latitudinal trend also emerges in the mean characteristics of the environments. As we go down (up) in latitude, it tends to increase (decrease) in PWAT and ECAPE (although very small change), which also slightly decreases (increases) the potential energy associated with downdrafts. This implies that hailstorms in HE1 occur under relatively drier conditions, whereas HE2 and HE3 are more moisture-rich. Mid-level dryness in HE1 may promote stronger evaporative cooling favorable for downdrafts, though mean DCAPE varies only modestly across environments (~50 J kg$^{-1}$). Grant and van den Heever (2014) found that the extent of mid-tropospheric moisture (such as mid-level dryness exists) affects storm structure and hail-growth regions within their simulated storms. Since the entrainment of dry and stable air into an updraft may dilute its buoyancy (Peters et al., 2019, 2023), this result may be observational evidence of the impacts of entrainment on hail growth (or, perhaps, storm formation and maintenance) – whether detrimental or aid in this production process. In addition, referring to Knight (1981), hailstone embryo type is associated with the temperature of the cloud base. Specifically, for cloud base (LCL) temperatures warmer than 10 °C, embryos are more likely to be frozen-drops which was seen across much of our hail-based Skew-Ts.

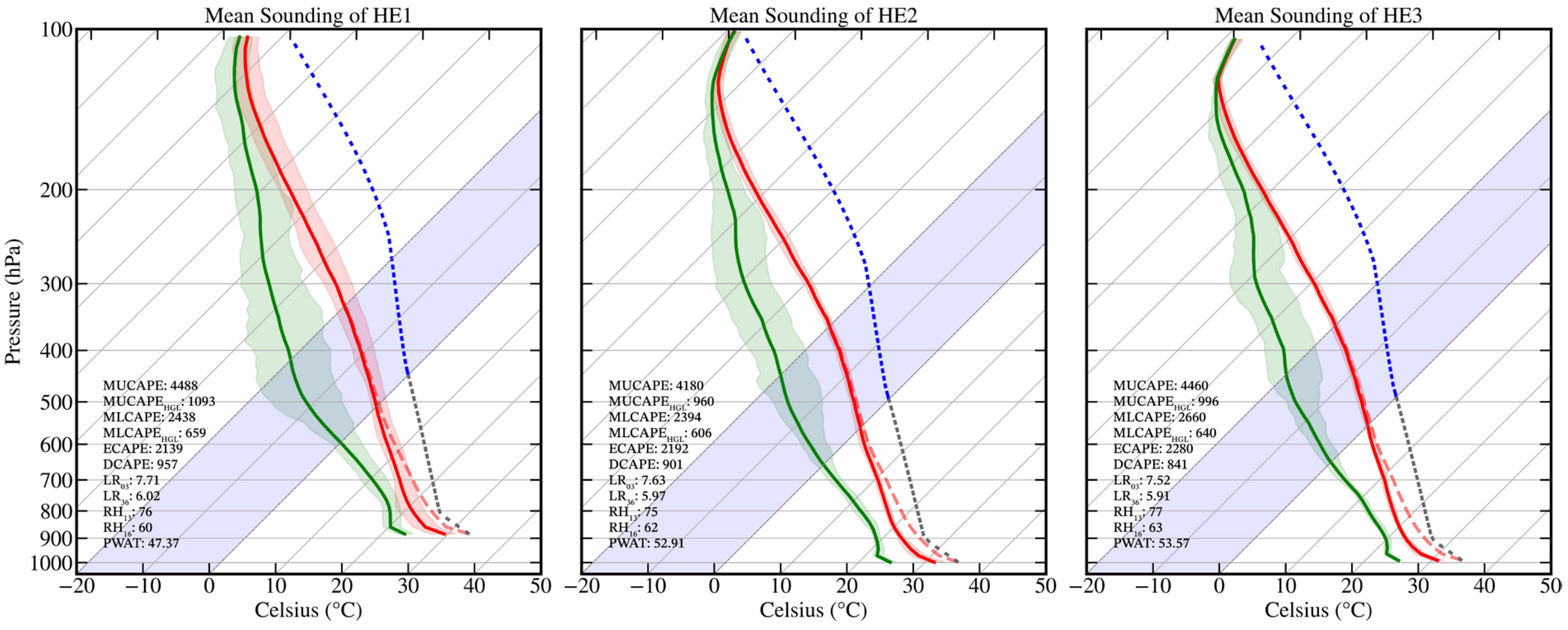

**Figure 14.** Mean thermodynamic profile for (a) HE1, (b) HE2, and (c) HE3. The Skew-T is composed of mean Dewpoint (green line) and interquartile range (green shading) for each sample, mean Temperature (red line) and interquartile range (red shading) for each sample. The most-unstable parcel profile (dashed gray line, changing to blue above the FZL) is also displayed, including the HGZ highlighted as the light blue region.

Although examining the mean characteristics of hail environments can prove useful in operations (e.g., Johnson and Sugden, 2014; Nixon and Allen, 2022), such averaging may obscure important parameter relationships that add considerable complexity. For example, the physical processes governing interactions such as those between CAPE and shear remain uncertain. Dennis and Kumjian (2017) demonstrated that modifying environmental wind shear in simulations produced substantial differences in hail growth and production. As we will see and also asserted in recent literature e.g., Nixon et al. (2023), incorporating wind profiles provides a broader perspective such as an interplay between buoyancy, shear, and moisture that ultimately modulates the likelihood of hail.

#### 3.6.2 Differing Hodographs

In the tropical setting such as Philippines, there are at least three dominant wind flow that affects the country; (1) Southwest monsoon associated with the Asian summer monsoon that typically spans from mid-to-late May to September (Murakami and Matsumoto, 1994), (2) Northeast monsoon during the latter period of the ber-months spanning early-to-mid November to February associated with Asian winter monsoon (Juneng and Tangang, 2010), and (3) Easterlies, dubbed as the pre-summer monsoon, associated with the tropical trade wind spanning from March to May, although may occur during *monsoon breaks* i.e., where either of the two mentioned seasonal wind flow has temporarily receded due to change of background synoptics (Wang and Xu, 1997; Yumul et al., 2010). Based on our previous analysis as depicted in Figs. 2 and 3, most of our thunderstorm (and so hail events) happened between March to October where both the southwesterlies and easterlies dominate. To dive deep into this matter, I clustered the Skew-T and Hodograph profiles of each hail environment in accordance with the dominant wind profile. For simplicity, I call hail events that occurred during easterlies as Easterly severe weather setups (E), while hail events that occurred during southwesterlies as Westerly severe weather setups (W). Figure 15 shows the mean Skew-T profile for the E and W setups, while Fig. 16 displays the mean Hodograph as classified according to the prevailing wind flow in each hail environment.

The mean thermodynamic characteristics of easterly and westerly setups reveal several important differences (Fig. 15). A key distinction is that easterly setups generally occur in moister environments, while westerly setups are typically influenced by warm air advection from southwesterly flow (e.g., monsoonal circulation). Both setups exhibit similar RH at 1-3 km ($RH_{13}$), as well as comparable values of MUCAPE, $MUCAPE_{HGZ}$, and mixed-layer parcel parameters. However, differences emerge at 1-6 km ($RH_{16}$), where easterly setups average 63% (Fig. 16a) compared to 58% for westerly setups (Fig. 16b). This slightly higher midlevel moisture in easterly cases reduces downdraft potential energy (813 J $kg^{-1}$) relative to the drier westerly environments (1003 J $kg^{-1}$). Column-integrated moisture also differs, with higher PWAT in easterly setups (52.17 mm) than in westerly setups (47.55 mm): a distinction that may be important given that most hail events over Luzon are associated with easterly modes (111 cases vs. 60 cases). At the same time, the drier environments of westerly setups favor larger CAPE, supporting stronger updraft potential. Another contrast lies in the LCL-FZL depth, which is generally greater in easterly setups. By comparison, some westerly setups show shallower LCL-FZL depths, likely reflecting their geographic occurrence, where hail embryos can be more efficiently lofted within the storm updraft.

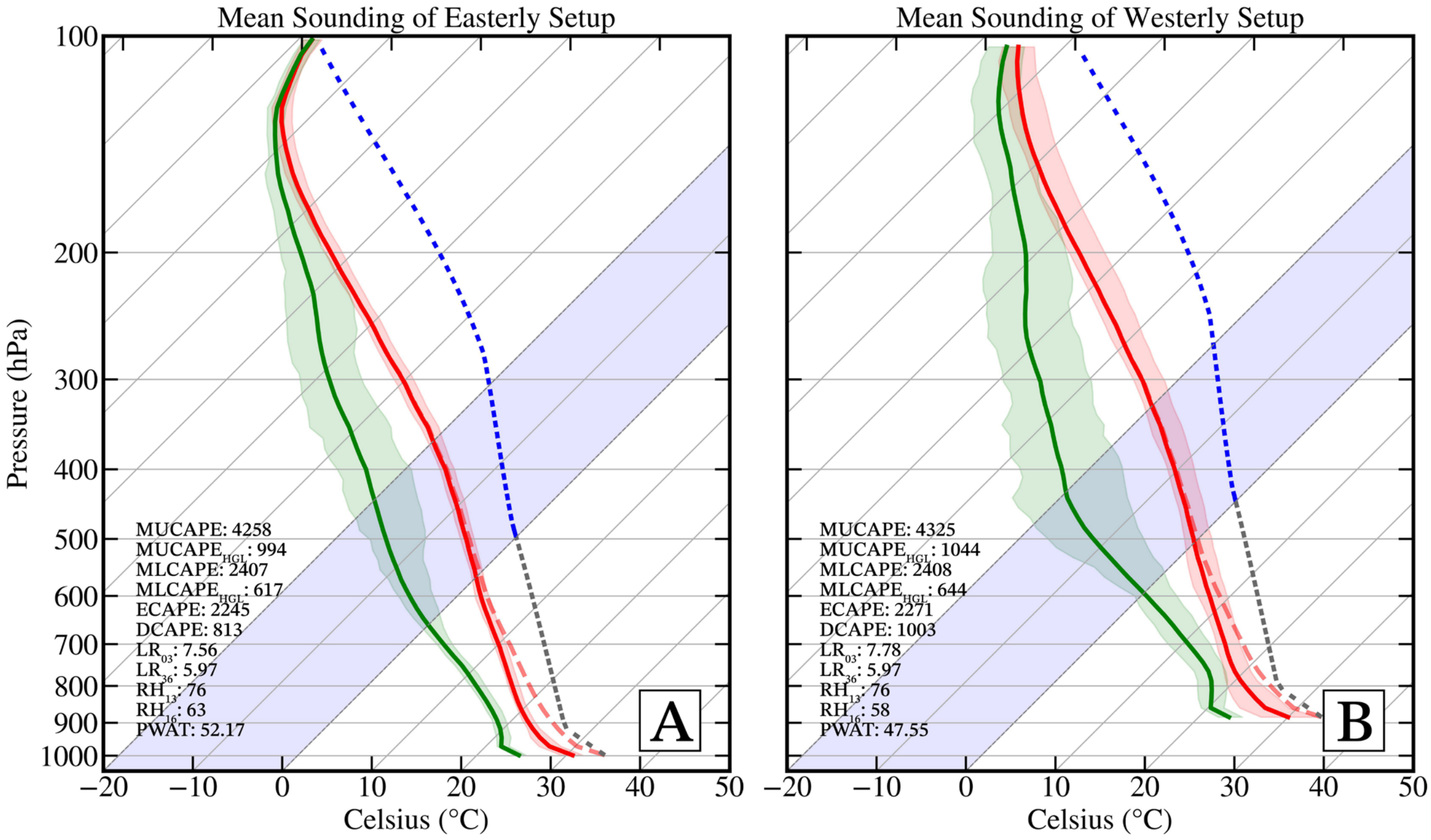

**Figure 15.** Mean thermodynamic profile for (a) easterly severe weather setup and (b) westerly severe weather setup. Similar components are found as in Fig. 15.

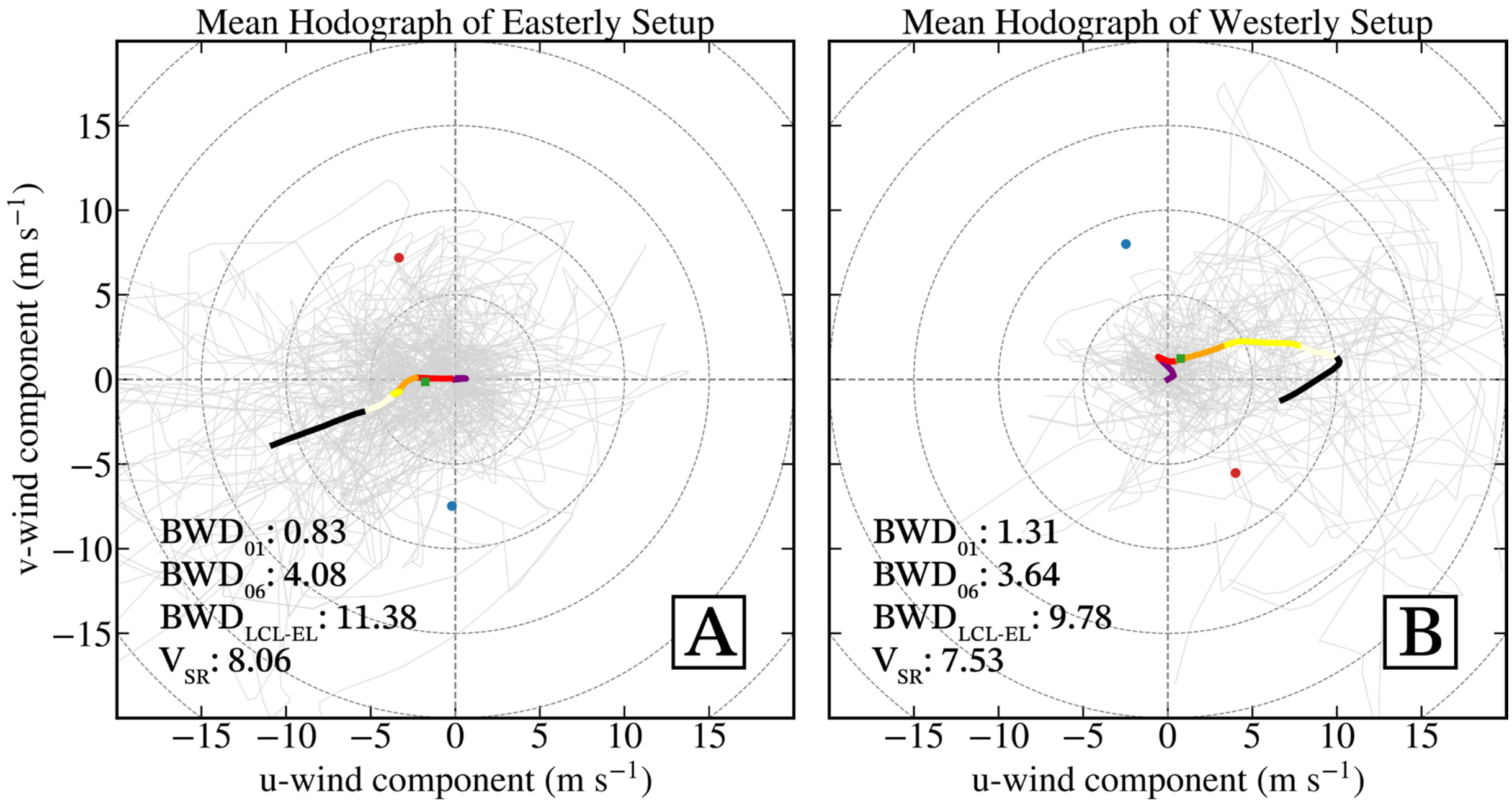

**Figure 16.** Mean kinematic profile for (a) easterly severe weather setup and (b) westerly severe weather setup. The mean profile is composed of the 0-1 km shear (purple), the 1-3 km shear (red), the 3-6 km shear (orange), the 6-9 km shear (gold), the 9-12 km shear (light yellow), 12-15 km shear (black), and the storm motion components such as right-moving storm motion (red circle), left-moving storm motion (blue circle), and mean wind storm motion (green box), with 5-20 m s$^{-1}$ wind range rings plotted.

In tandem with thermodynamic conditions, the average kinematic profile shown in the hodograph highlights further distinctions (Fig. 16). Both setups share weak LLS and DLS, consistent with the raincloud plots in Figs. 8a and 8b, with mean values near 1 m s$^{-1}$ and 4 m s$^{-1}$, respectively. The $V_{SR}$ also remains modest, averaging 7-8 m s$^{-1}$. The most striking feature, however, lies in the hodograph shapes, which display the hallmark of nearly all hail-producing environments: a predominantly unidirectional shear profile. In other words, the hodographs are largely straight, showing little clockwise turning in the lowest levels and thus a weak meridional component of the shear-relative winds. For easterly setups, the upper-level mean wind direction is ~73°, maintaining an easterly component from the surface to the upper troposphere. Westerly setups average ~254°, producing a *"square-root"*-like shape: minimal turning in the 0-1 km shear, transitioning to a more zonal component aloft. In addition, the mean $BWD_{LCL\text{-}EL}$ for easterly setups is 11.38 m s$^{-1}$, while westerly setups have slightly lower cloud depth shear magnitudes at 9.78 m s$^{-1}$. These hodograph shapes associated with the two severe weather steps align with previous findings (Johnson and Sugden, 2014; Dennis and Kumjian, 2017; Kumjian et al., 2019; Gutierrez and Kumjian, 2021) and fits the conceptual model for splitting severe thunderstorms e.g., such as supercells (Weisman and Klemp, 1986). Consequently, combined with our thermodynamic analysis above, hail environments in the Luzon feature weaker LLS, DLS, and $V_{SR}$ yet accompanied by stronger CAPE and both modest $RH_{13}$ and $RH_{16}$ that can favor hail production. Such balance in these environments, especially in the tropics, is consistent with the current literature standing in the environments of severe convective storms that produce hail (Nixon et al., 2023).

Further, the severe weather setup classification, based on the prevailing wind profile, was used in each hail environment to understand the thermo-kinematic variance across each sector. Fig. 17 shows the mean thermodynamic profile in each hail environment per severe weather setup, while the corresponding hodograph in these environments per severe weather setup is depicted in Fig. 18.

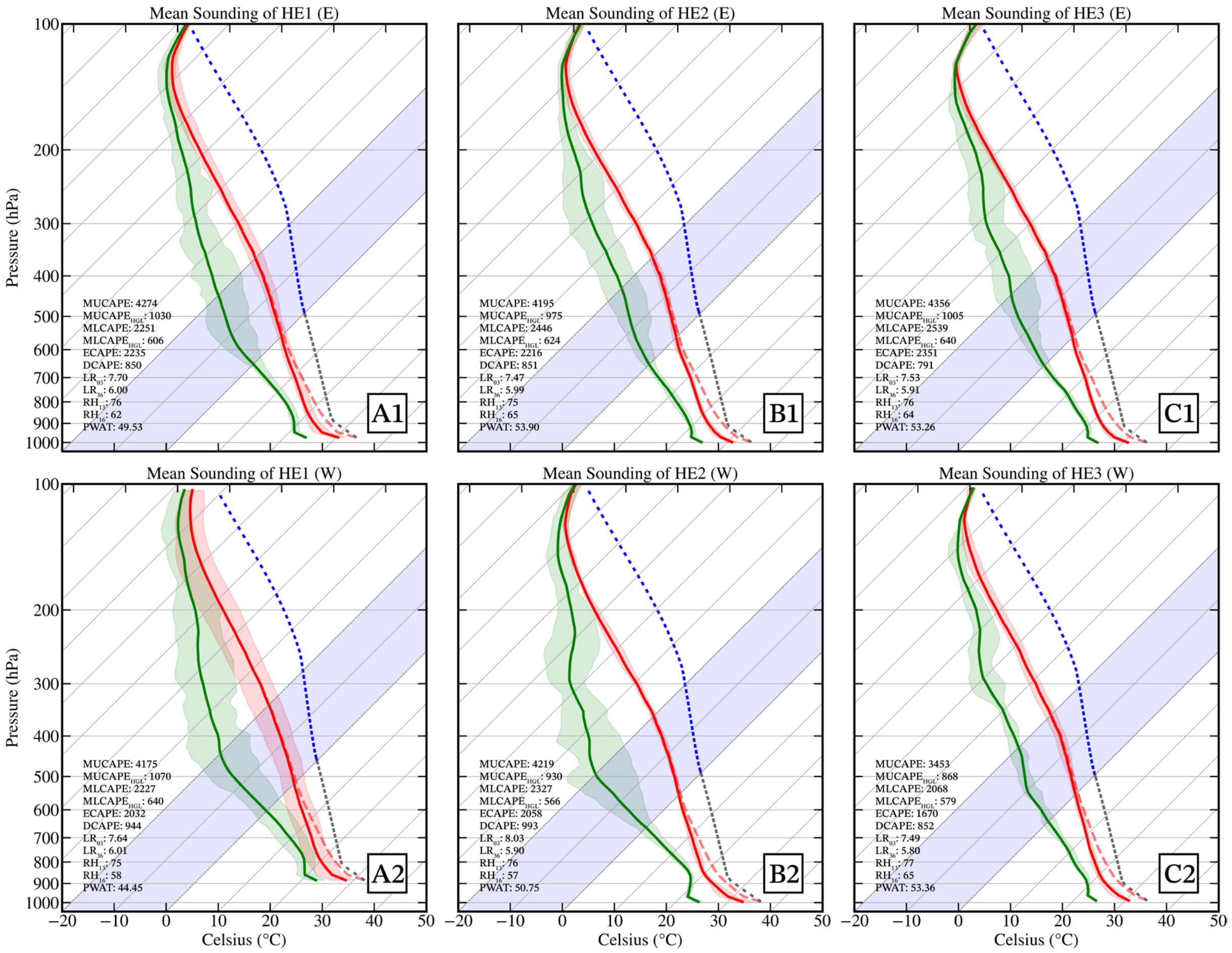

**Figure 17.** Similar to Fig. 15 but classified based on the hodograph profile for; (1) easterly severe weather setup and (2) westerly severe weather setup across (a) HE1, (b) HE2, and (c) HE3.

Thermodynamically, the HE1 (E), HE2 (E), and HE3 (E) have identical characteristics to another typically comprised by MUCAPE > 4000 J kg$^{-1}$ while also having an ECAPE > 2200 J kg$^{-1}$; (more than) half of the total instability is therefore realized by the hailstorms in these environments and setup (Fig. 17a1, 17b1, 17c1). HE1 (E) attains the steepest LLR at 7.70 °C km$^{-1}$, while also having less moisture across the atmospheric column at PWAT < 50 mm compared to other hail environments. It is worth mentioning that modest IQR was also present in the HE1 (E) temperature-dewpoint profile, compared to other easterly-driven hail environments with HE2 (E) and HE3 (E), particularly their dewpoint profile, diverges around 700-hPa.

For westerly setups, on the other hand, hail environments tend to be accompanied by drier-mid levels compared to their easterly counterparts while still attaining MUCAPE > 4000 J kg$^{-1}$, except for HE3 (W) with $RH_{16}$ of 65% and MUCAPE of 3453 J kg$^{-1}$ likely affected by the limited samples for such cases (Fig. 17a2, 17b2, 17c2). Similar to HE1 (E), the HE1 (W) also had an IQR that spans a wider range compared to its easterly counterpart. It is also accompanied by lesser PWAT among the rest at 44.45 mm, but with higher LCL and low FZL indicative of slightly shallow low-level depth. Meanwhile, HE2 (W) is accompanied by the steepest LLR among the rest of environmental setups at 8.01 °C km$^{-1}$ with moisture depth extending around 850-hPa, and slightly drier compared to HE1 (W) with $RH_{16}$ at 57%. Compared to other setups and environments, this sector also displays more well mixed low-levels owing to the steep lapse rates and large MUCAPE, combined with drier mid-levels, also results in large DCAPE close to 1000 J kg$^{-1}$.

In conjunction, the kinematics seen through the hodograph of each environmental setup is analyzed, as shown in Fig. 18. For easterly setups across all hail environments (Fig. 18a1, 18b1, 18c1), they are consistent with the recent results I also achieved where straight hodographs are seen, accompanied by weak LLS (≤1 m s$^{-1}$), including DLS (≤5 m s$^{-1}$). But composed by strong $BWD_{LCL\text{-}EL}$ > 10 m s$^{-1}$ due to unidirectional wind profile i.e., where speed shear exists but lacks directional shear, with HE2 (E) having a shear vector magnitude within this cloud depth of 14.73 m s$^{-1}$. The storm-relative winds in the lowest 1 km are also the same, leaning around 8 m s$^{-1}$ across all environmental setups. Meanwhile, the straight hodograph is still depicted for HE1 (W), except for both HE2 (W) (at some point) and HE3 (W) where it suffers due to the lack of samples (Fig. 18a2, 18b2, 18c2). Both LLS and DLS remained weak, including $BWD_{LCL\text{-}EL}$ for the last two westerly setups, except for HE1 (W) which attained a large $BWD_{LCL\text{-}EL}$ of 19.52 m $s^{-1}$ likely influenced by the geographic location and its proximity to sub-tropics where synoptic forcing i.e., frontal systems can exist. It is worth mentioning that all of our westerly setups, regardless of location in Luzon, all feature upper-level winds that suddenly veered to the east, and may have some influence on storm maintenance and hail production. Although such a feature was not analyzed in this matter as it can be linked to the synoptic pattern where upper-level northeasterly flow can exist in the tropics. Still, this suggest that these hodographs, particularly for the easterly severe weather setups – regardless of geographical sector, are favorable for hail-producing storms as these hail hodographs do not impart excessive momentum on air parcels thanks to the lack of storm inflow and LLS, hence will not shut down hail growth and/or limit the residence time in the HGZ (Kumjian et al., 2021; Lin and Kumjian, 2022; Nixon and Allen, 2022).

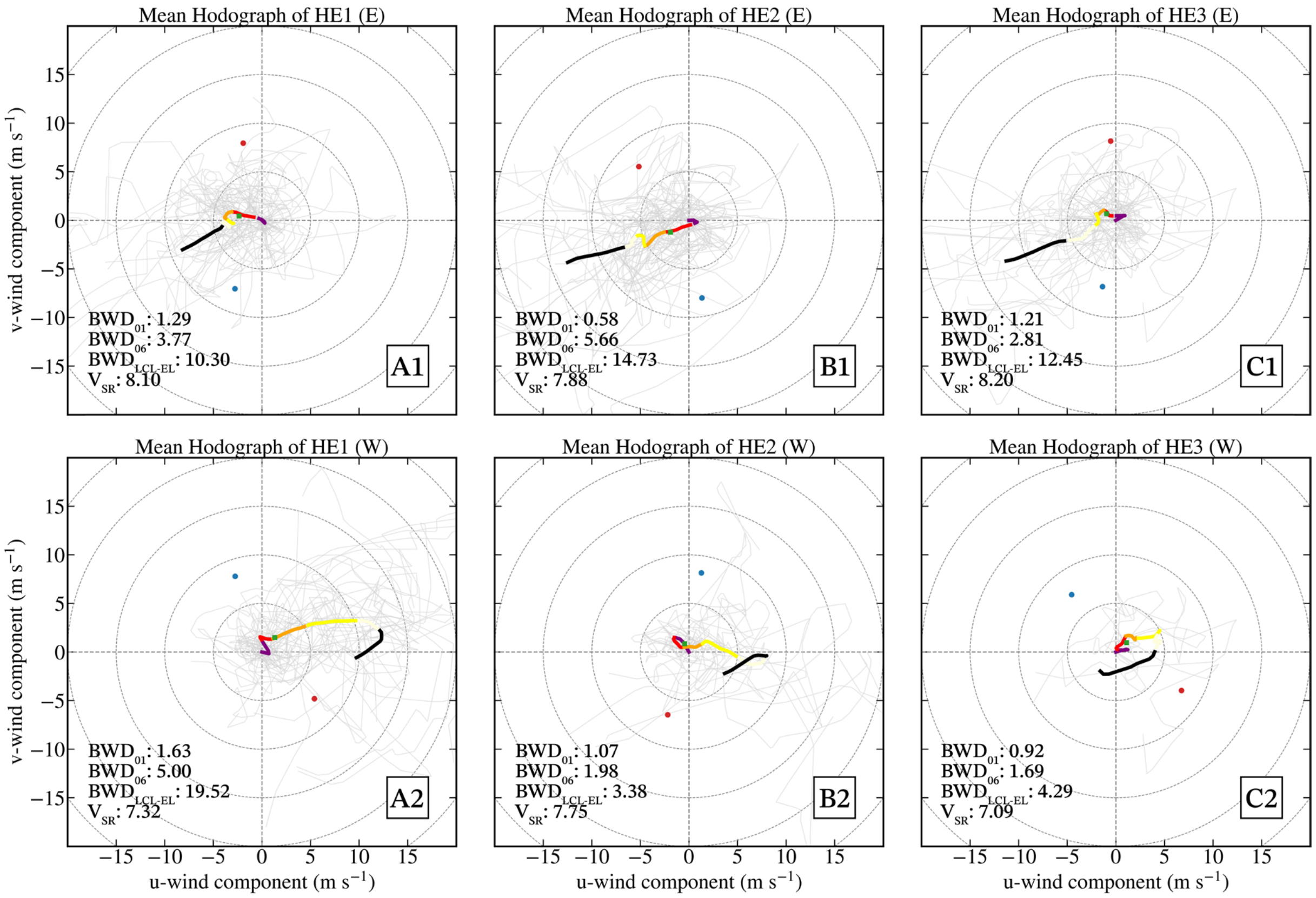

**Figure 18.** Similar to Fig. 17 as it was classified based on the hodograph profile for; (1) easterly severe weather setup and (2) westerly severe weather setup across (a) HE1, (b) HE2, and (c) HE3.

# 4. Discussion

The overarching goal of this study is to establish a baseline climatology of sounding-derived parameters relevant to hail-producing thunderstorms across the Luzon landmass. The focus lies in characterizing the types of regional hail environments while accounting for variations in topographic setting, and in assessing the climatological occurrence and distribution of parameters that are physically important for operational meteorology, particularly in hail forecasting. To this end, I identified climatologically large or extreme values of both thermodynamic and kinematic parameters, including the introduction of new parameters, that may prove useful for forecasting applications. Finally, I examined in detail the thermodynamic and kinematic structures associated with hail-supportive environments to place these climatological findings in the context of physical processes that govern hail development. Below, the most important details and their relationship/comparison to previous studies are highlighted.

## 4.1 Spatio-Temporal Analysis of Hail Events

Aside from mapping hail events across the Luzon landmass, a climatology of thunder hours (as a proxy for lightning and thunderstorm) was also constructed to identify hot spots that may harbor severe weather activity like hail. Results reveal that most of the hail events occur across the GMMR and the high plains of Cordillera Provinces and Northern Luzon. These hail events also coincide with increased thunderstorm activity within the regions of interest (>100 hr $yr^{-1}$), predominantly lying in the western and central section of Luzon. These areas are characterized by enhanced mesoscale variability, where convective initiation and development are strongly influenced by terrain-induced circulations, diurnal temperature oscillations, and land-sea interactions, especially along narrow coastlines (e.g., Rotunno and Ferretti, 2001).

The joint diurnal and monthly distribution suggest that hail activity peaks between 06-08 UTC (14-16 LST), with a clear increase beginning in April and a maximum in May. Consequently, the number of thunder events also coincides with hail activity across the Luzon landmass. This indicates, also verifies the previous in Capuli (2024), that hail occurrence is strongly tied to the broader convective regime of the pre-monsoon season where easterlies tend to dominate, when thermodynamic and kinematic environments become increasingly favorable for deep convection and severe weather. The transition from the dry to wet season provides enhanced surface heating, higher boundary-layer moisture, and sufficient wind shear, all of which collectively support storm intensification and hail production.

Not only that, but the diurnal distribution of hail and thunderstorm activity reported in Luzon matches the findings in several climatologies related to severe convective hazards (Groenemeijer and Kühne, 2014; Chen et al., 2018; Miglietta and Matsangouras, 2018; León-Cruz et al., 2022) and is also consistent with prior work in Capuli (2024). In Part I of Project SWAP, it was confirmed that hail events generally occur in the afternoon between 06-08 UTC, with secondary maxima toward 10 UTC. Moreover, the temporal findings, such as the monthly activity and diurnal window of hail occurrence, are in line with the climatological results of Mahavik et al. (2025) in Thailand. Both Project SWAP Part I and the recently published study by Mahavik et al. (2025) underscore the role of strong surface heating, convective instability, ample kinematics such as DLS, and local topographic influences as the primary ingredients driving hail formation and its observed diurnal cycle. The predominance of afternoon hailstorms suggests that localized convective developments, enhanced by orographic effects, play a central role in initiating hail-producing thunderstorms.

## 4.2 Thermodynamic Parameters

The probability for hail maximizes with high boundary layer moisture (analysis in Section 3.3, to be discussed below), steep mid- and low-level lapse rates (resulting in high CAPE), and a high LCL height. Low-level lapse rates are considerably higher over the Luzon landmass compared to Europe and the United States (Taszarek et al., 2020), consistent with results for 0-3 km CAPE. In addition, several authors i.e., Rasmussen and Blanchard (1998), Groenemeijer and van Delden (2007), Kaltenbock et al. (2009), Grams et al. (2012) and Púčik et al. (2015) also noted that LCL during hail days was higher compared with non-severe thunderstorms. This study confirms the notion that hail events are associated with higher LCL (>700-800 m) heights and higher CAPE; as caused by the steep lapse rates (whether MU or ML-parcel was used), compared to thunderstorms in general, seen across all hail environments in Luzon. Numerical simulations performed by McCaul and Cohen (2002) support this as they showed that the updraft intensity and storm diameter generally increased as LCL was increased. Not only the diameter of the severe storm, the storm's updraft area within the HGZ also expands with increased CAPE, as shown in previous studies (McCaul and Weisman, 2001; Markowski and Richardson, 2009; Kirkpatrick et al., 2011; Peters et al., 2019). Therefore, one of the physical mechanisms for this preference in a tropical environment is likely linked to updraft width and intensity.

Although the environmental instability as measured through MUCAPE and MLCAPE acquired in this study are quite extreme (>4000 J $kg^{-1}$ and > 2000 J $kg^{-1}$, respectively), these hail-producing storms may only realize around 50% of their CAPE due to entrainment effects diluting the updraft, as seen when the ECAPE-to-CAPE ratio was computed (Peters et al., 2023). It is worth noting that this 50% can act as a threshold, as Peters et al. (2023) noted that thunderstorms that realize 50% of their CAPE can be classified as severe thunderstorms, particularly supercells. Still, with large CAPE and even low DLS (as seen in Fig. 8b) conditions can be also related to local downbursts in deep and dry mid-level layers (Wakimoto, 1985; Holmes and Oliver, 2000). And this can be seen with large DCAPE distribution across all hail environments in Luzon with DCAPE > 800 J $kg^{-1}$, thus has a predictive value not only for hail environments, but for severe wind events i.e., downbursts/downdrafts. Therefore, one of the physical mechanisms for this preference in a tropical environment is likely linked to updraft width and intensity.

## 4.3 Kinematic Parameters

Strong vertical shear has long been associated with both convective organization and a corresponding severity (Brooks et al., 2003; Trapp et al., 2007; Allen et al., 2011; Brooks, 2013; Púčik et al., 2015; Taszarek et al., 2017). Firstly, both the LLS and $V_{SR}$ across all hail environments in Luzon are weak (LLS < 5 m $s^{-1}$; $V_{SR}$ ~7-8 m $s^{-1}$). The result is inline to the previous findings of Púčik et al. (2015) and Taszarek et al. (2020) who have found weak shear vector magnitudes for hail setups, whether in the United States and/or Europe. Typically, strong LLS and $V_{SR}$ are linked to increased probabilities for tornadoes such as that they require sufficient and ample streamwise vorticity in the lowest few kilometers of the troposphere e.g., either 500 m or 1 km (Coffer and Parker, 2018; Nixon and Allen, 2022). These parameters better distinguish between nonsevere thunderstorms and tornadoes based on the climatological analysis conducted by Taszarek et al. (2020). However, Nixon et al. (2023) have found out that weak LLS and $V_{SR}$ is required for hail growth and residence, especially in environments that have strong CAPE as found in this study.

Kaltenbock et al. (2009) defined DLS differently than I did, so that a comparison of absolute values is not possible: their bulk shear was computed between the lowest model level and the 500-hPa level, whereas we have computed it from 10 m AGL to 6 km AGL, inline to Púčik et al. (2015). This study shows that DLS does not discriminate hail environments given the fact that they are nearly identical across sectors i.e., weak DLS (<10 m $s^{-1}$). This result contradicts previous studies by Groenemeijer and van Delden (2007), Taszarek and Kolendowicz (2013), Púčik et al. (2015), Taszarek et al. (2017), and Nixon and Allen (2022) who all have found that hail environments tend to be associated with elevated DLS magnitude. The weak DLS observed in our cases can be explained by the inherently weak temperature-pressure gradients and mid-level winds in the tropics, which limit speed shear magnitude, except during the close approach of tropical cyclones.

However, given the generally weak nature of DLS in the Luzon landmass and tropics (León-Cruz, 2025, Sari and Lasher-Trapp, 2025), it is necessary to explore other wind-based parameters that may better discriminate hail-bearing storms from non-severe thunderstorms. Capuli (2025) proposed that the bulk wind difference between the LCL and EL shows forecasting skill in identifying hail-supporting environments, as hodographs associated with hail tend to exhibit strong speed shear rather than directional shear, often resembling nearly straight-line wind profiles. In a case study of the 13 August 2021 hailstorm, which produced hailstones measuring significant hailstones at 5-8 cm, the ambient $BWD_{LCL\text{-}EL}$ was approximately 30 m $s^{-1}$. Motivated by this, I examined climatological hail cases and found that more than half were associated with $BWD_{LCL\text{-}EL}$ values exceeding 10 m $s^{-1}$, a threshold consistent with severe convective storm classification and functionally analogous to the 'high CAPE, high shear' setups when conventional DLS is used. These findings align with the stipulations of Nixon et al. (2023), particularly for the context of the Luzon landmass, where hail environments are characterized by large undiluted CAPE and strong shear between the LCL and EL, but comparatively weak LLS and $V_{SR}$, which favor hail production without imparting excessive momentum to the rising air parcels. In fact, simulated storms conducted by Dennis and Kumjian (2017) show that a strong zonal component to the shear vector could potentially enhance hail production by elongating the storm updraft in the direction of the shear vector (in most cases, easterly), thereby increasing the volume in which hailstones can collect mass and increase in size.

## 4.4 Moisture Parameters

Evaluation of the moisture within the low-levels and mid-levels which is subject to entrainment effects ($RH_{13}$ and $RH_{16}$) and the vertically integrated moisture (PWAT) indicate considerable differences across hail environments. Taszarek et al. (2017) asserted that increasing moisture content generally favors increasing hail growth (and even size), especially when low-levels are considered. Further, when compared to Europe and United States, the PWAT of every hail environments across Luzon are generally much higher given that the Philippines is a tropical country that has a reservoir of moisture readily available, compared to midlatitude counterparts which requires synoptic low/patterns for moisture advection to the region (Carlson and Ludlam, 1968; Banacos and Ekster, 2010).

Meanwhile, the average moisture across several cloud depths (e.g., 1-3 km and 1-6 km) in Luzon's hail environments tends to be sub-saturated (>70%), while the 1-6 km layer moisture exhibits a wider range from relatively dry (<60%) to sub-saturated (>60%), likely influenced by prevailing winds. The mean moisture conditions, represented by $RH_{13}$, can be important in forecasting as Lin and Kumjian (2022) found that this specific layer can modulate the amount of LWC present in the HGZ. Afterall, air parcels originating in the low-levels have more vapor content available to condense into liquid. Meanwhile, when combined with weak DLS regimes, these high and low $RH_{16}$ conditions corroborate with the recent findings of Nixon et al. (2023), who noted that reduced $RH_{16}$ often coincides with stronger CAPE, as also observed in some of the hail events analyzed here. Likewise, some of the hail environments in Luzon tend to have large CAPE and high $RH_{16}$ which may counter entrainment effects. Since large CAPE is accompanied as well with stronger CAPE within and below HGZ; such that the ΔFZL is large, and weak $V_{SR}$/LLS, the likelihood of hail production increases substantially since the low-level inflow cannot induce erratic trajectories in the mid-levels of the updraft (HGZ). Thus, while this balance/relationship between buoyancy-low-level kinematics has been well established in the U.S. Great Plains, it also appears to manifest across hailstorm environments in Luzon landmass and, more broadly, in the tropics.

## 4.5 Composite Parameters

Considering hazards such as hail, the relative frequency over Luzon is typically 3-4 times lower than across the United States and Europe. Explanation for this reduced probability is showcased by the phase space of two severe thunderstorm predictors, CAPE and DLS, and their resulting combination of $W_{MAX}$SHEAR. Climatologically, as discussed by Púčik et al. (2015) and Taszarek et al. (2020) the occurrence of favorable instability-shear parameter phase space is much more frequent over the United States (6000-8000 J $kg^{-1}$) and Europe (3000-4000 J $kg^{-1}$), although Europe has generally higher probability for convective initiation. In our case, the $W_{MAX}$ distribution tends to be concentrated between 80-90 m $s^{-1}$, roughly 3000-4000 J $kg^{-1}$, comparable to that of Europe.

But, the weakness lies in the DLS where it is significantly weaker by a factor of 2 than the two midlatitude regions. It was found out by Taszarek et al. (2020) that severe thunderstorms are more likely to occur when shear is sufficient (e.g., at least 10-20 m $s^{-1}$) aside from just large instability. Our conflicting results of DLS say otherwise that these hail-bearing storms in Luzon landmass occur at much lower DLS. Lack of DLS can be explained by the weak temperature-pressure gradients in the tropics. This mirror conditions along the Gulf of Mexico coast and Florida (>2500 J $kg^{-1}$) accompanied by median 0-6 km shear between 5 and 10 m $s^{-1}$ (Brooks et al., 2003). Although, increased DLS for some events (which cannot be captured by ERA5) can also be explained by the proximity of mountain ranges (like the SMMR) where environmental wind shear is enhanced by interaction of the wind field with orography, similar to the cases where severe storms often initiate near the Alps (Kunz et al., 2018).

When modifying the $W_{MAX}$SHEAR parameter using $BWD_{LCL\text{-}EL}$, the phase space reveal that although probability typically increases in response to increasing instability, severe thunderstorms; in this case that can produce hail, are more likely to occur when the magnitude of the *deeper*-layer shear such as that the convective cloud depth is considered and sufficient (>10 m $s^{-1}$/20 knots), rather than when CAPE is large but shear is weak such as when DLS is instead in play (<10 m $s^{-1}$). As an example, an environment composed by $BWD_{LCL\text{-}EL}$ of 10 m $s^{-1}$ and $W_{MAX}$ of 90 m $s^{-1}$ (or around 3000 J $kg^{-1}$) means that the $W_{MAX}SHEAR_{LCL\text{-}EL}$ is at 900 $m^2$ $s^{-2}$. This is nearly comparable to the Great Plains setup that typically sees $W_{MAX}$SHEAR of 1100 $m^2$ $s^{-2}$ such that it is highly supportive for severe thunderstorms. Given that this comparison is just by substituting the shear vector parameter to identify hail environments along Luzon, this result is consistent with prior work (Brooks et al., 2003; Gensini and Ashley, 2011; Li et al., 2020). Combined with LCL (such as that the LCL > 700-900 m), the results indicate that the probability for hail increases along with both increasing (1) $W_{MAX}$SHEAR and (2) $W_{MAX}SHEAR_{LCL\text{-}EL}$, and along with LCL values.

Therefore, the phase space of the conventional $W_{MAX}$SHEAR (including the $W_{MAX}SHEAR_{LCL\text{-}EL}$) reveals that it is close to universally applicable not only for both European and the United States, but also for the tropics such as the Philippines with the added advantage of being a combination of only two parameters that captures much of the signal (Brooks, 2013; Taszarek et al., 2017). Good discriminators between hail-producing storms and non-severe thunderstorms considers both the $W_{MAX}$SHEAR and $W_{MAX}SHEAR_{LCL\text{-}EL}$. The combination of DLS and CAPE as a good hail forecasting tool was also indicated by Craven and Brooks (2004) and Púčik et al. (2015), and even when $BWD_{LCL\text{-}EL}$ was substituted.

## 4.6 Hail Environmental Profile

As shown consistently in HE1, the terrain's influence on hail occurrence is not without significance as it can modify the heat transport and may trigger convective initiation supportive of such SWE (Hohl et al., 2002; Barthlott et al., 2006; Changnon et al., 2009). The elevated terrain of HE1 means that they tend to have low FZLs, high LCLs that readily support hail embryos to be lofted to the HGZ. Meanwhile, HE2 and HE3 are the prototypical surface-based thunderstorms that tend to be severe especially during days when the environmental shear i.e., $BWD_{LCL\text{-}EL}$ is sufficient (>10 m $s^{-1}$). The identified hail environments achieve sufficient thermodynamic instability (and speed in $W_{MAX}$) due to the steep LLRs and MLRs aloft. Although, it should be noted that these hail-bearing storms only realize a half of the net CAPE due to the entrainment factors i.e. particularly lack of storm-relative inflow.

The analysis of the thermo-kinematic setups of hail environments across Luzon is inline to the previous findings by Nixon et al. (2023) who found a thermodynamic-kinematic balance that can allow hail growth in the updraft. Such as that large CAPE, albeit entrained, associated with weak LLS and storm inflow in the lowest 1 km is applicable and can harbor storms that produce hail in the identified sectors in Luzon, and likely across the Philippine archipelago. Furthermore, the wind profile displayed through the hodograph also confirms the presence of unidirectional shear i.e., sufficient speed shear as height increases but lacks directional shear and accompanied by weak LLS and $V_{SR}$. From here, hodographs that have this kind of characteristic; whether it be easterlies or westerlies are the prevailing wind, should be called hail hodographs to be consistent with current nomenclature. The hodograph shape this study achieved is also corroborates with the analysis performed by Nixon and Allen (2022) who differentiated hail-producing hodographs and tornado-producing hodographs. Furthermore, in the presence of unidirectional shear profiles, this creates symmetric vertical pressure gradients on the storm's flanks (essentially two low-pressure zones) and can support potential splitting of severe thunderstorms as the primary storm mode (Weisman and Klemp, 1986). Such a pattern is a simple but powerful supplement to currently used forecast parameters and is easy to implement in the forecast process.

Overall, the hail environments in Luzon, Philippines can be tied to storms that can produce large accumulations of small hail (SPLASH). As analyzed by Kumjian et al. (2019), these SCSs tend to harbor <2 cm hail whose environmental setups tend to have modest CAPE (>2000 J $kg^{-1}$), DLS that are not extreme (10-20 m $s^{-1}$), and modestly high moisture. Similarly, Philippine hail environments display comparable convective characteristics, except for generally weaker DLS, though compensated by sufficient wind shear within the convective cloud depth (from the LCL to EL). Synthesizing these factors, the physical mechanism driving Philippine hail environments is thermo-kinematic. Despite relatively weak LLS and DLS, the combination of extreme CAPE, high LCL, and adequate cloud-layer shear generates the intense vertical accelerations necessary to loft hydrometeors into the HGZ. Crucially, this weaker shear profile proves advantageous by preventing the rapid horizontal advection of hail embryos, thereby maximizing their residence time within the updraft core.

## 5. Conclusion

Here I focused on the baseline climatology of hail-bearing severe convective storm environments. ERA5 convective environments and severe weather reports from the Project SWAP (as in Part I) were combined on a common grid of 0.25° and 1-hr step for years 2005-2024. This is the first time that a baseline climatology was constructed to characterize hail environments across Luzon landmass by examining their topographic variations, climatological distributions of key severe storm parameters, extreme thermodynamic and kinematic conditions, and the thermodynamic and hodograph structures that support hail development for improved forecasting. High vertical resolution available with ERA5, including 28 levels up to 2 km AGL allowed exploration of convective parameters with greater confidence in their fidelity. This is especially important for variables that are sensitive to the number of available levels in a boundary layer such as CAPE, or LLS. Analysis yielded several findings, among which the most important are listed below:

1) Severe thunderstorms that produce hail (including non-severe ones) predominantly occur along the western portion of Luzon, including the Cordillera High Plains, the western Zambales Mountains, and the Greater Metro Manila Region (GMMR). These hailstorms are most frequent between April and October, peaking in May or June with diurnal hours between 06 and 08 UTC as the favored window of severe convective activity that entails a potential for hail. In particular, hail environments see average thunder hours >100 hr $yr^{-1}$, and can spike up to 200 hr $yr^{-1}$ annually.
2) Intensity of convective hazards increases with higher instability, considering hail events. Due to the steep LLRs and relatively moist cloud base, hail environments across Luzon tend to have MUCAPE > 4000 J $kg^{-1}$, and MLCAPE > 2000 J $kg^{-1}$. Luzon hail environments also tend to support downdraft events with DCAPE at excess of 800 J $kg^{-1}$ due to the dry mid-levels ($RH_{16}$ ~60-65%). Lastly, these hail environments are accompanied by relatively high LCLs and PWATs generally exceeding >45-50 mm, still indicative that moisture across the vertical column is sufficient for hail growth.
3) Along with convective support for storm initiation, the kinematics i.e., wind profile-based parameters are also in place for storm's severity and hail occurrence. While both LLS and $V_{SR}$ are generally weak, the $BWD_{LCL\text{-}EL}$ (substituting the conventional DLS product) is sufficient for elongating the updraft structure thereby increasing the volume over which hail growth processes occur. The role of weak LLS and $V_{SR}$ is important in keeping the balance with the ample instability, for hail embryos and its residence time within the updraft's HGZ.
4) For the composite parameters, although originally developed in Europe, the Hail Size Index (HSI) generally reflects the observed hail sizes in the Philippines, typically around 1 cm. Meanwhile, $W_{MAX}$SHEAR appears to be a reliable indicator of overall thunderstorm severity, particularly when applied to hail-producing events. While these parameters have been shown to yield stronger performance in regions such as the Great Plains of the United States and Europe, their calibration in the Philippine setting remains useful. To address the weak DLS commonly observed across Luzon hail environments since $W_{MAX}$ (or CAPE) often dominates the parameter space, a modified parameter, $W_{MAX}SHEAR_{LCL\text{-}EL}$, is introduced. This formulation incorporates the cloud depth between the LCL and EL, thereby allowing the representation of severe convective storm environments in Luzon to better align with those in regions more prone to severe hazards, such as the U.S. and Europe. This provides a new discriminator between hail-bearing and non-severe storms in Luzon (and/or in general, in the Philippines).
5) Luzon severe weather environments tend to have two modes: Easterly severe weather environments accompanied by prevailing easterlies, and Westerly severe weather environments aligned with the southwest monsoon and mid-level westerly wind flow. Most of the hail events are associated with easterly setups than westerly setups, accompanied by $BWD_{LCL\text{-}EL} \geqslant 10$ m s$^{-1}$.

6) The thermodynamic profiles across hail environments, whether under easterly or westerly severe weather regimes, are generally similar, with only minor differences observed in HE1 due to its elevated terrain setting. Complimentary, the hodograph shapes indicate weak LLS and $V_{SR}$, with little directional turning, but notable speed shear at the mid- and upper-levels. For easterly regimes, the resulting cloud-layer shear vector, as provided by the mid- and upper-level speed shear, tilts the storm's updraft and separates it from the downdraft, allowing hailstones to be re-ingested into the growth zone for multiple cycles. Meanwhile, hail environments in westerly regimes typically associated with low-level southwest monsoonal flow, are supported by mid-level westerly winds, which helps the storm maintain a vigorous updraft in a moisture-rich monsoon environment. Such as it displaces heavy water loading away from the updraft core preventing the weight of the accumulated rain from collapsing the storm and ensures that mid-level cloud hydrometeors remain suspended in the FZL long enough to grow into hail. These findings support the introduction of the $BWD_{LCL\text{-}EL}$ parameter as a useful predictor for hail potential in Luzon (and perhaps across the Philippines), while highlighting the role of straight-line hodograph shapes in characterizing hail-supportive environments.

In part, these results are broadly consistent with prior work on convective environments and their corresponding climatologies over the United States (Rasmussen and Blanchard, 1998; Brooks et al., 2003; Trapp et al., 2007; Gensini and Ashley, 2011; Thompson et al., 2012, 2013; Tippett et al., 2014; Allen et al., 2015; Li et al., 2020), Europe (Marsh et al., 2009; Mohr and Kunz, 2013; Púčik et al., 2015; Groenemeijer et al., 2017; Rädler et al., 2018; Taszarek et al., 2017, 2021a; Kunz et al., 2020), and other regions (Chen et al., 2018; Miglietta and Matsangouras, 2018; Ni et al., 2020; León-Cruz et al., 2022; Raupach et al., 2023; Sharma and Roy, 2023; Mahavik et al., 2025). Not only was it consistent, but it provided new insight into hail environments (both thermodynamics and kinematics in play) in the Philippines – a tropical country, particularly in Luzon. Also, it allows us to identify the different environments where hail-bearing storms can develop in the country, which is a first step for generating improvements in severe weather forecasting encompassing hailstorms, and even tornadic storms (subject for Part III). There are differences between environments obtained for the Luzon, Philippines and the aforementioned continents. For example, the environmental conditions across the U. S. Great Plains are associated with higher instability and LLS, thus conducive to tornadoes, as compared to Europe, and even more so in the Philippines.

Although this study utilized all available hailstorm samples and the most recent reanalysis data, further progress in developing more accurate and robust climatologies will benefit from prolonged measurement periods of lightning detection networks, improved storm data collection, and the higher spatial and temporal resolution expected from next-generation reanalysis datasets such as ERA6. Future work, particularly as SWEs continue to be archived under Project SWAP, should also examine the diurnal cycles of convective environments given the hourly resolution of ERA5 (and ERA6 in the near future). In addition, incorporating complementary datasets and methodologies such as numerical modeling or satellite-derived products that provide overshooting-top climatologies (e.g., Bedka et al., 2018; Giordani et al., 2024) will be valuable for advancing the understanding of severe weather in the Philippines. Despite the known limitations of ERA5, including its pressure-level constraints and inherent biases, its application remains essential in countries lacking advanced meteorological instrumentation (i.e., rawinsondes and weather radar data) is crucial to obtain information about convective environments. Such efforts are critical for improving knowledge of both the risks posed by severe convective storms and the potential changes in their characteristics under a warming climate.

As Brooks (2013) asserted, as climate changes, the magnitude of CAPE and shear is also changing. Given that different combinations of CAPE and shear favor the occurrence of different convective phenomena, this may provide insight into expected future changes in the distribution and nature of convective hazards. For example, if in the future both CAPE and shear increase, tornadoes and hailstorms will likely become more common. If low CAPE, high shear days will be more frequent, the number of severe convective wind gust events may increase. Bearing in mind these issues, I recommend that future studies address the question of how the distribution of severe thunderstorm ingredients will change, tailored in the context of the complex topography of the Philippines.

**Data Availability Statement.** Data used in this paper were derived from the ERA5 reanalysis (openly available through the **Climate Data Source**. The 2023 Philippines Administrative Level 0-4 shapefiles are available in **Humanitarian Data**

**Exchange**. The author's associated 1D vertical profile of standard ERA5 data and sounding data will be available soon through Project SWAP 4th Data Release (DR4) slated between Q1-Q2 of 2026. The Digital Elevation Model (DEM) is from Copernicus GLO-30 Digital Elevation Model distributed and available in **OpenTopography**. Finally, the Thunder Hour data is available at the **Earth Networks Thunder Hour Repository**. Proper attribution is required for these datasets. This paper has made use of the following Python packages: *Cartopy, GeoPandas, Matplotlib, MetPy, NumPy, Pandas, Rasterio, rioxarray, SciPy, SounderPy* (Gillett, 2025), *PtitPrince* (Allen et al., 2021), *and xarray*.

**Acknowledgements.** I am extensively grateful to the Philippine population for reporting the occurrence of the hailstones across the country. G. H. Capuli, the community project leader of Project SWAP, also appreciated the valuable comments of the three anonymous reviewers and editor, which helped to improve this manuscript. This work, as well as the project, received no funding, but was 'funded' by extensive and exhaustive effort, whose dedication and commitment to advancing our understanding of severe weather phenomena were indispensable. I am thankful to my family and loved ones for their unwavering support throughout this research.

## References

Albrecht, R. I., S. J. Goodman, D. E. Buechler, R. J. Blakeslee et al. (2016). Where are the lightning hotspots on Earth?, Bull. Amer. Meteor. Soc., 97, 11, 2051-2068, doi:10.1175/BAMS-D-14-00193.1.

Allen, J. T., D. J. Karoly and G. A. Mills (2011). A severe thunderstorm climatology for Australia and associated thunderstorm environments, Aust. Meteor. Oceanogr. J., 61, 3, 143, doi:10.22499/2.6103.001.

Allen, J. T., D. J. Karoly and K. J. Walsh (2014). Future Australian severe thunderstorm environments. Part II: The influence of a strongly warming climate on convective environments, J. Climate, 27, 10, 3848-3868, doi:10.1175/JCLI-D-13-00426.1.

Allen, J. T. and M. K. Tippett (2015). The characteristics of United States hail reports: 1955-2014, Electron. J. Severe Storms Meteor., 10, 3, doi:10.55599/ejssm.v10i3.60.

Allen, J. T., M. K. Tippett and A. H. Sobel (2015). An empirical model relating U.S. monthly hail occurrence to large-scale meteorological environment, J. Adv. Model. Earth Syst., 7, 226-243, doi:10.1002/2014MS000397.

Allen, J. T., I. M. Giammanco, M. R. Kumjian, H. J. Punge et al. (2020). Understanding hail in the Earth system, Rev. Geophys., 58, 1, e2019RG000665, doi:10.1029/2019RG000665.

Allen, M., D. Poggiali, K. Whitaker, T. R. Marshall et al. (2021). Raincloud plots: A multi-platform tool for robust data visualization, Wellcome Open Res., 4, 63, doi:10.12688/wellcomeopenres.15191.2.

Amburn, S. A. and P. L. Wolf (1997). VIL density as a hail indicator, Wea. Forecasting, 12, 3, 473-478, doi:10.1175/1520-0434(1997)012<0473:VDAAHI>2.0.CO;2.

Anderson, C. J., C. K. Wikle, Q. Zhou and J. A. Royle (2007). Population influences on tornado reports in the United States, Wea. Forecasting, 22, 3, 571-579, doi:10.1175/WAF997.1.

Ashley, W. S., A. J. Krmenec and R. Schwantes (2008). Vulnerability due to nocturnal tornadoes, Wea. Forecasting, 23, 5, 795-807, doi:10.1175/2008WAF2222132.1.

Atkins, N. T. and R. M. Wakimoto (1991). Wet microburst activity over the Southeastern United States: Implications for forecasting, Wea. Forecasting, 6, 4, 470-482, doi:10.1175/1520-0434(1991)006<0470:WMAOTS>2.0.CO;2.

Banacos, P. C. and M. L. Ekster (2010). The association of the elevated mixed layer with significant severe weather events in the Northeastern United States, Wea. Forecasting, 25, 4, 1082-1102, doi:10.1175/2010WAF2222363.1.

Barthlott, C., U. Corsmeier, C. Meißner, F. Braun et al. (2006). The influence of mesoscale circulation systems on triggering convective cells over complex terrain, Atmos. Res., 81, 2, 150-175, doi:10.1016/j.atmosres.2005.11.010.

Baumgardt, D. (2011). Hail estimation: How good are your spotters?, National Weather Service, La Crosse, WI, 24, https://www.weather.gov/media/arx/research/hail_size_MSP.pdf.

Bedka, K. M., J. T. Allen, H. J. Punge, M. Kunz et al. (2018). A long-term overshooting convective cloud-top detection database over Australia derived from MTSAT Japanese Advanced Meteorological Imager observations, J. Appl. Meteor. Climatol., 57, 4, 937-951, doi:10.1175/JAMC-D-17-0056.1.

Blair, S. F. and J. W. Leighton (2012). Creating high-resolution hail datasets using social media and post-storm ground surveys, Electron. J. Oper. Meteor., 13, 32-45, doi:10.15191/nwajom.

Blair, S. F., J. M. Laflin, D. E. Cavanaugh, K. J. Sanders et al. (2017). High-resolution hail observations: Implications for NWS warning operations, Wea. Forecasting, 32, 3, 1101-1119, doi:10.1175/WAF-D-16-0203.1.

Bourscheidt, V., K. L. Cummins, O. Pinto Jr. and K. P. Naccarato (2012). Methods to overcome lightning location system performance limitations on spatial and temporal analysis: Brazilian case, J. Atmos. Oceanic Technol., 29, 9, 1304-1311, doi:10.1175/JTECH-D-11-00213.1.

Brimelow, J. C., G. W. Reuter and E. R. Poolman (2002). Modeling maximum hail size in Alberta thunderstorms, Wea. Forecasting, 17, 5, 1048-1062, doi:10.1175/1520-0434(2002)017<1048:MMHSIA>2.0.CO;2.

Brimelow, J., W. Burrows and J. Hanesiak (2017). The changing hail threat over North America in response to anthropogenic climate change, Nat. Climate Change, 7, 516-522, doi:10.1038/nclimate3321.

Brooks, H. E., C. A. Doswell III and J. Cooper (1994). On the environments of tornadic and nontornadic mesocyclones, Wea. Forecasting, 9, 4, 606-618, doi:10.1175/1520-0434(1994)009<0606:OTEOTA>2.0.CO;2.

Browning, K. A. (1963). The growth of large hail within a steady updraught, Q. J. R. Meteorol. Soc., 89, 490-506, doi:10.1002/qj.49708938206.

Browning, K. A. and G. B. Foote (1976). Airflow and hail growth in supercell storms and some implications for hail suppression, Q. J. R. Meteorol. Soc., 102, 499-533, doi:10.1002/qj.49710243303.

Brooks, H. E., J. W. Lee and J. P. Craven (2003). The spatial distribution of severe thunderstorm and tornado environments from global reanalysis data, Atmos. Res., 67-68, 73-94, doi:10.1016/S0169-8095(03)00045-0.

Brooks, H. E. (2009). Proximity soundings for severe convection for Europe and the United States from reanalysis data, Atmos. Res., 93, 546-553, doi:10.1016/j.atmosres.2008.10.005.

Brooks, H. E. (2013). Severe thunderstorms and climate change, Atmos. Res., 123, 129-138, doi:10.1016/j.atmosres.2012.04.002.

Bunkers, M. J., B. A. Klimowski, J. W. Zeitler, R. L. Thompson et al. (2000). Predicting supercell motion using a new hodograph technique, Wea. Forecasting, 15, 1, 61-79, doi:10.1175/1520-0434(2000)015<0061:PSMUAN>2.0.CO;2.

Capuli, G. H. (2024). Project Severe Weather Archive of the Philippines (SWAP) Part 1: Establishing a baseline climatology for severe weather across the Philippine archipelago, Ann. Geophys., 67, 5, GC553, doi:10.4401/ag-9151.

Capuli, G. H. (2025). Friday the 13th hailstorm in the province of Bulacan, Philippines (13 August 2021): A case study, Asia-Pac. J. Atmos. Sci., 61, 13, doi:10.1007/s13143-025-00396-6.

Carlson, T. N. and F. H. Ludlam (1968). Conditions for the occurrence of severe local storms, Tellus, 20, 203-226, doi:10.1111/j.2153-3490.1968.tb00364.x.

Changnon, S. A., D. Changnon and S. D. Hilberg (2009). Hailstorms across the nation: An atlas about hail and its damages, Illinois State Water Survey, Contract Report No., CR-2009-12, https://www.isws.illinois.edu/pubdoc/cr/iswscr2009-12.pdf.

Chen, G. T.-J. and H.-C. Chou (2006). A summertime severe weather event occurring in the Taipei Basin, Terr. Atmos. Oceanic Sci., 17, 1, 3-22, doi:10.3319/TAO.2006.17.1.3(SWS).

Chen, J., X. Cai, H. Wang, L. Kang et al. (2018). Tornado climatology of China, Int. J. Climatol., 38, 5, 2478-2489, doi:10.1002/joc.5369.

Cifelli, R., N. Doesken, P. Kennedy, L. D. Carey et al. (2005). The Community Collaborative Rain, Hail, and Snow Network: Informal education for scientists and citizens, Bull. Amer. Meteor. Soc., 86, 8, 1069-1078, doi:10.1175/BAMS-86-8-1069.

Craven, J. P., H. E. Brooks and J. A. (2004). Baseline climatology of sounding derived parameters associated with deep moist convection, Natl. Wea. Dig., 28, 13-24.

Coffer, B. E. and M. D. Parker (2015). Impacts of increasing low-level shear on supercells during the early evening transition, Mon. Wea. Rev., 143, 5, 1945-1969, doi:10.1175/MWR-D-14-00328.1.

Coffer, B. E. and M. D. Parker (2018). Is there a "tipping point" between simulated nontornadic and tornadic supercells in VORTEX2 environments?, Mon. Wea. Rev., 146, 8, 2667-2693, doi:10.1175/MWR-D-18-0050.1.

Coffer, B. E., M. Taszarek and M. D. Parker (2020). Near-ground wind profiles of tornadic and nontornadic environments in the United States and Europe from ERA5 reanalyses, Wea. Forecasting, 35, 6, 2621-2638, doi:10.1175/WAF-D-20-0153.1.

Czernecki, B., M. Taszarek, M. Marosz, M. Półrolniczak et al. (2019). Application of machine learning to large hail prediction – The importance of radar reflectivity, lightning occurrence and convective parameters derived from ERA5, Atmos. Res., 227, 249-262, doi:10.1016/j.atmosres.2019.05.010.

Dennis, E. J. and M. R. Kumjian (2017). The impact of vertical wind shear on hail growth in simulated supercells, J. Atmos. Sci., 74, 3, 641-663, doi:10.1175/JAS-D-16-0066.1.

Diffenbaugh, N. S., M. Scherer and R. J. Trapp (2013). Robust increases in severe thunderstorm environments in response to greenhouse forcing, Proc. Natl. Acad. Sci. USA, 110, 16361-16366, doi:10.1073/pnas.1307758110.

DiGangi, E. A., M. Stock and J. Lapierre (2022). Thunder hours: How old methods offer new insights into thunderstorm climatology, Bull. Amer. Meteor. Soc., 103, 2, E548-E569, doi:10.1175/BAMS-D-20-0198.1.

Dobur, J. C. (2005). A comparison of severe thunderstorm warning verification statistics and population density within the NWS Atlanta county warning area, 4th Annu. Severe Storms Symp., Preprints Starkville, MS, D2-6.

Doswell, C. A. III and E. N. Rasmussen (1994). The effect of neglecting the virtual temperature correction on CAPE calculations, Wea. Forecasting, 9, 4, 625-629, doi:10.1175/1520-0434(1994)009<0625:TEONTV>2.0.CO;2.

Doswell, C. A. III, H. E. Brooks and R. A. Maddox (1996). Flash flood forecasting: An ingredients-based methodology, Wea. Forecasting, 11, 4, 560-581, doi:10.1175/1520-0434(1996)011<0560:FFFAIB>2.0.CO;2.

Doswell, C. A. III and J. S. Evans (2003). Proximity sounding analysis for derechos and supercells: an assessment of similarities and differences, Atmos. Res., 67-68, 117-133, doi:10.1016/S0169-8095(03)00047-4.

Duda, J. D. and W. A. Gallus Jr. (2010). Spring and summer Midwestern severe weather reports in supercells compared to other morphologies, Wea. Forecasting, 25, 1, 190-206, doi:10.1175/2009WAF2222338.1.

Elmore, K. L., Z. L. Flamig, V. Lakshmanan, B. T. Kaney et al. (2014). mPING: Crowd-sourcing weather reports for research, Bull. Amer. Meteor. Soc., 95, 9, 1335-1342, doi:10.1175/BAMS-D-13-00014.1.

Elmore, K. L., J. T. Allen and A. E. Gerard (2022). Sub-severe and severe hail, Wea. Forecasting, 37, 8, 1357-1369, doi:10.1175/WAF-D-21-0156.1.

Emanuel, K. A. (1994). Atmospheric Convection, Oxford University Press, New York, 580, ISBN:9780195066302.

Gensini, V. A. and W. S. Ashley (2011). Climatology of potentially severe convective environments from North American regional reanalysis, Electron. J. Severe Storms Meteor., 6, 8, 1-40, doi:10.55599/ejssm.v6i8.35.

Gillett, K. J. (2025). SounderPy: An atmospheric sounding visualization and analysis tool for Python, J. Open Source Softw., 10, 112, 8087, doi:10.21105/joss.08087.

Giordani, A., M. Kunz, K. M. Bedka, H. J. Punge et al. (2024). Characterizing hail-prone environments using convection-permitting reanalysis and overshooting top detections over south-central Europe, Nat. Hazards Earth Syst. Sci., 24, 7, 2331-2357, doi:10.5194/nhess-24-2331-2024.

Grams, J. S., R. L. Thompson, D. V. Snively, J. A. Prentice et al. (2012). A climatology and comparison of parameters for significant tornado events in the United States, Wea. Forecasting, 27, 1, 106-123, doi:10.1175/WAF-D-11-00008.1.

Grant, L. D. and S. C. van den Heever (2014). Microphysical and dynamical characteristics of low-precipitation and classic supercells, J. Atmos. Sci., 71, 7, 2604-2624, doi:10.1175/JAS-D-13-0261.1.

Groenemeijer, P. H. and A. Van Delden (2007). Sounding-derived parameters associated with large hail and tornadoes in the Netherlands, Atmos. Res., 83, 473-487, doi:10.1016/j.atmosres.2005.08.006.

Groenemeijer, P. and T. Kühne (2014). A climatology of tornadoes in Europe: Results from the European Severe Weather Database, Mon. Wea. Rev., 142, 12, 4775-4790, doi:10.1175/MWR-D-14-00107.1.

Groenemeijer, P., T. Púčik, A. M. Holzer, B. Antonescu et al. (2017). Severe convective storms in Europe: Ten years of research and education at the European Severe Storms Laboratory, Bull. Amer. Meteor. Soc., 98, 12, 2641-2651, doi:10.1175/BAMS-D-16-0067.1.

Gutierrez, R. E. and M. R. Kumjian (2021). Environmental and radar characteristics of gargantuan hail-producing storms, Mon. Wea. Rev., 149, 8, 2523-2538, doi:10.1175/MWR-D-20-0298.1.

Hersbach, H., B. Bell, P. Berrisford et al. (2020). The ERA5 global reanalysis, Q.J.R. Meteorol. Soc., 146, 1999-2049, doi:10.1002/qj.3803.

Hohl, R., H.-H. Schiesser and D. Aller (2002). Hailfall: the relationship between radar-derived hail kinetic energy and hail damage to buildings, Atmos. Res., 63, 3, 177-207, doi:10.1016/S0169-8095(02)00059-5.

Holmes, J. D. and S. E. Oliver (2000). An empirical model of a downburst, Eng. Struct., 22, 1167-1172, doi:10.1016/S0141-0296(99)00058-9.

Homeyer, C. R., E. M. Murillo and M. R. Kumjian (2023). Relationships between 10 years of radar-observed supercell characteristics and hail potential, Mon. Wea. Rev., 151, 10, 2609-2632, doi:10.1175/MWR-D-23-0019.1.

Hutchins, M. L., R. H. Holzworth, J. B. Brundell and C. J. Rodger (2012). Relative detection efficiency of the World Wide Lightning Location Network, Radio Sci., 47, RS6005, doi:10.1029/2012RS005049.

Jayaratne, E. R. and V. Ramachandran (1998). A five-year study of lightning activity using a CGR3 flash counter in Gaborone, Botswana, Meteor. Atmos. Phys., 66, 235-241, doi:10.1007/BF01026636.

Johns, R. H. and C. A. Doswell (1992). Severe local storms forecasting, Wea. Forecasting, 7, 4, 588-612, doi:10.1175/1520-0434(1992)007<0588:SLSF>2.0.CO;2.

Johnson, A. W. and K. E. Sugden (2014). Evaluation of sounding-derived thermodynamic and wind-related parameters associated with large hail events, Electron. J. Severe Storms Meteor., 9, 5, doi:10.55599/ejssm.v9i5.57.

Kaltenböck, R., G. Diendofer and N. Dotzek (2009). Evaluation of thunderstorm indices from ECMWF analyses, lightning data and severe storms reports, Atmos. Res., 93, 381-396, doi:10.1016/j.atmosres.2008.11.005.

Kaltenboeck, R. and A. Ryzhkov (2013). Comparison of polarimetric signatures of hail at S and C bands for different hail sizes, Atmos. Res., 123, 323-336, doi:10.1016/j.atmosres.2012.05.013.

Kirkpatrick, C., E. W. McCaul Jr. and C. Cohen (2011). Sensitivities of simulated convective storms to environmental CAPE, Mon. Wea. Rev., 139, 11, 3514-3532, doi:10.1175/2011MWR3631.1.

King, A. T. and A. D. Kennedy (2019). North American supercell environments in atmospheric reanalyses and RUC-2, J. Appl. Meteor. Climatol., 58, 1, 71-92, doi:10.1175/JAMC-D-18-0015.1.

Knight, N. C. (1981). The climatology of hailstone embryos, J. Appl. Meteor. Climatol., 20, 7, 750-755, doi:10.1175/1520-0450(1981)020<0750:TCOHE>2.0.CO;2.

Knight, C. A. and N. C. Knight (2001). Hailstorms, Meteor. Monogr., 28, 50, 223-248, doi:10.1175/0065-9401-28.50.223.

Knight, C. A. and N. C. Knight (2005). Very large hailstones from Aurora, Nebraska, Bull. Amer. Meteor. Soc., 86, 12, 1773-1782, doi:10.1175/BAMS-86-12-1773.

Kumjian, M. R. and A. V. Ryzhkov (2008). Polarimetric signatures in supercell thunderstorms, J. Appl. Meteor. Climatol., 47, 7, 1940-1961, doi:10.1175/2007JAMC1874.1.

Kumjian, M. R., Z. J. Lebo and A. M. Ward (2019). Storms producing large accumulations of small hail, J. Appl. Meteor. Climatol., 58, 2, 341-364, doi:10.1175/JAMC-D-18-0073.1.

Kumjian, M. R. and K. Lombardo (2020). A hail growth trajectory model for exploring the environmental controls on hail size: Model physics and idealized tests, J. Atmos. Sci., 77, 8, 2765-2791, doi:10.1175/JAS-D-20-0016.1.

Kumjian, M. R., K. Lombardo and S. Loeffler (2021). The evolution of hail production in simulated supercell storms, J. Atmos. Sci., 78, 11, 3417-3440, doi:10.1175/JAS-D-21-0034.1.

Kunz, M., U. Blahak, J. Handwerker, M. Schmidberger et al. (2018). The severe hailstorm in Southwest Germany on 28 July 2013: characteristics, impacts and meteorological conditions, Q. J. R. Meteorol. Soc., 144, 231-250, doi:10.1002/qj.3197.

Kunz, M., J. Wandel, E. Fluck, S. Baumstark et al. (2020). Ambient conditions prevailing during hail events in central Europe, Nat. Hazards Earth Syst. Sci., 20, 6, 1867-1887, doi:10.5194/nhess-20-1867-2020.

Lagmay, A. M. F., G. Bagtasa, I. A. Crisologo, B. A. B. Racoma et al. (2015). Volcanoes magnify Metro Manila's southwest monsoon rains and lethal floods, Front. Earth Sci., 2, 2014, doi:10.3389/feart.2014.00036.

León-Cruz, J. F., L. F. Pineda-Martínez and N. Carbajal (2022). Tornado climatology and potentially severe convective environments in Mexico, Clim. Res., 87, 147-165, doi:10.3354/cr01692.

León-Cruz, J. F. (2025). Tornadic environments in Mexico, Atmos. Res., 315, 107916, doi:10.1016/j.atmosres.2025.107916.

Li, F., D. R. Chavas, K. A. Reed and D. T. Dawson II (2020). Climatology of severe local storm environments and synoptic-scale features over North America in ERA5 reanalysis and CAM6 simulation, J. Climate, 33, 19, 8339-8365, doi:10.1175/JCLI-D-19-0986.1.

Lin, Y. and M. R. Kumjian (2022). Influences of CAPE on hail production in simulated supercell storms, J. Atmos. Sci., 79, 1, 179-204, doi:10.1175/JAS-D-21-0054.1.

Mahavik, N., A. Kangerd, J. Kunwilai and S. Arthayakun (2025). Spatiotemporal characteristics and diurnal patterns of pre-monsoon hailstorms in northern Thailand, Nat. Hazards, 121, 18057-18089, doi:10.1007/s11069-025-07505-8.

Marion, G. R. and R. J. Trapp (2019). The dynamical coupling of convective updrafts, downdrafts, and cold pools in simulated supercell thunderstorms, J. Geophys. Res. Atmos., 124, 664-683, doi:10.1029/2018JD029055.

Markowski, P. M. and Y. P. Richardson (2009). Tornadogenesis: Our current understanding, forecasting considerations, and questions to guide future research, Atmos. Res., 93, 3-10, doi:10.1016/j.atmosres.2008.09.015.

Marsh, P. T., H. E. Brooks and D. J. Karoly (2009). Preliminary investigation into the severe thunderstorm environment of Europe simulated by the Community Climate System Model 3, Atmos. Res., 93, 607-618, doi:10.1016/j.atmosres.2008.09.014.

McCaul, E. W. Jr. and M. L. Weisman (2001). The sensitivity of simulated supercell structure and intensity to variations in the shapes of environmental buoyancy and shear profiles, Mon. Wea. Rev., 129, 4, 664-687, doi:10.1175/1520-0493(2001)129<0664:TSOSSS>2.0.CO;2.

McCaul, E. W. Jr. and C. Cohen (2002). The impact on simulated storm structure and intensity of variations in the mixed layer and moist layer depths, Mon. Wea. Rev., 130, 7, 1722-1748, doi:10.1175/1520-0493(2002)130<1722:TIOSSS>2.0.CO;2.

Miglietta, M. M. and I. T. Matsangouras (2018). An updated "climatology" of tornadoes and waterspouts in Italy, Int. J. Climatol., 38, 9, 3667-3683, doi:10.1002/joc.5526.

Miller, L. J., J. D. Tuttle and C. A. Knight (1988). Airflow and hail growth in a severe Northern High Plains supercell, J. Atmos. Sci., 45, 4, 736-762, doi:10.1175/1520-0469(1988)045<0736:AAHGIA>2.0.CO;2.

Mohr, S. and M. Kunz (2013). Recent trends and variabilities of convective parameters relevant for hail events in Germany and Europe, Atmos. Res., 123, 211-228, doi:10.1016/j.atmosres.2012.05.016.

Nelson, S. P. (1983). The influence of storm flow structure on hail growth, J. Atmos. Sci., 40, 8, 1965-1983, doi:10.1175/1520-0469(1983)040<1965:TIOSFS>2.0.CO;2.

Ni, X., A. Muehlbauer, J. T. Allen, Q. Zhang et al. (2020). A climatology and extreme value analysis of large hail in China, Mon. Wea. Rev., 148, 4, 1431-1447, doi:10.1175/MWR-D-19-0276.1.

Nixon, C. J. and J. T. Allen (2022). Distinguishing between hodographs of severe hail and tornadoes, Wea. Forecasting, 37, 10, 1761-1782, doi:10.1175/WAF-D-21-0136.1.

Nixon, C. J., J. T. Allen and M. Taszarek (2023). Hodographs and skew Ts of hail-producing storms, Wea. Forecasting, 38, 11, 2217-2236, doi:10.1175/WAF-D-23-0031.1.

Peters, J. M., C. J. Nowotarski and H. Morrison (2019). The role of vertical wind shear in modulating maximum supercell updraft velocities, J. Atmos. Sci., 76, 10, 3169-3189, doi:10.1175/JAS-D-19-0096.1.

Peters, J. M., C. J. Nowotarski, J. P. Mulholland and R. L. Thompson (2020). The influences of effective inflow layer streamwise vorticity and storm-relative flow on supercell updraft properties, J. Atmos. Sci., 77, 9, 3033-3057, doi:10.1175/JAS-D-19-0355.1.

Peters, J. M., J. P. Mulholland and D. R. Chavas (2022). Generalized lapse rate formulas for use in entraining CAPE calculations, J. Atmos. Sci., 79, 3, 815-836, doi:10.1175/JAS-D-21-0118.1.

Peters, J. M., D. R. Chavas, C. Su, H. Morrison et al. (2023). An analytic formula for entraining CAPE in midlatitude storm environments, J. Atmos. Sci., 80, 9, 2165-2186, doi:10.1175/JAS-D-23-0003.1.

Pilguj, N., M. Taszarek, J. T. Allen and K. A. Hoogewind (2022). Are trends in convective parameters over the United States and Europe consistent between reanalyses and observations?, J. Climate, 35, 12, 3605-3626, doi:10.1175/JCLI-D-21-0135.1.

Picca, J. and A. Ryzhkov (2012). A dual-wavelength polarimetric analysis of the 16 May 2010 Oklahoma City extreme hailstorm, Mon. Wea. Rev., 140, 4, 1385-1403, doi:10.1175/MWR-D-11-00112.1.

Pilorz, W., M. Zięba, J. Szturc and E. Łupikasza (2022). Large hail detection using radar-based VIL calibrated with isotherms from the ERA5 reanalysis, Atmos. Res., 274, 106185, doi:10.1016/j.atmosres.2022.106185.

Potvin, C. K., K. L. Elmore and S. J. Weiss (2010). Assessing the impacts of proximity sounding criteria on the climatology of significant tornado environments, Wea. Forecasting, 25, 3, 921-930, doi:10.1175/2010WAF2222368.1.

Prein, A. F. and A. J. Heymsfield (2020). Increased melting level height impacts surface precipitation phase and intensity, Nat. Climate Change, 10, 771-776, doi:10.1038/s41558-020-0825-x.

Púčik, T., P. Groenemeijer, D. Rýva and M. Kolář (2015). Proximity soundings of severe and nonsevere thunderstorms in central Europe, Mon. Wea. Rev., 143, 12, 4805-4821, doi:10.1175/MWR-D-15-0104.1.

Púčik, T., C. Castellano, P. Groenemeijer, T. Kühne et al. (2019). Large hail incidence and its economic and societal impacts across Europe, Mon. Wea. Rev., 147, 11, 3901-3916, doi:10.1175/MWR-D-19-0204.1.

Rädler, A. T., P. Groenemeijer, E. Faust and R. Sausen (2018). Detecting severe weather trends using an additive regressive convective hazard model (AR-CHaMo), J. Appl. Meteor. Climatol., 57, 3, 569-587, doi:10.1175/JAMC-D-17-0132.1.

Rasmussen, R. M. and A. J. Heymsfield (1987). Melting and shedding of graupel and hail. Part I: Model physics, J. Atmos. Sci., 44, 19, 2754-2763, doi:10.1175/1520-0469(1987)044<2754:MASOGA>2.0.CO;2.

Rasmussen, E. N. and D. O. Blanchard (1998). A baseline climatology of sounding-derived supercell and tornado forecast parameters, Wea. Forecasting, 13, 4, 1148-1164, doi:10.1175/1520-0434(1998)013<1148:ABCOSD>2.0.CO;2.

Raupach, T. H., J. S. Soderholm, R. A. Warren and S. C. Sherwood (2023). Changes in hail hazard across Australia: 1979-2021, npj Clim. Atmos. Sci., 6, 1, 143, doi:10.1038/s41612-023-00454-8.

Reges, H. W., N. Doesken, J. Turner, N. Newman et al. (2016). CoCoRaHS: The evolution and accomplishments of a volunteer rain gauge network, Bull. Amer. Meteor. Soc., 97, 10, 1831-1846, doi:10.1175/BAMS-D-14-00213.1.

Rodger, C. J., J. B. Brundell, R. H. Holzworth, E. D. Douma et al. (2017). The World Wide Lightning Location Network (WWLLN): Update on new dataset and improved detection efficiencies, 32nd URSI General Assembly and Scientific Symp., Montreal, QC, Canada, EFGH28-1.

Rotunno, R. and R. Ferretti (2001). Mechanisms of intense alpine rainfall, J. Atmos. Sci., 58, 13, 1732-1749, doi:10.1175/1520-0469(2001)058<1732:MOIAR>2.0.CO;2.

Sari, F. P. and S. Lasher-Trapp (2025). Hailstorm events over a maritime tropical region: Storm environments and characteristics, J. Geophys. Res. Atmos., 130, e2024JD042718, doi:10.1029/2024JD042718.

Selga, M. (1929). Hail in the Philippines, Bureau of Printing, Manila, 20.

Schaefer, J. T. and J. Galway (1982). Population biases in tornado climatology, Preprints, 12th Conf. on Severe Local Storms, San Antonio, TX, 51-54.

Sharma, P. and S. Sen Roy (2023). Hailstorms over India during the summer season, Meteor. Atmos. Phys., 135, 41, doi:10.1007/s00703-023-00980-3.

Smith, B. T., R. L. Thompson, J. S. Grams, C. Broyles et al. (2012). Convective modes for significant severe thunderstorms in the contiguous United States. Part I: Storm classification and climatology, Wea. Forecasting, 27, 5, 1114-1135, doi:10.1175/WAF-D-11-00115.1.

Snyder, J. C., A. V. Ryzhkov, M. R. Kumjian, A. P. Khain et al. (2015). A ZDR column detection algorithm to examine convective storm updrafts, Wea. Forecasting, 30, 6, 1819-1844, doi:10.1175/WAF-D-15-0068.1.

Tan, M. L., D. Hoffmann, E. Ebert, A. Cui et al. (2022). Exploring the potential role of citizen science in the warning value chain for high impact weather, Front. Commun., 7, 2022, doi:10.3389/fcomm.2022.949949.

Tang, B. H., V. A. Gensini and C. R. Homeyer (2019). Trends in United States large hail environments and observations, npj Clim. Atmos. Sci., 2, 45, doi:10.1038/s41612-019-0103-7.

Taszarek, M. and L. Kolendowicz (2013). Sounding-derived parameters associated with tornado occurrence in Poland and Universal Tornadic Index, Atmos. Res., 134, 186-197, doi:10.1016/j.atmosres.2013.07.016.

Taszarek, M., H. E. Brooks and B. Czernecki (2017). Sounding-derived parameters associated with convective hazards in Europe, Mon. Wea. Rev., 145, 4, 1511-1528, doi:10.1175/MWR-D-16-0384.1.

Taszarek, M., J. T. Allen, T. Púčik, K. A. Hoogewind et al. (2020). Severe convective storms across Europe and the United States. Part II: ERA5 environments associated with lightning, large hail, severe wind, and tornadoes, J. Climate, 33, 23, 10263-10286, doi:10.1175/JCLI-D-20-0346.1.

Taszarek, M., N. Pilguj, J. T. Allen, V. Gensini et al. (2021a). Comparison of convective parameters derived from ERA5 and MERRA-2 with rawinsonde data over Europe and North America, J. Climate, 34, 8, 3211-3237, doi:10.1175/JCLI-D-20-0484.1.

Taszarek, M., J. T. Allen, M. Marchio et al. (2021b). Global climatology and trends in convective environments from ERA5 and rawinsonde data, npj Clim. Atmos. Sci., 4, 35, doi:10.1038/s41612-021-00190-x.

Taszarek, M., J. T. Allen, H. E. Brooks, N. Pilguj et al. (2021c). Differing trends in United States and European severe thunderstorm environments in a warming climate, Bull. Amer. Meteor. Soc., 102, 2, E296-E322, doi:10.1175/BAMS-D-20-0004.1.

Tippett, M. K., A. H. Sobel, S. J. Camargo and J. T. Allen (2014). An empirical relation between U.S. tornado activity and monthly environmental parameters, J. Climate, 27, 8, 2983-2999, doi:10.1175/JCLI-D-13-00345.1.

Thompson, R. L., R. Edwards, J. A. Hart, K. L. Elmore et al. (2003). Close proximity soundings within supercell environments obtained from the Rapid Update Cycle, Wea. Forecasting, 18, 6, 1243-1261, doi:10.1175/1520-0434(2003)018<1243:CPSWSE>2.0.CO;2.

Thompson, R. L., B. T. Smith, J. S. Grams, A. R. Dean et al. (2012). Convective modes for significant severe thunderstorms in the contiguous United States. Part II: Supercell and QLCS tornado environments, Wea. Forecasting, 27, 5, 1136-1154, doi:10.1175/WAF-D-11-00116.1.

Thompson, R. L., B. T. Smith, A. R. Dean and P. T. Marsh (2013). Spatial distributions of tornadic near-storm environments by convective mode, Electron. J. Severe Storms Meteor., 8, 5, doi:10.55599/ejssm.v8i5.50.

Trapp, R. J., N. S. Diffenbaugh, H. E. Brooks, M. E. Baldwin et al. (2007). Changes in severe thunderstorm environment frequency during the 21st century caused by anthropogenically enhanced global radiative forcing, Proc. Natl. Acad. Sci. USA, 104, 50, 19719-19723, doi:10.1073/pnas.0705494104.

Van Den Broeke, M. S. (2020). A preliminary polarimetric radar comparison of pretornadic and nontornadic supercell storms, Mon. Wea. Rev., 148, 1567-1584, doi:10.1175/MWR-D-19-0296.1.

Virman, M., M. Bister, J. Räisänen, V. A. Sinclair et al. (2021). Radiosonde comparison of ERA5 and ERA-Interim reanalysis datasets over tropical oceans, Tellus A, 73, 1, 1929752, doi:10.1080/16000870.2021.1929752.

Wakimoto, R. M. (1985). Forecasting dry microburst activity over the High Plains, Mon. Wea. Rev., 113, 7, 1131-1143, doi:10.1175/1520-0493(1985)113<1131:FDMAOT>2.0.CO;2.

Warren, R. A., H. Richter and R. L. Thompson (2021). Spectrum of near-storm environments for significant severe right-moving supercells in the contiguous United States, Mon. Wea. Rev., 149, 10, 3299-3323, doi:10.1175/MWR-D-21-0006.1.

Weisman, M. L. and J. B. Klemp (1982). The dependence of numerically simulated convective storms on vertical wind shear and buoyancy, Mon. Wea. Rev., 110, 6, 504-520, doi:10.1175/1520-0493(1982)110<0504:TDONSC>2.0.CO;2.

Zhou, Z., Q. Zhang, J. T. Allen, X. Ni et al. (2021). How many types of severe hailstorm environments are there globally?, Geophys. Res. Lett., 48, e2021GL095485, doi:10.1029/2021GL095485.

***CORRESPONDING AUTHOR: Generich H. CAPULI**,
Philippine Atmospheric, Geophysical and Astronomical Services Administration (PAGASA),
Research and Development and Training Division, Numerical Modelling Section, Quezon City, Philippines
e-mail: genhcapuli@rtu.edu.ph/genecap89@gmail.com
https://orcid.org/0000-0003-1253-7043